\documentclass[final,5p,times,twocolumn]{elsarticle}
\usepackage[utf8]{inputenc}
\usepackage{subfig}
\usepackage{multirow}
\usepackage{amsmath}
\usepackage{changepage}
\usepackage{algorithm,algorithmic}
\usepackage[T1]{fontenc}
\usepackage{hyperref}
\usepackage[table,xcdraw]{xcolor}
\usepackage{soul}
\hypersetup{
    colorlinks=true,
    linkcolor=red,
    filecolor=magenta,      
    urlcolor=blue,
    citecolor=green,
}
\usepackage{soul}
\usepackage{svg}
\newcommand{\orcid}[1]{\href{https://orcid.org/#1}{\includegraphics[scale=0.01]{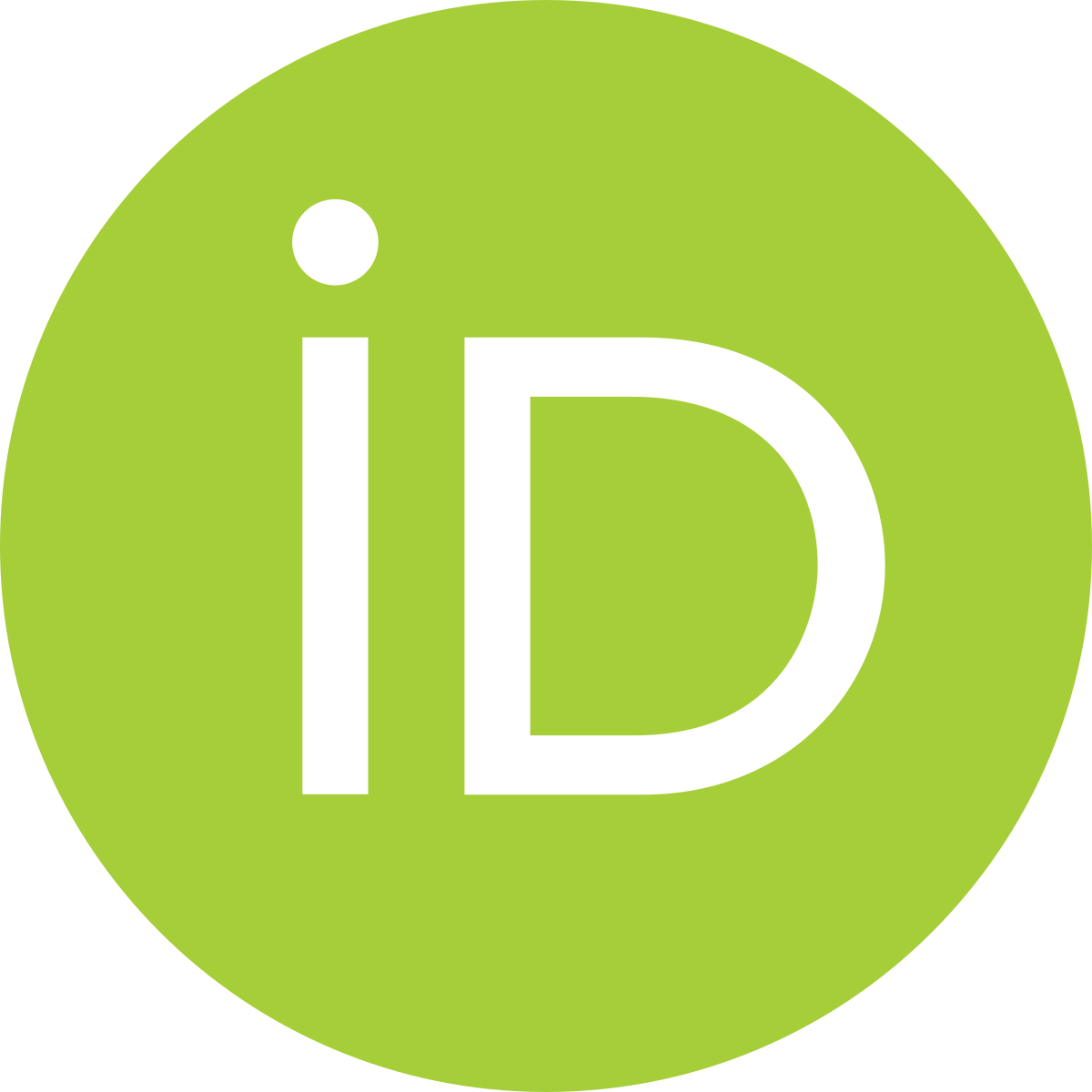}}}

\begin{document}

\begin{frontmatter}

\title{Segmentation of Brain MRI using an Altruistic Harris Hawks' Optimization algorithm\fnref{f1}}
\fntext[f1]{Article published in \url{https://doi.org/10.1016/j.knosys.2021.107468}}

\author[a1]{Rajarshi Bandyopadhyay \orcid{0000-0003-1379-4738}}
\author[a2]{Rohit Kundu \orcid{0000-0001-8665-8898}}
\author[b,c]{Diego Oliva \orcid{0000-0001-8781-7993}\corref{cor1}}
\cortext[cor1]{Corresponding author}
\ead{diego.oliva@cucei.udg.mx}
\author[a1]{Ram Sarkar \orcid{0000-0001-8813-4086}}

\address[a1]{Department of Computer Science \& Engineering, Jadavpur University, INDIA}
\address[a2]{Department of Electrical Engineering, Jadavpur University, INDIA}
\address[b]{Depto. de Ciencias Computacionales, Universidad de Guadalajara, CUCEI, Guadalajara, Mexico}

\address[c]{School of Computer Science \& Robotics, Tomsk Polytechnic University, Tomsk, Russia}

\section*{Segmentation of Brain MRI using an Altruistic Harris Hawks' Optimization algorithm}

\textbf{Rajarshi Bandopadhyay}
\begin{center}
    Department of Computer Science and Engineering, Jadavpur University\\
    188, Raja S.C. Mullick Road, Kolkata-700032, West Bengal, INDIA\\
    Email: \href{mailto:rajarshibanerjee03@gmail.com}{rajarshibanerjee03@gmail.com}
\end{center}

\textbf{Rohit Kundu}
\begin{center}
    Department of Electrical Engineering, Jadavpur University\\
    188, Raja S.C. Mullick Road, Kolkata-700032, West Bengal, INDIA\\
    Email: \href{mailto:rohitkunduju@gmail.com}{rohitkunduju@gmail.com}
\end{center}

\textbf{Diego Oliva*}
\begin{center}
    Depto. Universidad de Guadalajara, CUCEI,\\
    Av. Revolucion 1500, Guadalajara, Jal, MEXICO\\
     Email: \href{mailto:diego.oliva@cucei.udg.mx}{diego.oliva@cucei.udg.mx}\\
    and\\
    School of Computer Science \& Robotics,\\
    Tomsk Polytechnic University, Tomsk, RUSSIA\\
   \end{center}

\textbf{Ram Sarkar}
\begin{center}
    Department of Computer Science and Engineering, Jadavpur University\\
    188, Raja S.C. Mullick Road, Kolkata-700032, West Bengal, INDIA\\
    Email: \href{mailto:ramjucse@gmail.com}{ramjucse@gmail.com}
\end{center}

\begin{center}
    *Corresponding author: Diego Oliva\\
    *Corresponding author email: \href{mailto:diego.oliva@cucei.udg.mx}{diego.oliva@cucei.udg.mx}
\end{center}

\begin{abstract}
\textcolor{black}{Segmentation is an essential requirement in medicine when digital images are used in illness diagnosis, especially, in posterior tasks as analysis and disease identification. An efficient segmentation of brain Magnetic Resonance Images (MRIs) is of prime concern to radiologists due to their poor illumination and other conditions related to de acquisition of the images.} Thresholding is a popular method for segmentation that uses the histogram of an image to label different homogeneous groups of pixels into different classes. However, the computational cost increases exponentially according to the number of thresholds. In this paper, we perform the multi-level thresholding using an evolutionary metaheuristic. It is an improved version of the Harris Hawks Optimization (HHO) algorithm that combines the chaotic initialization and the concept of altruism. Further, for fitness assignment, we use a hybrid objective function where along with the cross-entropy minimization, we apply a new entropy function, and leverage weights to the two objective functions to form a new hybrid approach. The HHO was originally designed to solve numerical optimization problems.  Earlier, the statistical results and comparisons have demonstrated that the HHO  provides very promising results compared with well-established metaheuristic techniques. In this article, the altruism has been incorporated into the HHO algorithm to enhance its exploitation capabilities. \textcolor{black}{We evaluate the proposed method over 10 benchmark images from the WBA database of the Harvard Medical School and 8 benchmark images from the Brainweb dataset using some standard evaluation metrics. On the Harvard WBA dataset, a Peak Signal to Noise Ratio (PSNR) of 26.61 and a Structural Similarity Index (SSIM) of 0.92 are achieved using 5 thresholds. For the same scenario, using the Brainweb dataset, a PSNR of 24.77 and SSIM of 0.86 are obtained.} The obtained results justify the superiority of the proposed approach compared to existing state-of-the-art methods and baseline methods. The relevant codes for the proposed approach are available at: \url{https://github.com/Rohit-Kundu/Segmentation-HHO_Altruism}.
\end{abstract}

\begin{keyword}
Magnetic Resonance Imaging (MRI) \sep Brain MRI Segmentation \sep Metaheuristic \sep Altruism \sep Hybrid Objective Function
\end{keyword}
\end{frontmatter}

\section{Introduction}\label{intro}
Recent advances in biomedical technologies allow the acquisition of medical images in different modalities, like Computer Tomography, Ultrasound and Magnetic Resonance Imaging (MRI). Brain MRI provides a high-resolution spatial image of the brain through a non-invasive test detailing its anatomical features. However, analyzing brain MRIs is an esoteric and time-consuming task that requires expert clinicians to manually extract important information. Moreover, such a method is prone to subjective variability which might delay or provide a wrong diagnosis of an obscure disease. Computer-Aided Diagnosis (CAD) tools are alternative solutions that can significantly reduce the time for such insidious disease identification.

Image segmentation is one of the most important preprocessing steps for medical image diagnosis \cite{pham2000current, feng2017multi} which separates the object of interest from the whole image required for analyzing behavioural and morphological changes of the object to detect the disease. Image segmentation, in essence, classifies each pixel in an image into homogeneous groups, each of them shares some common characteristic such as intensity or texture. Brain MRI segmentation has influential clinical implications for early disease detection and prognosis. Accurate MRI segmentation leads to the visualization of different brain structures to delineate lesions, analyze brain development, plan surgeries and identify diseases such as dementia, schizophrenia, multiple sclerosis, and Alzheimer etc.

Thresholding is a simple, but efficient image segmentation procedure where the histogram of a grayscale image is used to distribute its pixels into different classes. For this, a distinct threshold ($th$) is set for establishing intensity values. The thresholding process can be distinguished into two categories, namely, binary thresholding, which uses a single $th$ value and multi-level thresholding, which uses multiple $th$ values. Binary (bi-level) thresholding is used for object recognition tasks, where the objects need to be separated from their background, whereas multi-level thresholding divides parts of an image into several classes. Thresholding methods can also be categorized into parametric and non-parametric approaches. Parametric approaches aim to estimate the parameters from a probability density function to separate the classes, while non-parametric techniques use intrinsic features of the image such as entropy and variance, which are optimized to find the ideal threshold values.

Unsupervised segmentation of brain MRIs is a popular area of research since in the biomedical domain, acquiring ground truth information for the images is difficult since it requires the expertise of trained medical professionals and preparation of ground truth is very tiresome. Baur et al. \cite{baur2021modeling} used spatial auto-encoders with skip-connections for the unsupervised anomaly detection from brain MRI images. Singh et al. \cite{singh2021unsupervised} proposed the ``local Zernike moment and unbiased non-local means-based bias-corrected fuzzy C-means" method for unsupervised segmentation of brain MRIs which could deal with both noisy images and images containing intensity non-homogeneity artefacts. Bercea et al. \cite{bercea2021feddis} proposed a federated-learning based model, named ``FedDis" for disentangling the parameter space into appearance and shape and share the shape parameter with the different parties which collaboratively train a machine learning model.

Several thresholding methods have also been proposed in the related literature like \cite{abutaleb1989automatic, chang1997fast, kittler1986minimum, oliva2020combining}. Otsu's thresholding method \cite{otsu1979threshold} is popularly used for histogram thresholding which uses the between-class variance for computing the optimal threshold value. Kapur's entropy \cite{kapur1985new} thresholding method maximizes the entropy of the object and the background pixels to compute the optimal threshold. The minimum error thresholding method proposed by Cho et al. \cite{cho1989improvement}, assumes that the object and background pixels are normally distributed, thus defining and optimizing an objective function related to the Bayes risk. However, in such classical methods for threshold determination, the computational cost increases exponentially with the increase in thresholding levels \cite{oliva2019multilevel}. Thus these classical methods are suitable only for a small number of thresholding levels like binary thresholding or up to two thresholds.

As an alternative to the classical methods, evolutionary meta-heuristics and swarm intelligence based algorithms have been widely used to address the multi-level thresholding problem \cite{akay2013study, sathya2011optimal, song2021mppcede,singh2021nature}. For example, Genetic Algorithm (GA) has been used by Lai et al. \cite{lai2004hybrid} where GA was coupled with Gaussian models, and Yin et al. \cite{yin1999fast} proposed a learning strategy for faster convergence to compute the optimal threshold values. Aranguren et al. \cite{aranguren2021improving} used the Success History-based Adaptive Differential Evolution with a linear population size reduction (L-SHADE) algorithm and Oliva et al. \cite{oliva2017cross} used the non-greedy Crow Search Algorithm \cite{askarzadeh2016novel} for multi-level thresholding.

\textcolor{black}{Deng et al.} \cite{deng2021quantum} \textcolor{black}{incorporated the divide-and-conquer method of the cooperative co-evolution evolutionary algorithm and quantum computing attributes of the quantum evolutionary algorithm into the Differential Evolution algorithm to improve the diversity of its solutions, augment the convergence speed, and get over the low solution efficiency of the algorithm. Deng et al.} \cite{deng2020enhanced} \textcolor{black}{introduced a new strategy for differential mutation of the difference vector to augment the search capacity as well as the descent capacity of the quantum-inspired differential evolution algorithm. Cai et al.} \cite{cai2021improved} \textcolor{black}{combined the approaches of Hamming adaptive rotation, random rotation direction as well as co-operative co-evolution to address the shortcomings of the quantum-inspired evolutionary algorithm such as the poor convergence speed, below par global search capacity, and complex designing of the rotation angle. Deng et al.} \cite{deng2020novel} \textcolor{black}{proposed a three-objective model for gate allocation to address the quick growth of the air traffic demands and prevent the gate resource from limiting airport development. The authors have introduced an improved version of the quantum-inspired evolutionary algorithm incorporating the niche co-evolution approach as well as the enhanced Particle Swarm Optimization to solve the gate allocation model.}

Significant contributions have been made in the literature for brain MRI thresholding segmentation using evolutionary meta-heuristics. For example, Kaur et al. \cite{kaur2016optimized} used Particle Swarm Optimization (PSO) with a two-dimensional minimum cross-entropy, and Ramakrishnan et al. \cite{ramakrishnan2017professional} used the Grey Wolf Optimization (GWO) algorithm for optimal threshold selection. Tang et al. \cite{tang2017improved} applied Bacterial Foraging Optimization (BFO) with Tsallis entropy for multi-level thresholding of brain MRIs, Kadry et al. \cite{kadry2021evaluation} used a modified Moth-flame Optimization (MFO) algorithm. Panda et al. \cite{panda2021novel} recently proposed a new technique based on a hybrid of Adaptive Cuckoo Search and Squirrel Search algorithm (ACS-SSA), using a row-class entropy method for the thresholding based segmentation of brain MRIs, wherein they aimed to preserve the spatial information of the images.

In this paper, we propose an unsupervised brain MRI segmentation framework employing multi-level thresholding using the Harris Hawks Optimization (HHO) algorithm \cite{heidari2019harris} embedded with chaotic initialization and the concept of altruism. The proposed mechanism does not require any ground truth to extract the objects contained in the image. The proposed algorithm uses a hybrid of objective functions, where along with the minimum cross-entropy  \cite{li1993minimum}, we propose another entropy-based objective function and assign weights to the two criteria for asserting fitness to the Harris Hawks population in each generation. HHO is inspired by how Harris Hawks chase down prey and the cooperation between them. HHO can emulate myriad chasing patterns based on many dynamic scenarios and numerous escaping patterns of the prey. However, it suffers from local optimum drawbacks. To counter this, altruism has been incorporated which elevates the exploitation capabilities of the algorithm. This prevents the algorithm from getting stuck in some local optimum and thus enhances the exploitation property of the algorithm. Also, to increase the diversity of the solutions, chaotic initialization has been done to the feature vectors using Logistic chaotic maps. Along with the hybrid objective function, this algorithm yields much better solutions (measured in terms of performance metrics such as Peak Signal to Noise Ratio or Structural Similarity Index Metric) concerning the other meta-heuristics used earlier in this domain.

\subsection{Motivation and Contributions}
Several meta-heuristic algorithms have been proposed for solving medical image segmentation problems over the years. Each one of the algorithms has its own merits and demerits while dealing with a particular problem at hand. The performance depends largely on the exploration and exploitation capabilities of the method under consideration. Most of the time, new concepts are incorporated into the algorithm to deal with the shortcomings of the basic version of the algorithm. For example, a method that enhances the exploitation capability can be incorporated into an existing algorithm having good exploration capability to balance both the capabilities. This has encouraged researchers to improve upon the previous work that was done and incorporate new techniques into the existing ones. In this paper, we have proposed the altruism that can be incorporated into the existing HHO to enhance its exploitation. Also, a new fitness function has been introduced that gives better results in terms of some standard metrics for image thresholding. Previously, HHO and other optimization algorithms have been used in the domain of image segmentation \cite{rodriguez2020efficient}. The modifications that are incorporated into the basic HHO yields significantly better results than many of the already proposed algorithms. Thus, combining the HHO algorithm with the altruism concept and chaotic maps can lead to increased exploitation of the search space and an overall improvement in the performance of the hybrid algorithm. 

\textcolor{black}{In a nutshell, the main contributions of this work are listed as follows:}
\begin{enumerate}
    \item \textcolor{black}{We incorporate the concept of altruism into the basic HHO algorithm for the first time to the best of our knowledge.}
    
    \item \textcolor{black}{We introduce a new entropy function and use a hybrid of this with the minimum cross-entropy as the objective function for the HHO algorithm.}
    
    \item \textcolor{black}{We make use of the Logistic chaotic map for initialization of Harris Hawks in the HHO algorithm to increase the diversity of the solutions.}
    
    \item \textcolor{black}{We validate the effectiveness of the proposed algorithm by applying it on 8 standard brain MRIs from the Brainweb dataset} \cite{cocosco1997brainweb} \textcolor{black}{and 10 MRIs from the Harvard Whole Brain Atlas (WBA) dataset} \cite{sutton1999whole}.
    
    \item \textcolor{black}{We compare the proposed method with many state-of-the-art metaheuristic algorithms and conventional methods in the literature for unsupervised brain MRI segmentation which shows that our method outperforms those, justifying the effectiveness of the approach.}
\end{enumerate}

The rest of the paper has been organized as follows: Section~\ref{sec_thresh} describes the multi-level thresholding using a hybrid of two objective functions; Section~\ref{sec_hho} describes the improved Harris Hawks Optimization algorithm used in this research; Section~\ref{results} evaluates the performance of the proposed method on two publicly available brains MRI datasets and finally, Section~\ref{conclusions} concludes the findings from this research.

\section{Multi-level Thresholding}\label{sec_thresh}
Multi-level thresholding segments a grayscale image into several distinct regions by determining more than one threshold value. The image is divided into separate homogeneous regions which correspond to a background and more than one object. This is useful for objects with coloured or complex backgrounds on which binary thresholding fails to produce satisfactory results.

\textcolor{black}{The multi-level thresholding problem can be defined as follows: a gray-scale image $I$ consists of $N+1$ number of classes and requires $N$ thresholds to split the image into segments.} \autoref{mth} \textcolor{black}{mathematically formulates this problem, where $C_p$ denotes the $p^{th}$ class of the image $I$; $th_p$ represents the threshold value for $p \in \{1, 2 , 3, \hdots, N\}$; $g(x,y)$ represents the gray-level intensity of $(x,y)$ and $G$ represents the gray-level intensity of $I$ in the range $(1,2,\hdots,G)$.}

\begin{equation}\label{mth}
    \begin{split}
        C_0 = \{g(x,y) \in I\text{  } |\text{  } 0\leq g(x,y) \leq th_1-1\}\\
        C_1 = \{g(x,y) \in I\text{  } |\text{  } th_1\leq g(x,y) \leq th_2-1\}\\
        C_2 = \{g(x,y) \in I\text{  } |\text{  } th_2\leq g(x,y) \leq th_3-1\}\\
        \vdots\\
        C_N = \{g(x,y) \in I\text{  } |\text{  } th_N\leq g(x,y) \leq G\}\\
    \end{split}
\end{equation}

Thus, the main objective of multi-level thresholding is to find the threshold values that divide the pixels of an image into sub-parts or segments.

Depending on the intensity levels in an input image, entropy establishes an index of statistical diversity. Entropy-based thresholding is a robust method, since the selected threshold is determined based on a global and objective property of the image histogram, and is independent of small variations in the input image \cite{pun1980new}. Thus entropy-based thresholding method is preferred in the present research for the brain MRI segmentation problem. For this, we introduce a new entropy function and formulate a hybrid of this new function with the traditional minimum cross-entropy to form the final objective function that is used for the fitness assignment of the Harris Hawks in the HHO algorithm.

\subsection{Minimum Cross-entropy}
Entropy computes the uncertainty related to a particular dataset and the cross-entropy \cite{kullback1997information} thesholding formulates the optimal thresholding value as the minimization of a data theoretic distance ($D$). If $X = \{x_1,x_2,x_3,\hdots,x_Z\}$ and $Y = \{y_1,y_2,y_3,\hdots,y_Z\}$ be two probability distributions on the same space, then the information theoretic distance $D(X,Y)$ between them is given by \autoref{dxy}.

\begin{equation}\label{dxy}
    D(X,Y) = \sum_{z=1}^{Z} x_z \ln \left(\frac{x_z}{y_z}\right)
\end{equation}

The minimum cross-entropy \cite{li1993minimum} computes the threshold by minimizing the cross-entropy between the input grayscale image $I$ with histogram $h_I^{r} = 1,2,\hdots, G$ ($G$ being the number of values of grey intensities present in the image) and the thresholded image $I_{th}$ employing the threshold $th$ as a single threshold value (for binary thresholding) that segments image $I$ into two distinct areas: foreground and background. This thresholded image $I_{th}$ is given by \autoref{thresh}.

\begin{equation}\label{thresh}
    I_{th}(p,q) = \begin{cases}
    \mu(1,th), \text{  } I(p,q)<th\\
    \mu(th,G+1), \text{  } I(p,q)\geq th
    \end{cases}
\end{equation}

where $\mu(th_a,th_b)$ is given by \autoref{mu}.

\begin{equation}\label{mu}
    \mu(th_a,th_b) = \frac{\sum_{i = th_a}^{th_b-1} i\times h_I^{r}(i)}{\sum_{i = th_a}^{th_b-1} h_I^{r}(i)}
\end{equation}

The cross-entropy objective function ($F_{CE}^{binary}$) for the binary thresholding scenario can thus be expressed as \autoref{bth}.

\begin{equation}\label{bth}
\begin{split}
    F_{CE}^{binary}(th) = \sum_{i=1}^{th-1}i\times h_I^{r}(i)\times \ln \left( \frac{i}{\mu(1,th)} \right) \\ 
    + \sum_{i=th}^{G}i\times h_I^{r}(i)\times \ln \left( \frac{i}{\mu(th,G+1)} \right)
\end{split}
\end{equation}

Extending, \autoref{bth} to the multi-level thresholding problem, the objective function ($F_{CE}^{multi}$) is given as in \autoref{multith}.
\begin{equation}\label{multith}
\begin{split}
    F_{CE}^{multi}(th) = \sum_{i=1}^{G}i\times h_I^{r}(i)\times \ln \left(i\right) - \sum_{i=1}^{th-1}i\times h_I^{r}(i)\times \ln \left[ \mu(1,th) \right] \\
    - \sum_{i=th}^{G}i\times h_I^{r}(i)\times \ln \left[ \mu(th,G+1) \right]
\end{split}
\end{equation}

This multi-level objective function has been formulated based on the use of a vector \textbf{th}$=\left[th_1,th_2,\hdots,th_N\right]$ consisting of $N$ different thresholding values. Thus the multi-level thresholding objective function can be expressed as \autoref{multi_vec}.
\begin{equation}\label{multi_vec}
    F_{CE}^{multi}(\textbf{th}) = \sum_{i=1}^{G}i\times h_I^{r}(i)\times \ln \left(i\right) - \sum_{i=1}^{N}S_i
\end{equation}
where $S_i$ denotes the entropy for threshold value $th_i$ and is computed by \autoref{entropy}.
 
\begin{equation}\label{entropy}
\begin{split}
     S_1 = \sum_{i = 1}^{th_1-1} i\times h_I^{r}(i)\times \ln \left[ \mu(1,th_1) \right] \\
     S_i = \sum_{i = th_{i-1}}^{th_i-1} i\times h_I^{r}(i)\times \ln \left[ \mu(th_{i-1},th_i) \right], \text{  }\forall 1<i<N \\
     S_N = \sum_{i = th_N}^{G} i\times h_I^{r}(i)\times \ln \left[ \mu(th_N,G+1) \right]
\end{split}
\end{equation}

\subsection{Proposed Entropy Function}
The formulation of the objective function is similar to the minimum cross-entropy explained in the previous section, with a change only in the entropy expression. The proposed entropy function (PEF) is mathematically formulated in \autoref{prop_entropy}.

\begin{equation}\label{prop_entropy}
\begin{split}
     S_1^{PEF} = -\sum_{i = 1}^{th_1-1}i\times\bigg\{ h_I^{r}(i)\times \left[1-\mu(1,th_1)\right] \\
     + \ln \left[ h_I^{r}(i) \right] \times \left(1-\ln \left[\mu(1,th_1)\right] \right) \bigg\}\\
     S_i^{PEF} = -\sum_{i = th_{i-1}}^{th_i-1} i\times\bigg\{ h_I^{r}(i)\times\left[1- \mu(th_{i-1},th_i) \right]\\
     + \ln \left[h_I^{r}(i) \right]\times\left( 1-\ln \left[\mu(th_{i-1},th_i)\right] \right)\bigg\}, \text{  }\forall 1<i<N \\
     S_N^{PEF} = -\sum_{i = th_N}^{G}i\times\bigg\{ h_I^{r}(i)\times \left[1-\mu(th_N,G+1)\right] \\
     + \ln \left[ h_I^{r}(i) \right] \times \left(1-\ln \left[\mu(th_N,G+1)\right] \right) \bigg\}\\
\end{split}
\end{equation}

Thus, the proposed multi-level thresholding objective function ($F_{PEF}^{multi}$) for threshold vector \textbf{th} $=\left[ th_1,th_2,th_3,\hdots,th_N \right]$ is given by \autoref{newobj}.

\begin{equation}\label{newobj}
    F_{PEF}^{multi}(\textbf{th}) = \sum_{i=1}^{G}i\times h_I^{r}(i)\times \ln \left(i\right) - \sum_{i=1}^{N}S_i^{PEF}
\end{equation}

\subsection{Hybrid Objective Function}
A hybrid objective function performs more robustly than its constituent objective functions since they get simultaneously optimized. The hybrid objective function used in this research for the fitness assignment of the Harris hawks in the HHO algorithm is formulated as in \autoref{finalobj}, where $\alpha$ and $\beta$ are weights associated with the two objective functions.

\begin{equation}\label{finalobj}
     F_{obj}^{multi} = \alpha F_{CE}^{multi} + \beta F_{PEF}^{multi}
\end{equation}

\section{Harris Hawks Optimization algorithm}\label{sec_hho}

\begin{figure*}
    \centering
    \resizebox{0.8\textwidth}{!}{
    \includegraphics{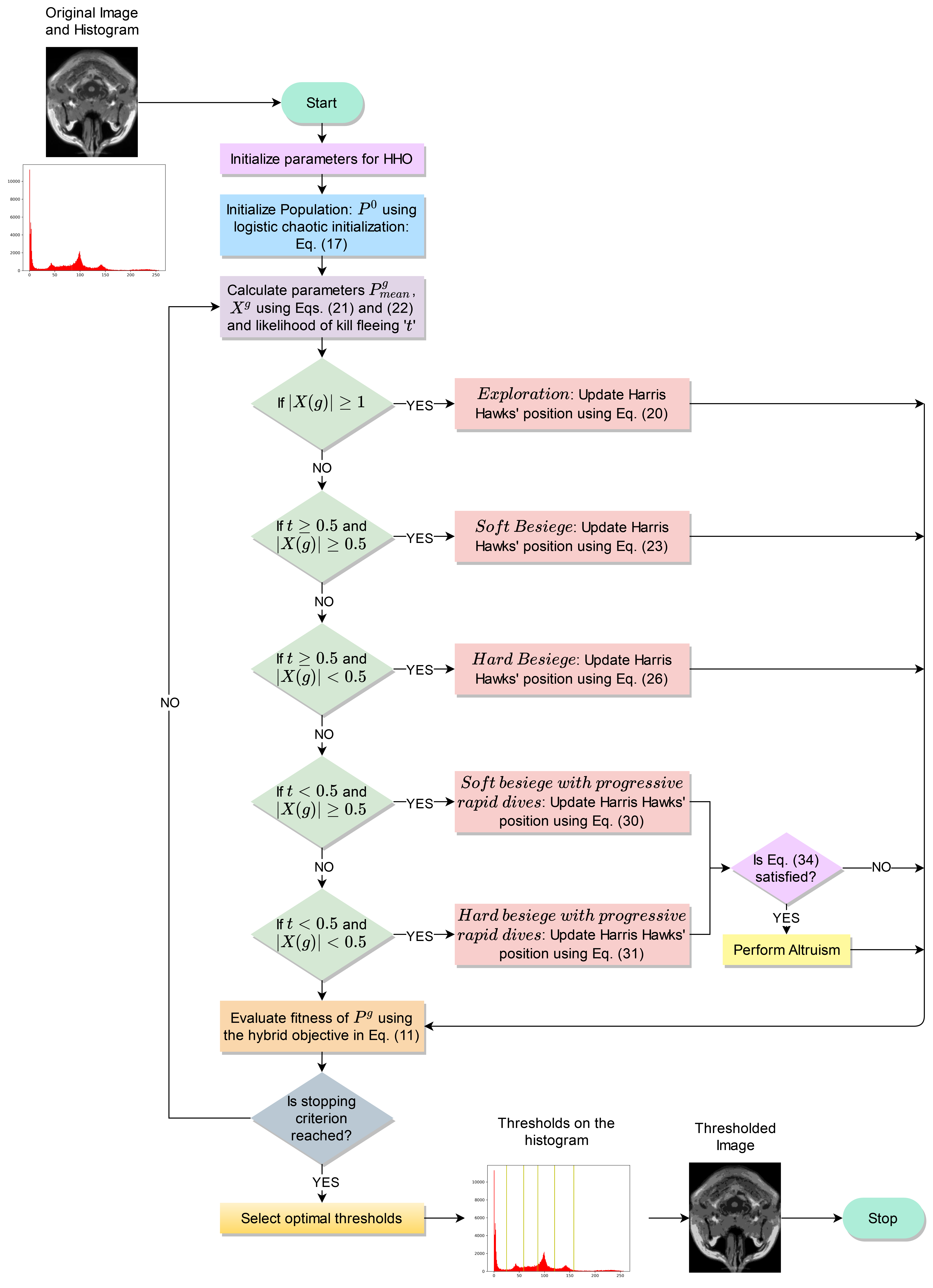}
    }
    \caption{\textcolor{black}{Flowchart of the proposed improved HHO algorithm used in this research for brain MRI segmentation using image thresholding.}}
    \label{fig:hho_flowchart}
\end{figure*}

The HHO algorithm \cite{heidari2019harris} is a population-based, gradient-free optimization approach. It is inspired by how Harris Hawks' track prey in the living world. The Harris Hawks' demonstrate a special kind of approach to hunt the prey named surprise pounce. This algorithm portrays the separate exploration and exploitation strategies of the Harris Hawks, which are affected by exploring prey, the surprise pounce or the seven kills, and finally attacking the prey in a trademark style. A thorough mathematical model is portrayed to recreate how Harris Hawks' perform hunting in nature. The myriad moves that this algorithm consists of, to imitate the pouncing technique of Harris Hawks' are stated hereafter considering the problem of brain MRI segmentation. \textcolor{black}{The flowchart of the proposed improved HHO algorithm used in this study is shown in} \autoref{fig:hho_flowchart}.

\subsection{Initialization: Chaotic Map}
Each of the Harris Hawks is denoted by a feature vector. The dimension of each of the Harris Hawks is equal to one added to the number of thresholds that are selected for the process of segmentation. For the first feature vector, the initialization is done randomly given that those elements of the feature vector are assigned values within the pixel range of the grayscale images. For the other vectors, the initialization is done using chaotic maps \cite{Ewees2020, SMenesy2020} and it is ensured that the values lie in the pixel range of the grayscale images as before. A chaotic map is used while doing so. The other parameters are initialized to some starting value after tuning. 

% Chaotic initialization is done when allocating values to the vectors initially. The feature vectors are defined using the Logistic chaotic map. The first feature vector is initialised randomly and the other feature vectors are derived from the first one by using the Logistic chaotic map. Additionally, all the other parameters are set to some starting values.

% The Chaotic maps diversify the range of solutions that we are looking for, to near the value of the optimal fitness. 

%Using Chaotic maps \cite{Ewees2020}\cite{SMenesy2020} to look for global optimum solutions aids in enhancing the diversity of the solutions.

During chaotic initialization, let $f_v$ be the representation of the $v^{th}$ feature vector. So, the formulae for deriving $f_{v+1}$, the representation of the $(v+1)^{th}$ feature vector, if the value of the $v^{th}$ feature vector is given by the following: 
\begin{enumerate}
    \item Sine Chaotic Map: 
        \begin{equation} 
         f_{v+1} = \frac{k}{4} sin(\pi f_v)\, ,\, k=4  
        \end{equation} 
    
    \item Singer Chaotic Map:
        \begin{gather}
         f_{v+1} = \mu( 7.86 f_v - 23.31 {f_v}^2 + 28.75 f_v ^3 - 13.302875 f_v ^4)\\
         \nonumber \mu= 1.07      
        \end{gather} 
         
    \item Sinusoidal Chaotic Map:
        \begin{equation} 
         f_{v+1} = c f_v ^2 sin(\pi f_v)\, ,\, c=2.3  
        \end{equation} 
    
    \item Chebyshev Chaotic Map:
        \begin{equation} 
         f_{v+1} = cos( \arccos{f_v} )  
        \end{equation} 
    
    \item Tent Chaotic Map:
        \begin{equation} 
         f_{v+1} = 
         \begin{cases}
          \frac{f_v}{0.7} & f_v < 0.7 \\
          \frac{10}{3} (1-f_v) & f_v \geq 0.7
        \end{cases}
        \end{equation} 
    
    \item Logistic Chaotic Map:
        \begin{equation} \label{eqa1}
         f_{v+1} = c f_v (1- f_v)\, ,\, c=4  
        \end{equation} 
    
    \item Iterative Chaotic Map:
        \begin{equation} 
         f_{v+1} =  sin(\frac{c \pi}{f_v})\, ,\, c=0.7  
        \end{equation} 
    
    \item Gauss/Mouse Chaotic Map:
        \begin{equation} 
        f_{v+1} = e^{-\alpha f_v ^2}+\beta , \alpha = 4.90, \beta = -0.58
        %  \begin{cases}
        %   1 & f_v = 0 \\
        %   \frac{1}{mod(f_v)} & otherwise
        % \end{cases}
        \end{equation} 

\end{enumerate}
The method of chaotic initialization is dynamic. It is one of the most modern methods to search for the global optimum solution in a search space. 
The choice of a particular chaotic map has been justified in \autoref{chaotic_map_table} and its corresponding reference in the results section.

\subsection{Exploration}
The Harris Hawks are one of the most clever and sharp birds and can strategically track down a target with their keen eyes and astute moves. The exploration is a phase of the HHO algorithm, where each of the Harris Hawks is considered a viable solution to the problem of identifying thresholds to optimize the value returned by the cost function when that particular feature vector is passed. From the current set of Harris Hawks, the prey is selected to be the one that yields the least value when passed to the objective function (the best solution among the present set of solutions is the one with minimum fitness value). To mimic the exploration phase of the Harris Hawks, there exist two plausible ways.  

%In this phase, all the Harris Hawks are considered as possible solutions to the FS problem. The Harris Hawks' are one of the most intelligent birds and they can effortlessly trace a prey with their keen eyes but at times, the prey cannot be seen. 
%In HHO, each Harris Hawk is considered a possible solution and the prey is taken to be the Harris Hawk which yields the minimum value when passed to the objective function (the best possible solution amongst the current set of Harris Hawks, one which yields maximum classification accuracy selecting the minimum number of features possible). There are two possible ways to imitate the exploration strategy of Harris Hawks.
%formula
\begin{equation}\label{eqa2}
P(g+1)=\begin{cases}
    P_{rand}(g)-e_1|P_{rand}(g) - 2e_2 P(g)|,\\ when\; r\geq 0.5 \\
    (P_{prey}(g)-P_{mean}(g))-e_3(LL+e_4(UL-LL)),\\when\; r < 0.5
\end{cases}
\end{equation}

\textcolor{black}{In} \autoref{eqa2}, \textcolor{black}{$P(g)$ stands for the position of the Harris Hawk in the $g^{th}$ iteration (or generation), where $g=\{1,2,\hdots,g_{max}\}$, `$g_{max}$' being the maximum number of generations. $P(g+1)$ signifies the location of the Harris Hawk at the end of the $g^{th}$ iteration. The place where the prey is located is designated by $P_{prey}(g)$. $P_{rand}(g)$ is a haphazard solution selected from the present set of Harris Hawks.} Also, $r$, $e_1, e_2, e_3, e_4$ are randomly selected values within the range (0, 1). $LL$ and $UL$ designate the lower bound and the upper bound respectively of the solutions. \textcolor{black}{$P_{mean}$ is used to represent the mean or average position of the present set of Harris Hawks. $P_{mean}(i)$ is given by} \autoref{eqa3}.

%Here, $P(g)$ denotes the position of a Harris Hawk in the $i^{th}$ iteration. $P(g+1)$ is the new position of the Harris Hawk at the end of the $i^{th}$ iteration. $P_{prey}(t)$ is the location of the prey. $P_{rand}(t)$ signifies an arbitrary solution chosen from the current population of the Harris Hawks. Here, $r$, $e_1, e_2, e_3, e_4$ are numbers that are randomly chosen from the range (0, 1).$LL$ denotes the lower limit or bound and $UL$ stands for the upper limit or bound.$P_m$ stands for the average position of the current generation of hawks. And the formula for $P_m$ is: 

\begin{equation}\label{eqa3}
    P_{mean}(g) =\frac{1}{M} \sum^{M}_{j=1} P_j(g)
\end{equation}
In \autoref{eqa3}, $M$ stands for the size of the present generation of Harris Hawks and the location of the $j^{th}$ Harris Hawk in the $g^{th}$ iteration is denoted by $P_j(g)$.
%where the size of the current population is M and the position of every hawk in the $i^{th}$ iteration is given by $P_j(i)$.

The former viewpoint generates solutions to the image segmentation problem based on the arbitrarily selected Harris Hawks' position as well as the position of the present set of Harris Hawks. The latter gives rise to solutions depending on the target's location, the average or mean location of the set of Harris Hawks at the beginning of the present iteration and some factors which are randomly scaled between 0 and 1. Among these, $e_3$ acts as a scaling factor. As $e_4$ approaches 1, it increases the randomness of the method. An arbitrary length is added to the lower bound to make sure that possible solutions in the feature space are less prone to remaining undiscovered. 

%This first perspective gives rise to solutions based on the position of a randomly chosen hawk and the locations of other hawks. The second perspective initiates solutions based on the position of the prey, the mean position of the hawks at the beginning of the current iteration, and other arbitrary scaling factors. Here, $e_3$ is a scaling factor and as the value of $e_4$ tends towards 1, it aids in enhancing the randomness of this method. To the Lower Limit ($LL$), a random length is added. It ensures that more domains of the feature space are discovered.

\subsection{Changeover between Exploration and Exploitation}
HHO algorithm displays a changeover between the two phases: exploration and exploitation. It also displays myriad exploitative behaviour depending on the prey's energy to escape and its likelihood of escaping. The escaping energy of the prey goes on declining as it attempts to escape. \textcolor{black}{Let's assume that initially, the kill or the prey (the rabbit) has energy denoted by $X_0$, the present iteration be indicated by `$g$' and the net number of iterations by `$g_{max}$', then the escaping energy (say, $X(g)$) of the kill in the $g^{th}$ iteration is given by} \autoref{eqa4}.

%The HHO algorithm can exhibit a changeover between the exploration and exploitation aspects. It can also show alteration between various exploitative actions based on the escaping energy of the rabbit and the probability of the rabbit escaping. As the prey attempts to escape, its escaping energy diminishes. If the initial energy of the prey is depicted by $X_0$, the current iteration be denoted by m and the total iterations be denoted by $M$, then the escaping energy of prey in the $m^{th}$ iteration (say, X) can be depicted by

\begin{equation}\label{eqa4}
    X(g) = 2X_0 (1-\frac{g}{g_{max}} )
\end{equation}

$X_0$ lies in the scope (-1, 1) arbitrarily. As $X_0$ diminishes from 0 to -1, the kill gets devitalized and as $X_0$ increases from 0 to 1, the prey gets re-energized. Commonly, $X$ diminishes as the number of iterations increases. When $|X|\geq 1$ the algorithm goes for the exploration strategy and when $|X|<1$ the exploitation strategy is opted for. 

%The value of $X_0$ varies in the range (-1, 1) randomly. When the value of $X_0$ decreases from 0 to -1, it signifies that the prey gets tired and when it increases from 0 to 1, it means that the prey gets revitalised. Normally, $X$ decreases as the iterations progress. Exploration is favoured when $|X|\geq 1$ and exploitation is done when $|X|<1$. 

\subsection{Exploitation}
In the exploitation phase, the algorithm displays the surprise pounce (or the seven kills strategy \cite{bednarz1988cooperative}) by attacking the kill detected in the phase before. But the prey tries to escape from threatening situations. So, the Harris Hawks emulate myriad chasing styles in real-life situations. The neighbourhood of the Harris Hawks is exploited in this phase. Four different approaches are manifested based on the position of the kill detected in the stage before. 

%In this stage of the HHO, the neighbourhood of the feature vectors are exploited. The Harris Hawks' execute the surprise pounce ( seven kills strategy \cite{bednarz1988cooperative} )  by charging on the target spotted in the exploration phase. It utilises four separate approaches based on the location of the chase that was determined in the previous stage. 

The approaches are ``\textit{soft besiege}", ``\textit{hard besiege}", ``\textit{soft besiege with progressive rapid dives}" and ``\textit{hard besiege with progressive rapid dives}".
The Harris Hawks choose out of soft and hard besiege approaches based on the value of `$X$' as stated before, where $X$ denotes the present amount of energy left in the kill. When $|X| \geq 0.5$, it goes for soft besiege else it selects hard besiege. Another variable `$t$' that denotes the likelihood of the prey escaping is taken arbitrarily in the span (0,1). A lesser value of `$t$' indicates that the kill has more chances to escape. The four different approaches are described hereafter. 

%The four approaches comprise soft besiege, hard besiege, soft besiege with progressive rapid dives and hard besiege with progressive rapid dives. When $|X| \geq 0.5$, then it is opted for soft besiege else hard besiege, where $X$ stands for the current energy of the prey. There is also another parameter $t$, which designates the probability of the prey escaping and it is taken randomly in the interval (0,1). A lower value of $t$ means that the prey has higher chances of fleeing. The four methods are discussed below.

The algorithm chooses the soft besiege approach when $t \geq 0.5$ and $|X| \geq 0.5 $. The kill has sufficient energy but it is encircled by the Harris Hawks when it attempts to mislead them and flee. Eventually, the Harris Hawks execute the surprise pounce (or the seven kills) on the prey. It can be formulated mathematically as stated in Equations \ref{eqa5}, \ref{eqa6} and \ref{eqa7}, where $P(g)$ is the population at $g^{th}$ iteration and $P(g+1)$ is the population at $(g+1)^{th}$ iteration.

%Soft besiege is opted for when $t \geq 0.5$ and $|X| \geq 0.5 $. The prey or the rabbit has ample energy but the hawks surround the rabbit when it tries to deceive the Harris Hawks and get away. Finally, the hawks carry out the surprise dive on the prey. This can be drafted mathematically as follows:\\

\begin{equation}\label{eqa5}
P(g+1)= \Delta P(g) - X |JS\times P_{prey}(g)-P(g)| 
\end{equation}

\begin{equation}\label{eqa6}
\Delta P(g) = P_{prey}(g) - P(g)
\end{equation}

\begin{equation}\label{eqa7}
JS = 2(1-e_5)
\end{equation}

In these equations, $e_5$ is an arbitrary number in the span 0 to 1, and $JS$ stands for the kill's arbitrary Jump Strength while it attempts to flee. $\Delta P(g)$ signifies the gap between the location of the kill and that of the present Harris Hawk in the $g^{th}$ iteration.

%Here, $e_5$ is a random number between 0 and 1, and $JS$ depicts the random Jump Strength of the prey while it is striving to get away. $\Delta P(g)$ stands for the difference between the location of the prey and the position of the current Harris Hawk in the $i^{th}$ iteration.

The algorithm goes for hard besiege when $t \geq 0.5$ and $|X| < 0.5 $. The prey is fatigued and it has less energy to escape the predators. The Harris Hawks hardly require to perform the surprise dive. The mathematical formulation is stated in \autoref{eqa8}.
%Hard besiege is selected when $t \geq 0.5$ and $|X| < 0.5 $. The rabbit is worn out and the amount of energy it has to escape is low. The Harris Hawks barely need to perform the surprise dive. 
\begin{equation}\label{eqa8}
P(g+1)=P_{prey}(g) - X |\Delta P(g)| 
\end{equation}

The algorithm opts for soft besiege with continuous or progressive rapid dives if the kill possesses $|X| \geq 0.5 $ and the value of $t < 0.5$. The kill has sufficient energy to flee and a higher likelihood of running away. Harris Hawks accomplish the surprise pounce in two steps. 
%Soft besiege with continuous or progressive rapid dives is the next technique. When the prey has $|X| \geq 0.5 $ and the value of $t < 0.5$, it has more energy to get away, and also good chances of escaping. Here the Harris Hawks carry out the surprise dive in two moves.\\ 
\textcolor{black}{The Harris Hawks encircle the kill and move to a new location after estimating the following movement of the kill (denoted by `$K$') emulating the formulation in} \autoref{eqa9}.

%In the first move, the Hawks surround the prey and move to a location after assessing the next move of the prey. 
\begin{equation}\label{eqa9}
    K= P_{prey}(g) - X|JS\times P_{prey}(g) - P(g)|
\end{equation}

The Harris Hawks select whether to perform the pounce after juxtaposing it with the foregoing dive and the corresponding outcome. If it is gauged that performing the pounce is unreasonable, it executes non-uniform irregular plunges based on the concept of L\'evy Flight (LF). The LF concept mimics the deceptive behaviour which preys manifest to get away from their predators. The Harris Hawks display rapid and uneven plunges around the kill when it endeavours to flee. Myriad real-life observations back the LF concept, for example, how animals like monkeys and sharks chase their targets \cite{Sims2008, Viswanathan2000}. 

%In the next move, the Harris Hawks choose whether to perform the jump after contrasting with the result of the preceding dive and its corresponding result and if it is seen that such a dive is not reasonable, then based on the Levy Flight idea, uneven irregular dives are performed. The Levy Flight (LF) idea is used to imitate the misleading manner in which preys behave while trying to escape and the swift, non-uniform descents of Hawks around the prey while it strives to get away. It is backed by observations in real life in the way in which various animals such as monkeys and sharks chase preys\cite{Sims2008} \cite{Viswanathan2000} .% give ref

\begin{equation} \label{eqa10}
    L= K+R\times levy(N)
\end{equation}

In \autoref{eqa10}, $R$ is an arbitrary vector of the dimensions $1 \times N$, where the dimension of the solution is depicted by $N$. Also, $levy$ is the function \cite{yang2010nature} that is used to replicate the LF concept and $N$ is calculated by \autoref{eqa9}. The mathematical formulation of this function is stated in \autoref{eqa11}.
%$D$ denotes the dimension of the solution. $R$ is a random vector of the dimensions $1 \times D$. Here, $levy$ is the Levy Flight function\cite{yang2010nature}, calculated using the equation written below. 
\begin{equation} \label{eqa11}
levy(x) = 0.01\times \frac{v_1 \times \sigma }{|v_2|^{\frac{1}{\beta}}} \;, 
\sigma = \left( \frac{\Gamma(1+\beta) \times \sin({\frac{\pi \beta}{2} })}{\Gamma(\frac{1+\beta}{2}) \times \beta \times 2^{(\frac{\beta -1 }{2})}}\right) ^ {1/ \beta}
\end{equation}
In \autoref{eqa11}, $\beta$ is initialized to 1.5. Variables $v_1\; and\; v_2$ are arbitrarily generated in the span 0 to 1, both limits are inclusive. In this approach, the location of the Harris Hawk at the close of this iteration can be updated following the mathematical formulation given in \autoref{eqa12}.

%where the value of $\beta$ is set to 1.5 and the other two parameters, $v_1\;and\; v_2$ are generated randomly in the range 0 to 1 (both inclusive). In the method of soft besiege with progressive rapid dives, the position of the Harris Hawk at the end of the iteration can be updated according to the formula stated below:
\begin{equation}\label{eqa12}
P(g+1)=\begin{cases}
    K, \text{  if  }fitness(K) < fitness(P(g))\\
    L, \text{  if  }fitness(L) < fitness(P(g))
\end{cases}
\end{equation}
In \autoref{eqa12}, $fitness(x)$ depicts the value returned when x is passed as parameter to the fitness function. 
%where $fitness(x)$ denotes the value obtained when x is passed as a parameter to the fitness or the objective function. 

The algorithm picks the hard besiege with continuous rapid dives when $t < 0.5\; and\;|X|<0.5$. The kill does not have sufficient energy left in it to flee. The predators, the Harris Hawks, exhibit swift plunges before manifesting the surprise pounce on the kill. The new position of the Harris Hawks can be formulated mathematically as in \autoref{eqa13}.

%The last technique is Hard besiege with continuous or progressive rapid dives, when $t < 0.5\;and\;|X|<0.5$. There is not much energy left in the prey to escape and the hawks implement rapid jumps before exhibiting the surprise dive on the prey. The motion of the Hawks can be mathematically shown as:
\begin{equation}\label{eqa13}
P(g+1)=\begin{cases}
    K,\text{  if  }fitness(K) < fitness(P(g))\\
    L,\text{  if  }fitness(L) < fitness(P(g))
\end{cases}
\end{equation}
Here, $K$ and $L$ are obtained from the equations noted in Equations \ref{eqa14} and \ref{eqa15}.
\begin{equation}\label{eqa14}
    K= P_{prey}(g) - X|JS\times P_{prey}(g) - P_{mean}(g)|
\end{equation}
\begin{equation}\label{eqa15}
    L= K+R\times levy(N)
\end{equation}
Here, $P_{mean}(g)$ is obtained as stated in \autoref{eqa3}.

\subsection{Altruism in HHO}
Altruism \cite{stich2010altruism} is a virtue that benefits other individuals at a cost to oneself. It is a state with the noble aim of increasing the welfare of another even at the cost of sacrificing its own. In the field of biology \cite{prudkov2016altruism}, it is the behaviour that increments the fitness of an individual while decrementing the fitness of the other participants. The virtue of altruism is generally observed in kin relations such as among the family members (for instance, from parent to child) but it has also been observed in larger social groups (for example, in social insects). In biology, the cost and benefit of altruism are measured in terms of reproductive fitness or the anticipated number of offspring. An individual thus reduces the number of children it is likely to give rise to and enhances the probability of another individual producing an offspring. In a gene pool, altruism evolves if the potential donor can compensate for its loss of offspring by enhancing the population by some other number of offspring having some fraction of its genes. 

In the case of the HHO, each of the Harris Hawks is considered a feasible solution to the segmentation problem. The exploitation phase of HHO comprises four phases. In the latter two phases, when a Harris Hawk exhibits the progressive rapid dives, altruism has been incorporated. In each of the two cases, the feature vectors which are derived from the original feature vector are considered viable solutions. The original feature vector behaves as the altruist and the derived ones as beneficiaries. A Harris Hawk shows an altruistic nature in this case by decrementing or incrementing the value at each position by a random amount generated such that it is in between the pixel ranges of grayscale images. Similarly, for the beneficiary Harris Hawk, the value at every position is incremented or decremented respectively by the same amount with which it is decremented or incremented for the altruist provided it lies within the pixel ranges of grayscale images for the operation. Altruism is performed if \autoref{eqa16} is satisfied.

\begin{equation}\label{eqa16}
    r\times B'>C'
\end{equation}

\autoref{eqa16} is known as Hamilton's rule \cite{waibel2011quantitative}. Here, $B'$ is the benefit or the decrease in fitness of the derived feature vector in this case. $C'$ stands for the cost or increase in fitness of the potential altruist or the original feature vector. Also, $r$ is the relatedness of the altruist to the beneficiary or the factor by which the latter depends on the former. If this equation is not obeyed or the benefit of the recipient is less than zero, there is no altruism and the feature vectors are reverted to their earlier values. 

\subsection{Proposed algorithm}
The proposed method wherein the HHO algorithm is combined with chaotic map initialization and altruism is shown in Algorithm \ref{HHO}. It is ensured that the values of the feature vectors are unique throughout the process. The uniqueness is checked while initialising the vectors, as well as, whenever one of the values of the feature vector gets updated, it is checked with other values inside the feature vector to ensure it is distinct, else that possible solution of the feature vector is discarded since the threshold values should be discrete for a particular feature vector.

\subsection{Computational Complexity}
The algorithm for the procedure of thresholding is HHO along with altruism. Let the number of initial features vectors be V. \textcolor{black}{The computational complexity for the process of initialization will be $O(V)$. If the number of iterations is `$g_{max}$' and the dimension of the search space is $N$, then the complexity for the process of updating the position of each vector is $O(V \times g_{max}) + O(V \times g_{max} \times N)$ for the HHO algorithm. This is also the lower bound of the computational complexity of the algorithm since it is possible that the algorithm does not opt for either soft besiege with progressive rapid dives or hard besiege with progressive rapid dives in the exploitation phase throughout the iterations owing to probabilistic constraints. For altruism, the computational complexity is $O(g_{max} \times V \times g_{Altr} \times N)$, where $g_{Altr}$ is the number of iterations taking place in the function for altruism. For the thresholding process, the upper bound of the computational complexity, hence, is $O(V+V \times g_{max}(1+N)+ g_{max} \times V \times g_{Altr} \times N)$.}

\begin{algorithm}[tbp]
    \caption{{The Pseudo-code for HHO with chaotic initialization and altruism}}\label{HHO}
    \textbf{Input}: \textcolor{black}{Population size `$M$' and maximum number of iterations `$g_{max}$'}\\
    \textbf{Output}: Location of the kill and the value obtained when it is passed as parameter to the fitness function
    
    \begin{algorithmic}[]
        \STATE Set the values of the first Harris Hawk randomly
        \STATE Set the remaining Harris Hawks by applying the Logistic chaotic map - using \autoref{eqa1}
        \FOR{$g$ \textbf{in} \{$1$ \textbf{to} $g_{max}$\}}
            \STATE Determine the fitness values of the Harris Hawks by passing the particular feature vector as parameter to the fitness function
            \STATE The Harris Hawk with best fitness value is set to be the prey, denoted by $P_{prey}(g)$
            \FOR{\textbf{each} Harris Hawk}
                \STATE  For the exploitation phase, set Jump Strength (JS):  $JS= 2 (1- rand() )$
                \STATE Determine the present energy X using \autoref{eqa3}
                \IF{$|X(g)| \geq 1$}
                    \STATE Perform exploration following \autoref{eqa2} and correspondingly, update the Harris Hawk's position
                \ELSIF{$|X(g)| < 1$}
                    \STATE Say, X and t stand for the present energy of the kill and likelihood of the kill fleeing, respectively
                    \IF{$t \geq 0.5$ and $|X(g)| \geq 0.5 $}
                    \STATE Opt for Soft besiege in accordance with equations \ref{eqa5} to \ref{eqa7} and update position of the Harris Hawk
                    \ELSIF{$t \geq 0.5$ and $|X(g)| < 0.5 $}
                    \STATE Opt for Hard besiege in accordance with \autoref{eqa8} and update position of the Harris Hawk
                    \ELSIF{$t < 0.5$ and $|X(g)| \geq 0.5 $}
                    \STATE Opt for Soft besiege with progressive rapid dives in accordance with Equations \ref{eqa9} to \ref{eqa12} and update position of the Harris Hawk and similarly check for Altruism in each of the derived feature vectors from the original Harris Hawk such that \autoref{eqa16} is obeyed 
                    \ELSIF{$t < 0.5$ and $|X(g)| < 0.5 $}
                    \STATE Opt for Hard besiege with progressive rapid dives in accordance with Equations \ref{eqa13} to \ref{eqa15} and update position of the Harris Hawk and check for Altruism in each of the derived feature vectors from the original Harris Hawk such that \autoref{eqa16} is obeyed. 
                    \ENDIF
                \ENDIF
                
            \ENDFOR
            \IF{Fitness of best solution == Concluding Condition}
            \STATE Break out of the for loop
            \ENDIF
            %\STATE Apply SA algorithm to the current set of solutions
        \ENDFOR
        \STATE Return the best solution along with the value obtained when it is passed as a parameter to the fitness function
    \end{algorithmic}
    
    \label{HSpseudocode}
\end{algorithm}

\section{Results and Discussion}\label{results}
In this section, we describe the datasets used in this research and the evaluation outcomes of the proposed model on these datasets. Further, we discuss the implications of the results obtained and compare the proposed model to state-of-the-art methods and establish the reliability of the approach.

\subsection{Datasets Used}
The proposed method has been evaluated on two datasets: (1) Ten T2-weighted images via magnetic resonance transaxial cut brain obtained from the WBA database developed by the Harvard Medical School \cite{sutton1999whole} and (2) Eight benchmark MRIs extracted from Brainweb database \cite{cocosco1997brainweb}.

\subsection{Evaluation Metrics}
To evaluate the performance of the proposed method the following six popularly used metrics have been considered:

\begin{enumerate}
    \item Peak Signal to Noise Ratio (PSNR) \cite{avcibas2002statistical}: The mathematical formula for PSNR is given in \autoref{psnr}, where $Max_I$ is the maximum pixel value of image $I$ and $MSE$ denotes the Mean Squared Error.
    \begin{equation}\label{psnr}
        PSNR = 10\log_{10}\left( \frac{Max_I^2}{MSE} \right)
    \end{equation}
    
    \item Structural Similarity Index (SSIM) \cite{wang2004image}: The SSIM is calculated using \autoref{ssim} between two images (say $I_a$ and $I_b$).
    \begin{equation}\label{ssim}
        SSIM(I_a,I_b) = \frac{(2\mu_{I_a}\mu_{I_b}+c_1)\times(2\sigma_{I_aI_b})}{(\mu_{I_a}^2+\mu_{I_b}^2+c_1)\times(\sigma_{I_a}^2+\sigma_{I_b}^2+c_2)}
    \end{equation}
    where, $\mu$ and $\sigma$ are the mean and standard deviations of pixel values of the images, $\sigma_{I_aI_b}$ is the covariance of $I_a$ and $I_b$, and $c_1$ and $c_2$ are two variables for weak denominator stabilization.
    
    \item Feature Similarity Index (FSIM) \cite{zhang2011fsim}: Mathematically, the FSIM metric is formulated as in \autoref{fsim}, where, $S_L(x)$ denotes the overall similarity between two images $I_a$ and $I_b$; $PC$ denotes the phase congruence between the images and $\Omega$ denotes the whole spatial domain of the image.
    \begin{equation}\label{fsim}
        FSIM = \frac{\sum_{x\in\Omega}S_L(x)\times PC_m(x)}{\sum_{x\in\Omega}PC_m(x)}
    \end{equation}
    
    \item Haar wavelet-based Perceptual Similarity Index (HPSI) \cite{reisenhofer2018haar}: The HPSI for two images $X$, $Y$ is given as \autoref{hpsi}.
    \begin{equation}\label{hpsi}
        HPSI(X,Y) = \frac{1}{l_{\alpha}} \left( \frac{  \sum_x\sum_{k=1}^{2}HS_{X,Y}^{(k)}[x]\times W_{X,Y}^{(k)}[x]  }{  \sum_x\sum_{k=1}^{2}  W_{X,Y}^{(k)}[x]   }   \right)^2
    \end{equation}
    where, $l_{\alpha}$ is a logistic function for a parameter $\alpha>0$ used to model the perceptual similarity; $HS$ is the local similarity measure based on Haar wavelet transform and $W_{X,Y}^{(k)}$ is a weight map given by \autoref{w_map}.
    \begin{equation}\label{w_map}
        W_{X,Y}^{(k)}[x] = max(W_{X}^{(k)}[x],W_{Y}^{(k)}[x])
    \end{equation}
    
    \item Quality Index based on Local Variance (QILV) \cite{aja2006image}: The QILV metric appraises the changes in the non-stationary behaviour of images and is calculated by \autoref{qilv} between images $X$ and $Y$.
    \begin{equation}\label{qilv}
        QILV(X,Y) = \left( \frac{2\mu_{V_X}\mu_{V_Y}}{\mu_{V_X}^2+\mu_{V_Y}^2} \right)\times \left( \frac{2\sigma_{V_X}\sigma_{V_Y}}{\sigma_{V_X}^2+\sigma_{V_Y}^2} \right) \times \left( \frac{\sigma_{V_X V_Y}}{\sigma_{V_X}\sigma{V_Y}} \right)
    \end{equation}
    $\mu$ and $\sigma$ are the global mean and standard deviations of the local variances $V_X$ and $V_Y$ respectively of images $X$ and $Y$.
    
    \item Universal Image Quality Index (UIQI) \cite{wang2002universal}: The UIQI evaluates the distortion of an image as a combination of correlation loss, luminance distortion, and contrast distortion and is calculated by \autoref{uiqi}.
    \begin{equation}\label{uiqi}
        UIQI(X,Y) = \frac{4\sigma_{XY}\mu_X\mu_Y}{(\sigma_X^2+\sigma_Y^2)(\mu_X^2+\mu_Y^2)}
    \end{equation}
    For the images $X$ and $Y$, $\mu$ and $\sigma$ denote their mean and standard deviations and $\sigma_XY$ denotes the covariance of the images.
    
\end{enumerate}

\subsection{Implementation}

\begin{table*}[]
\centering
\caption{Results obtained on the Harvard WBA brain MRI dataset where the number of the slice is mentioned.}
\label{res_harv}
\begin{tabular}{|c|c|c|c|c|c|c|c|c|}
\hline
\textbf{Slice} & \textbf{No. of Thresholds} & \textbf{PSNR} & \textbf{SSIM} & \textbf{FSIM} & \textbf{UIQI} & \textbf{QILV} & \textbf{HPSI} & \textbf{Time(s)} \\ \hline
\multirow{4}{*}{22}  & 2 & 23.1170 & 0.8175 & 0.7964 & 0.4396 & 0.7452 & 0.6905 & 22.6105 \\ \cline{2-9} 
                     & 3 & 24.8599 & 0.8656 & 0.8622 & 0.5448 & 0.8396 & 0.7912 & 26.8900 \\ \cline{2-9} 
                     & 4 & 26.3159 & 0.8956 & 0.8961 & 0.6211 & 0.8917 & 0.8389 & 30.6944 \\ \cline{2-9} 
                     & 5 & 27.6278 & 0.9146 & 0.9232 & 0.6752 & 0.9246 & 0.8784 & 34.4573 \\ \hline
\multirow{4}{*}{32}  & 2 & 22.5471 & 0.8239 & 0.8025 & 0.4308 & 0.7302 & 0.6808 & 23.4577 \\ \cline{2-9} 
                     & 3 & 24.2922 & 0.8698 & 0.8625 & 0.5517 & 0.8348 & 0.7869 & 27.6399 \\ \cline{2-9} 
                     & 4 & 25.7215 & 0.8943 & 0.8991 & 0.6140 & 0.8867 & 0.8399 & 31.2163 \\ \cline{2-9} 
                     & 5 & 26.9763 & 0.9183 & 0.9239 & 0.6784 & 0.9137 & 0.8808 & 35.2593 \\ \hline
\multirow{4}{*}{42}  & 2 & 21.7223 & 0.8148 & 0.8098 & 0.4277 & 0.7197 & 0.6542 & 23.1100 \\ \cline{2-9} 
                     & 3 & 23.2718 & 0.8550 & 0.8519 & 0.5065 & 0.8180 & 0.7519 & 27.5122 \\ \cline{2-9} 
                     & 4 & 24.7889 & 0.8865 & 0.8884 & 0.5750 & 0.8749 & 0.8130 & 31.4134 \\ \cline{2-9} 
                     & 5 & 25.9590 & 0.9013 & 0.9120 & 0.6168 & 0.9115 & 0.8532 & 35.9043 \\ \hline
\multirow{4}{*}{52}  & 2 & 21.7282 & 0.8242 & 0.8024 & 0.4698 & 0.7663 & 0.6601 & 22.6908 \\ \cline{2-9} 
                     & 3 & 23.7954 & 0.8802 & 0.8635 & 0.5540 & 0.8304 & 0.7739 & 27.4299 \\ \cline{2-9} 
                     & 4 & 25.0807 & 0.9011 & 0.8964 & 0.6188 & 0.8873 & 0.8275 & 31.5897 \\ \cline{2-9} 
                     & 5 & 26.0764 & 0.9168 & 0.9124 & 0.6730 & 0.9156 & 0.8595 & 35.2184 \\ \hline
\multirow{4}{*}{62}  & 2 & 20.4246 & 0.7986 & 0.7597 & 0.4108 & 0.7426 & 0.5920 & 23.2198 \\ \cline{2-9} 
                     & 3 & 22.6551 & 0.8587 & 0.8229 & 0.5106 & 0.8108 & 0.7063 & 27.4539 \\ \cline{2-9} 
                     & 4 & 24.1148 & 0.8882 & 0.8593 & 0.5747 & 0.8765 & 0.7788 & 31.1962 \\ \cline{2-9} 
                     & 5 & 25.4218 & 0.9085 & 0.8910 & 0.6403 & 0.9136 & 0.8341 & 35.5530 \\ \hline
\multirow{4}{*}{72}  & 2 & 19.6996 & 0.7935 & 0.7426 & 0.3672 & 0.6857 & 0.5737 & 23.0623 \\ \cline{2-9} 
                     & 3 & 21.8478 & 0.8370 & 0.8043 & 0.4747 & 0.8137 & 0.6733 & 27.1111 \\ \cline{2-9} 
                     & 4 & 24.0422 & 0.8891 & 0.8534 & 0.5519 & 0.8614 & 0.7711 & 32.4240 \\ \cline{2-9} 
                     & 5 & 25.3464 & 0.9118 & 0.8891 & 0.6214 & 0.9065 & 0.8212 & 36.0055 \\ \hline
\multirow{4}{*}{82}  & 2 & 19.9938 & 0.8001 & 0.7420 & 0.3571 & 0.6513 & 0.5857 & 23.2232 \\ \cline{2-9} 
                     & 3 & 22.5303 & 0.8441 & 0.8081 & 0.4840 & 0.7994 & 0.6851 & 27.2413 \\ \cline{2-9} 
                     & 4 & 24.3912 & 0.8929 & 0.8591 & 0.5587 & 0.8414 & 0.7928 & 31.6598 \\ \cline{2-9} 
                     & 5 & 25.5322 & 0.9115 & 0.8935 & 0.6294 & 0.8855 & 0.8367 & 36.3615 \\ \hline
\multirow{4}{*}{92}  & 2 & 21.1111 & 0.8243 & 0.7745 & 0.3936 & 0.7317 & 0.6148 & 23.1599 \\ \cline{2-9} 
                     & 3 & 23.9383 & 0.8822 & 0.8408 & 0.4671 & 0.8201 & 0.7396 & 27.5972 \\ \cline{2-9} 
                     & 4 & 25.7974 & 0.9074 & 0.8801 & 0.5743 & 0.8922 & 0.8265 & 31.6055 \\ \cline{2-9} 
                     & 5 & 26.9757 & 0.9194 & 0.9083 & 0.6359 & 0.9243 & 0.8608 & 35.6105 \\ \hline
\multirow{4}{*}{102} & 2 & 21.7561 & 0.8646 & 0.8454 & 0.4135 & 0.7308 & 0.6522 & 22.7122 \\ \cline{2-9} 
                     & 3 & 24.6831 & 0.9082 & 0.8842 & 0.5011 & 0.8255 & 0.7473 & 27.5126 \\ \cline{2-9} 
                     & 4 & 26.0502 & 0.9284 & 0.9173 & 0.5917 & 0.8868 & 0.8270 & 31.3697 \\ \cline{2-9} 
                     & 5 & 27.3213 & 0.9386 & 0.9349 & 0.6701 & 0.9236 & 0.8618 & 36.0524 \\ \hline
\multirow{4}{*}{112} & 2 & 24.1399 & 0.8948 & 0.9012 & 0.4069 & 0.7199 & 0.6601 & 23.4277 \\ \cline{2-9} 
                     & 3 & 25.8955 & 0.9214 & 0.9306 & 0.4962 & 0.8272 & 0.7435 & 27.7060 \\ \cline{2-9} 
                     & 4 & 27.3790 & 0.9382 & 0.9533 & 0.6142 & 0.8854 & 0.8294 & 31.5294 \\ \cline{2-9} 
                     & 5 & 28.8877 & 0.9545 & 0.9664 & 0.6740 & 0.9156 & 0.8834 & 37.0770 \\ \hline
\end{tabular}
\end{table*}

\begin{table*}[]
\centering
\caption{Results obtained on the Brainweb database where the number of the image title indicates the z-plane of the analyzed image}
\label{res_mri}
\begin{tabular}{|c|c|c|c|c|c|c|c|c|}
\hline
\textbf{Image} & \textbf{No. of Thresholds} & \textbf{PSNR} & \textbf{SSIM} & \textbf{FSIM} & \textbf{UIQI} & \textbf{QILV} & \textbf{HPSI} & \textbf{Time (s)} \\ \hline
\multirow{4}{*}{Z1}   & 2 & 17.9955 & 0.6164 & 0.7138 & 0.2213 & 0.7264 & 0.4953 & 33.4307 \\ \cline{2-9} 
                      & 3 & 21.0102 & 0.7479 & 0.8063 & 0.3283 & 0.8204 & 0.7042 & 38.1369 \\ \cline{2-9} 
                      & 4 & 22.756  & 0.7897 & 0.8491 & 0.4104 & 0.8586 & 0.7651 & 43.716  \\ \cline{2-9} 
                      & 5 & 24.0058 & 0.8031 & 0.875  & 0.4352 & 0.92   & 0.8058 & 49.2387 \\ \hline
\multirow{4}{*}{Z2}   & 2 & 17.9547 & 0.6129 & 0.7104 & 0.2168 & 0.7225 & 0.493  & 33.3904 \\ \cline{2-9} 
                      & 3 & 21.0538 & 0.7466 & 0.8049 & 0.3262 & 0.8203 & 0.6958 & 37.7423 \\ \cline{2-9} 
                      & 4 & 22.8181 & 0.7885 & 0.8475 & 0.4087 & 0.8599 & 0.7592 & 42.9616 \\ \cline{2-9} 
                      & 5 & 24.0395 & 0.8059 & 0.8758 & 0.4532 & 0.9232 & 0.8094 & 49.4712 \\ \hline
\multirow{4}{*}{Z5}   & 2 & 18.9015 & 0.6563 & 0.7091 & 0.2391 & 0.6544 & 0.5286 & 33.3372 \\ \cline{2-9} 
                      & 3 & 21.0832 & 0.742  & 0.8117 & 0.3244 & 0.8088 & 0.7012 & 43.45   \\ \cline{2-9} 
                      & 4 & 22.9353 & 0.7831 & 0.8559 & 0.4081 & 0.8577 & 0.7632 & 48.9929 \\ \cline{2-9} 
                      & 5 & 24.0307 & 0.7956 & 0.8777 & 0.4299 & 0.9162 & 0.8058 & 53.3716 \\ \hline
\multirow{4}{*}{Z10}  & 2 & 19.3314 & 0.6553 & 0.7256 & 0.2403 & 0.6215 & 0.5284 & 30.8795 \\ \cline{2-9} 
                      & 3 & 21.3817 & 0.7293 & 0.818  & 0.3179 & 0.7853 & 0.707  & 37.8957 \\ \cline{2-9} 
                      & 4 & 23.2543 & 0.7712 & 0.864  & 0.3943 & 0.8458 & 0.7751 & 42.3333 \\ \cline{2-9} 
                      & 5 & 24.3016 & 0.7916 & 0.8844 & 0.4518 & 0.8866 & 0.8022 & 47.4749 \\ \hline
\multirow{4}{*}{Z36}  & 2 & 19.1749 & 0.7004 & 0.7682 & 0.2678 & 0.6464 & 0.5364 & 31.8485 \\ \cline{2-9} 
                      & 3 & 21.3443 & 0.7676 & 0.8396 & 0.3484 & 0.8013 & 0.7007 & 37.8283 \\ \cline{2-9} 
                      & 4 & 23.1914 & 0.8074 & 0.8747 & 0.4207 & 0.8582 & 0.7661 & 44.1279 \\ \cline{2-9} 
                      & 5 & 24.2994 & 0.8191 & 0.8953 & 0.4383 & 0.9183 & 0.8011 & 50.1885 \\ \hline
\multirow{4}{*}{Z72}  & 2 & 20.7149 & 0.8165 & 0.8393 & 0.357  & 0.8435 & 0.7331 & 32.9571 \\ \cline{2-9} 
                      & 3 & 23.165  & 0.8626 & 0.8815 & 0.4451 & 0.9135 & 0.7991 & 37.2743 \\ \cline{2-9} 
                      & 4 & 23.6653 & 0.8702 & 0.8968 & 0.4541 & 0.9266 & 0.807  & 43.7322 \\ \cline{2-9} 
                      & 5 & 25.1448 & 0.8918 & 0.9179 & 0.5177 & 0.9514 & 0.8441 & 50.1932 \\ \hline
\multirow{4}{*}{Z108} & 2 & 21.4394 & 0.8398 & 0.8218 & 0.3448 & 0.8326 & 0.7452 & 31.6718 \\ \cline{2-9} 
                      & 3 & 22.0079 & 0.8493 & 0.8412 & 0.36   & 0.8612 & 0.7573 & 41.4043 \\ \cline{2-9} 
                      & 4 & 24.1245 & 0.8885 & 0.8882 & 0.4418 & 0.9092 & 0.8222 & 43.3232 \\ \cline{2-9} 
                      & 5 & 25.6447 & 0.9055 & 0.9127 & 0.5066 & 0.9387 & 0.8598 & 50.112  \\ \hline
\multirow{4}{*}{Z144} & 2 & 21.4795 & 0.8435 & 0.8334 & 0.2816 & 0.6312 & 0.5945 & 32.7324 \\ \cline{2-9} 
                      & 3 & 23.9674 & 0.8854 & 0.8909 & 0.3892 & 0.7954 & 0.7624 & 38.6617 \\ \cline{2-9} 
                      & 4 & 25.5294 & 0.9025 & 0.9164 & 0.4516 & 0.8637 & 0.8273 & 46.4189 \\ \cline{2-9} 
                      & 5 & 26.6825 & 0.9134 & 0.9285 & 0.5031 & 0.8991 & 0.8541 & 50.1067 \\ \hline
\end{tabular}
\end{table*}

\begin{figure*}
    \centering
    \subfloat[Original]{\includegraphics[scale=0.3]{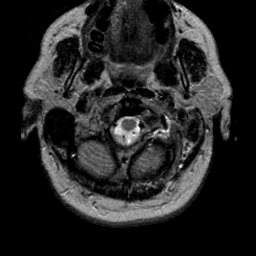}}\;\;
    \subfloat[2 Thresholds]{\includegraphics[scale=0.3]{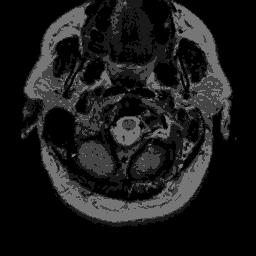}}\;\;
    \subfloat[3 Thresholds]{\includegraphics[scale=0.3]{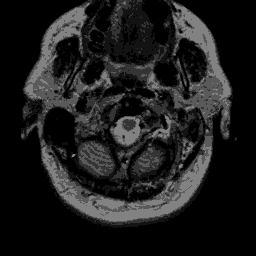}}\;\;
    \subfloat[4 Thresholds]{\includegraphics[scale=0.3]{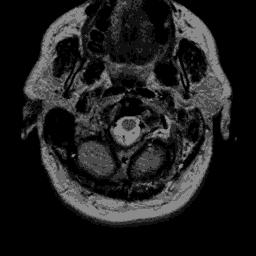}}\;\;
    \subfloat[5 Thresholds]{\includegraphics[scale=0.3]{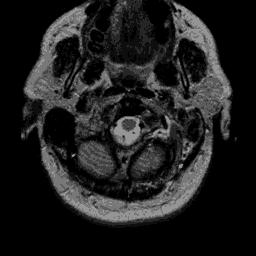}}\\
    \subfloat[Original]{\includegraphics[scale=0.2]{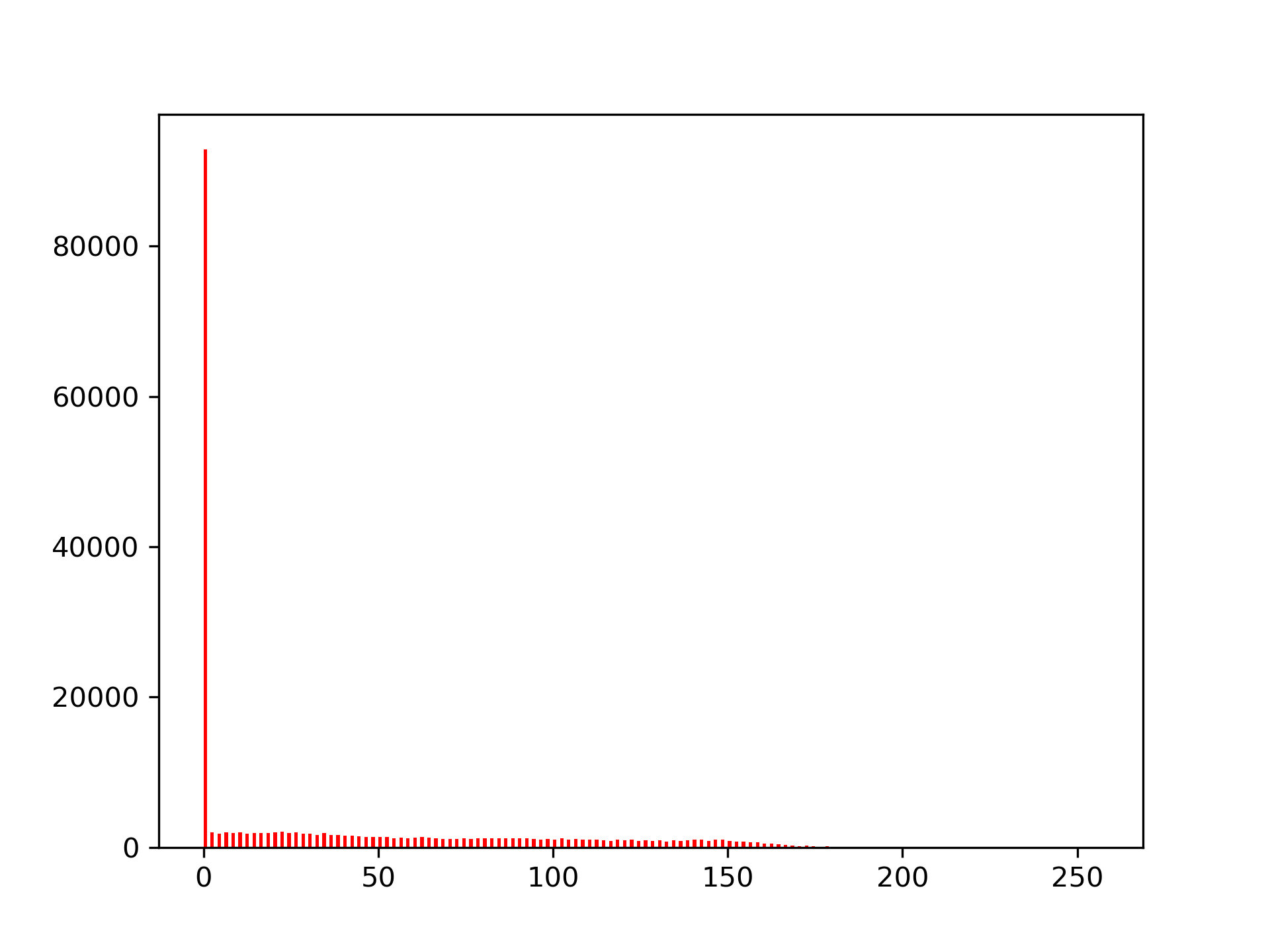}}
    \subfloat[2 Thresholds]{\includegraphics[scale=0.2]{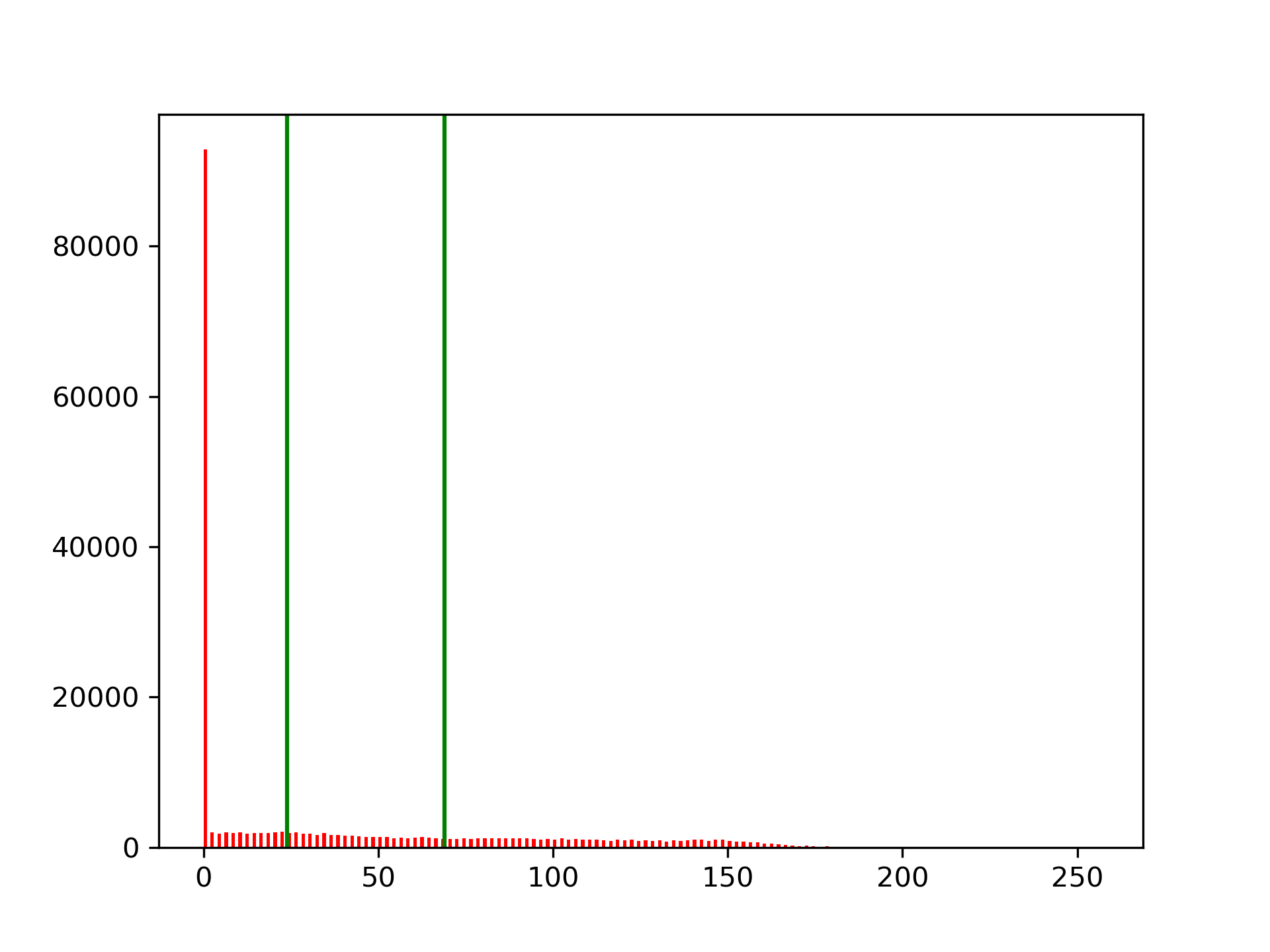}}
    \subfloat[3 Thresholds]{\includegraphics[scale=0.2]{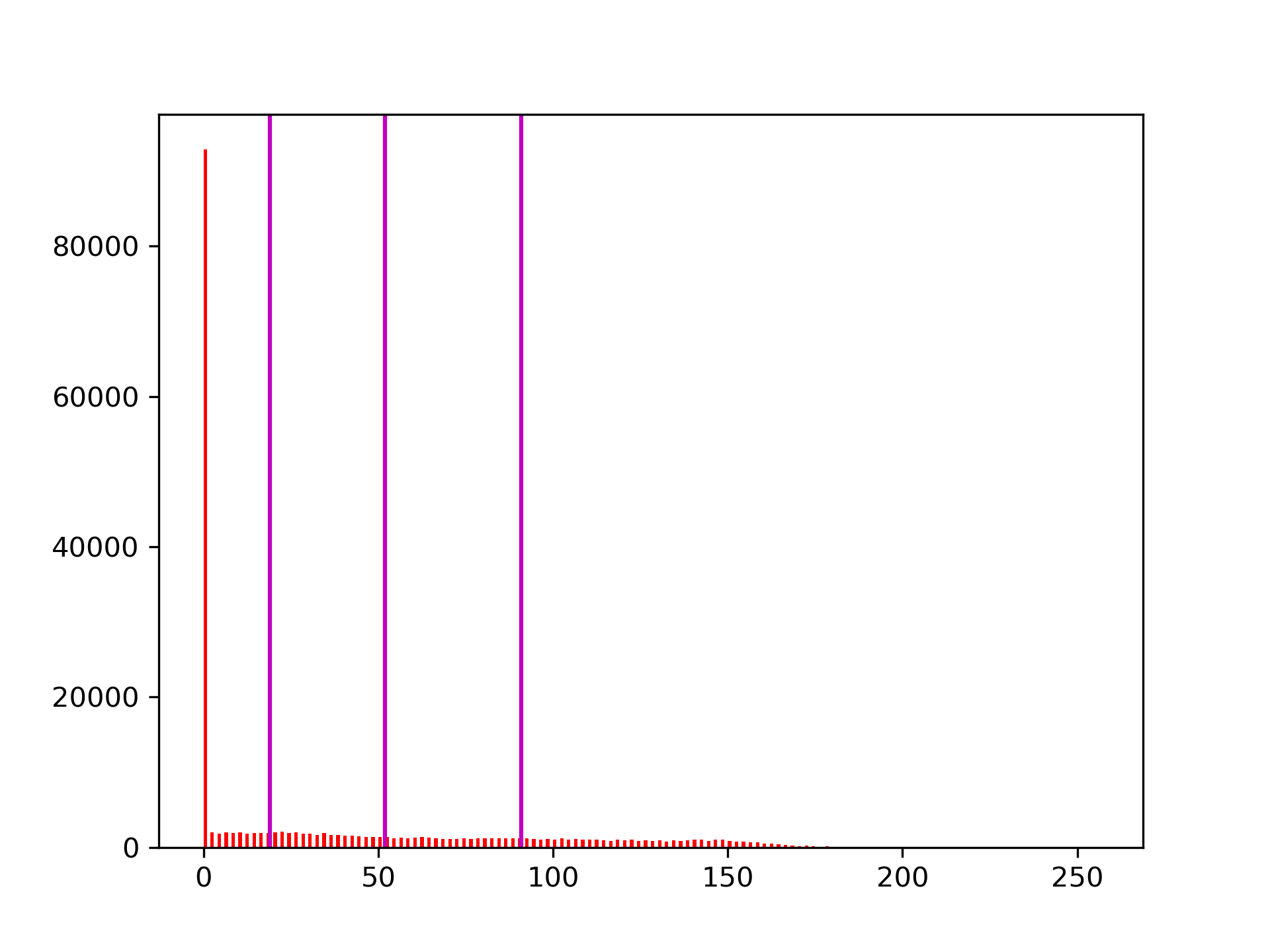}}
    \subfloat[4 Thresholds]{\includegraphics[scale=0.2]{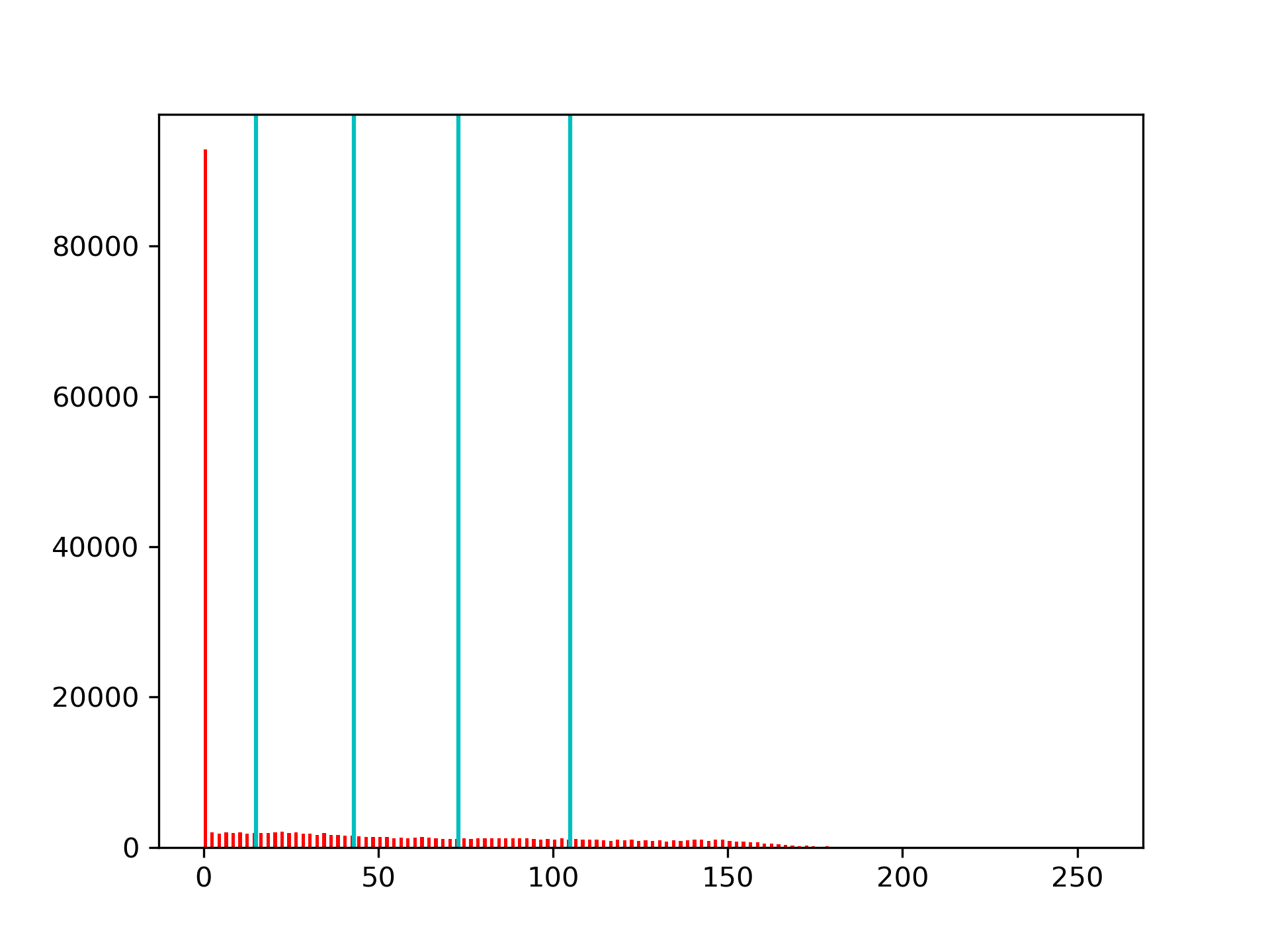}}
    \subfloat[5 Thresholds]{\includegraphics[scale=0.2]{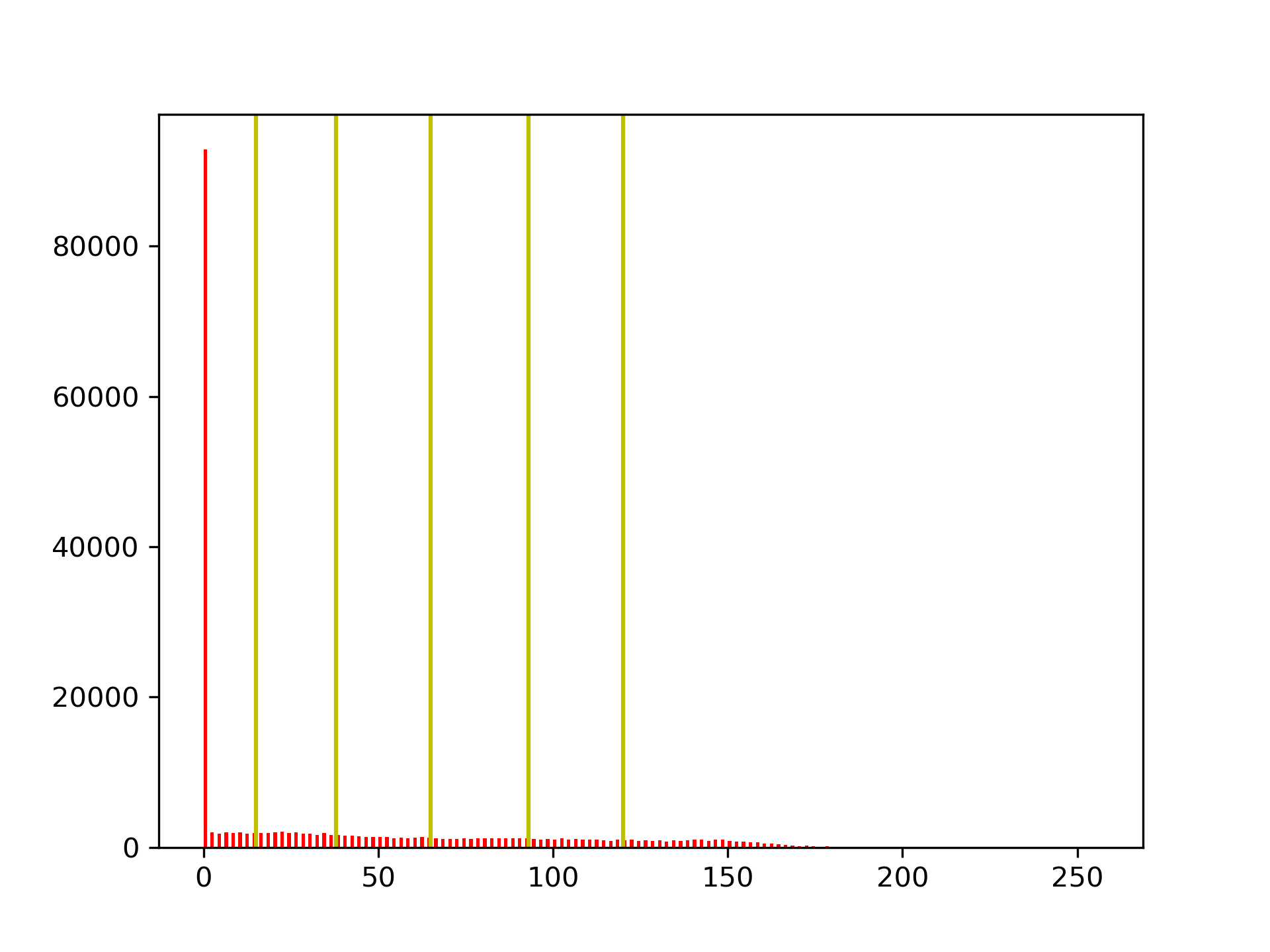}}
    \caption{Original image and thresholded images for slice 22 of the Harvard WBA dataset with number of thresholds: 2, 3, 4 and 5, with their corresponding histograms and thresholds marked with vertical lines.}
    \label{h22}
\end{figure*}

\begin{figure*}
    \centering
    \subfloat[Original]{\includegraphics[scale=0.3]{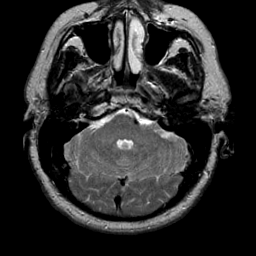}}\;\;
    \subfloat[2 Thresholds]{\includegraphics[scale=0.3]{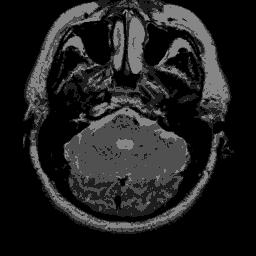}}\;\;
    \subfloat[3 Thresholds]{\includegraphics[scale=0.3]{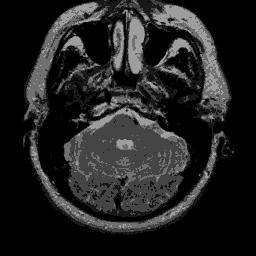}}\;\;
    \subfloat[4 Thresholds]{\includegraphics[scale=0.3]{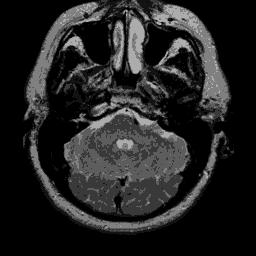}}\;\;
    \subfloat[5 Thresholds]{\includegraphics[scale=0.3]{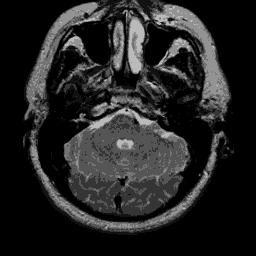}}\\
    \subfloat[Original]{\includegraphics[scale=0.2]{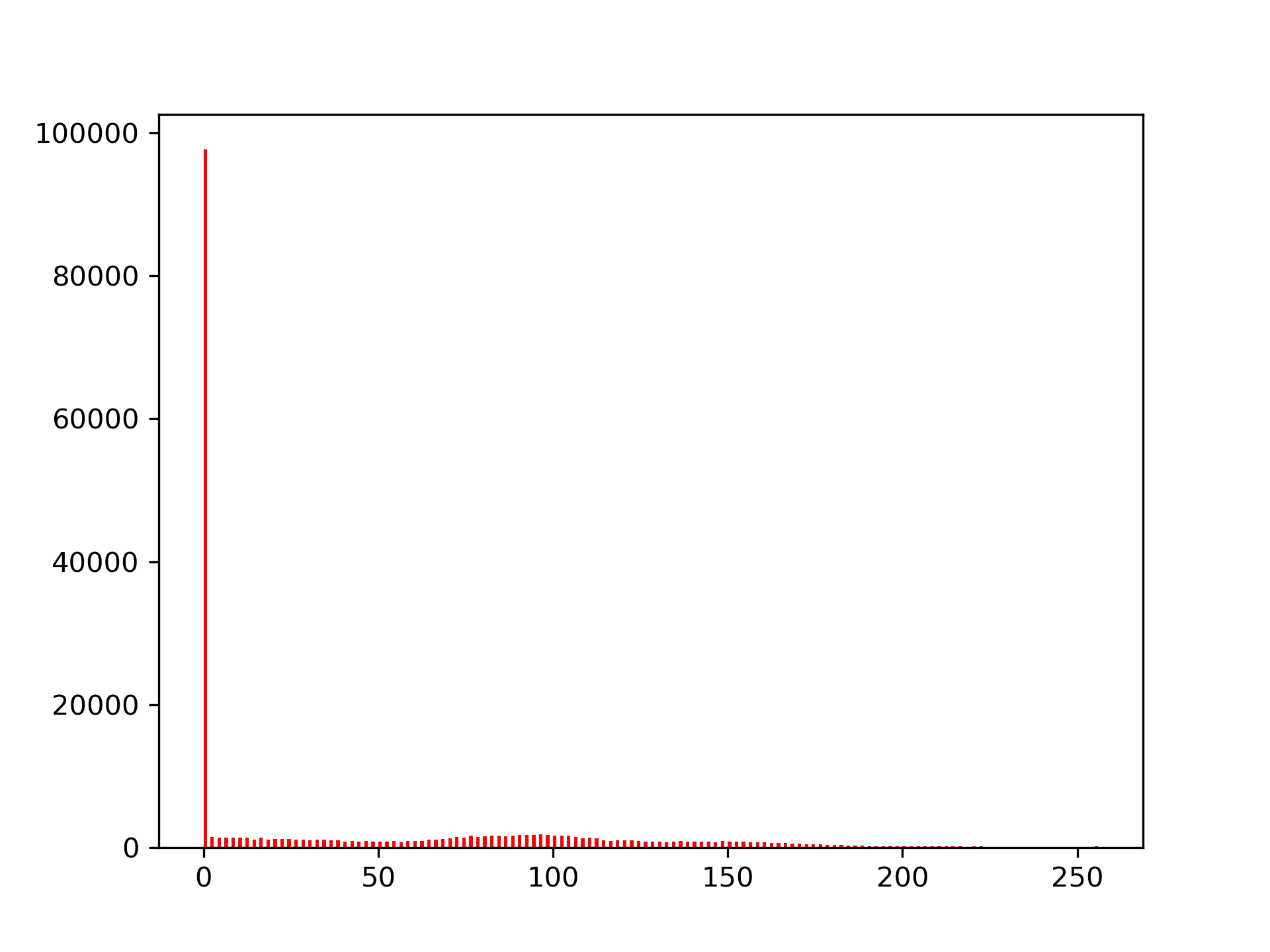}}
    \subfloat[2 Thresholds]{\includegraphics[scale=0.2]{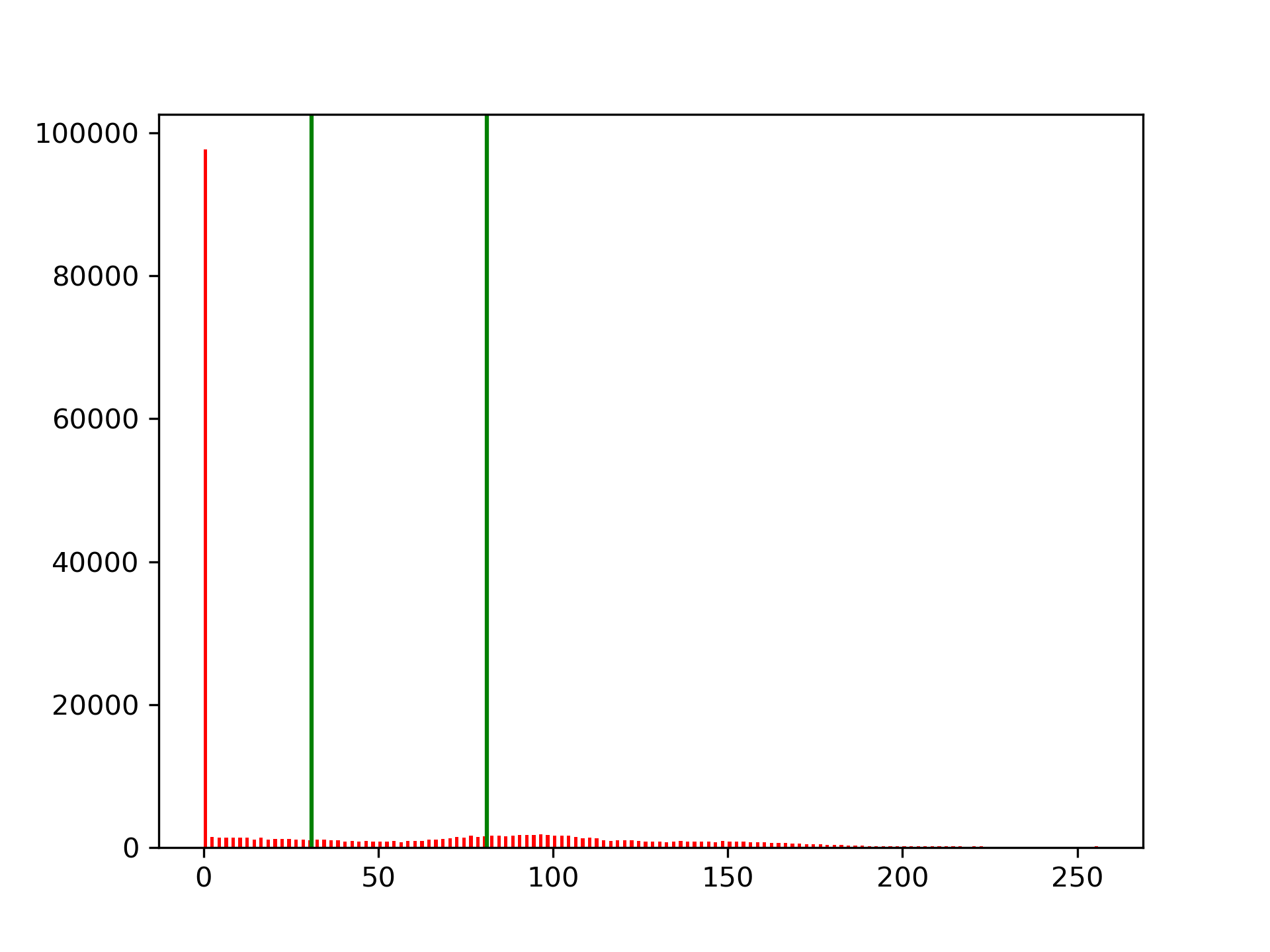}}
    \subfloat[3 Thresholds]{\includegraphics[scale=0.2]{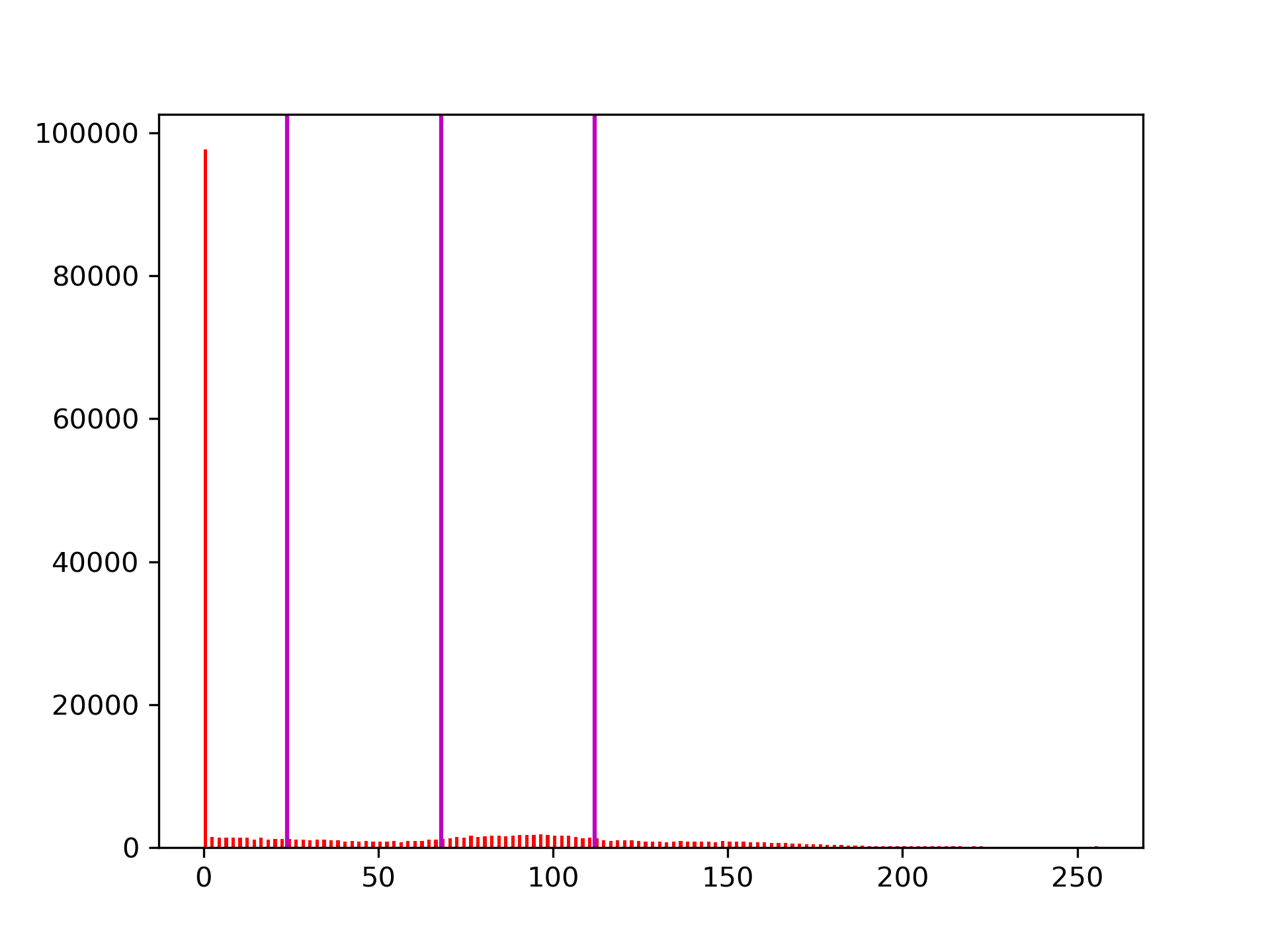}}
    \subfloat[4 Thresholds]{\includegraphics[scale=0.2]{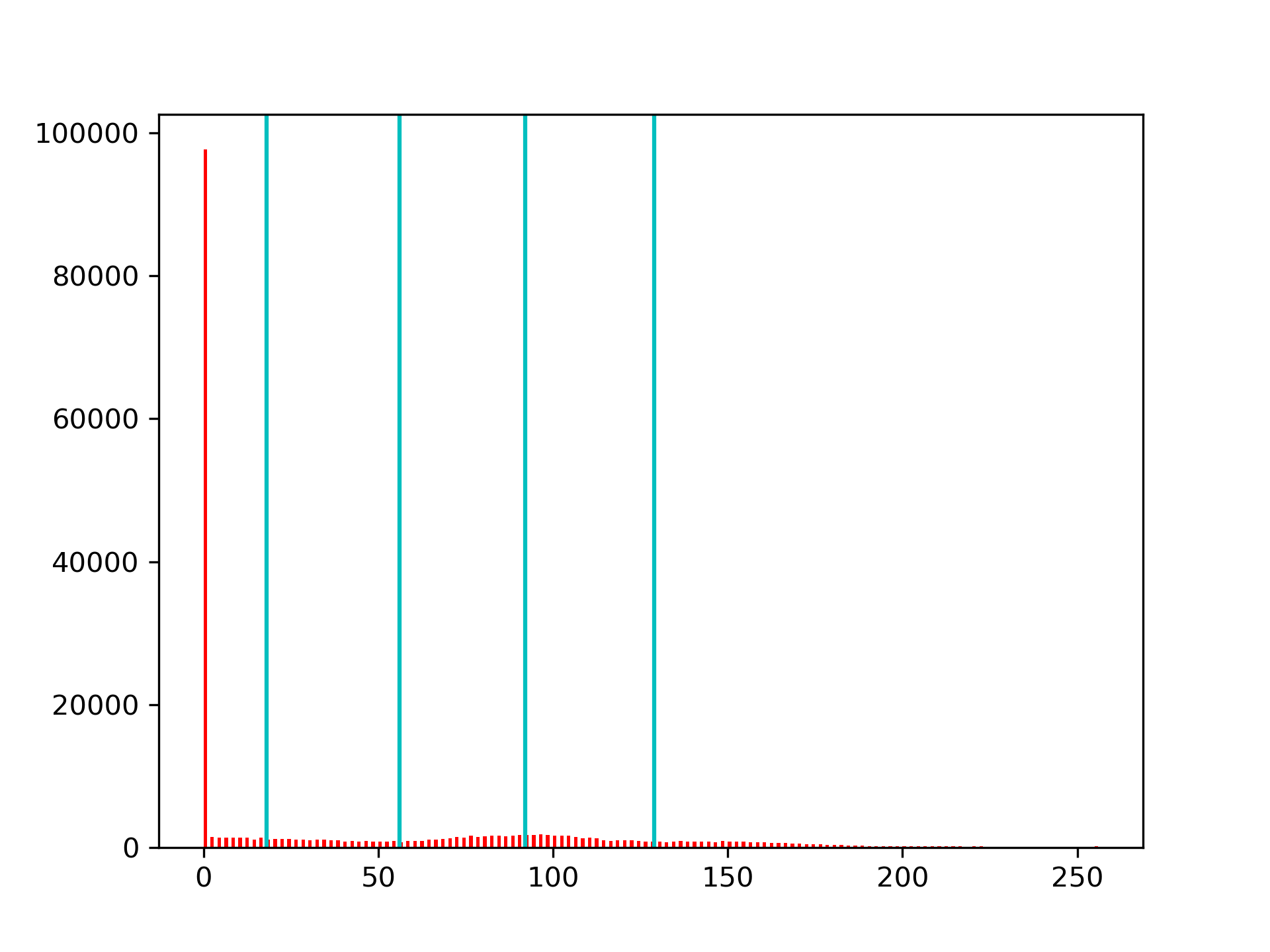}}
    \subfloat[5 Thresholds]{\includegraphics[scale=0.2]{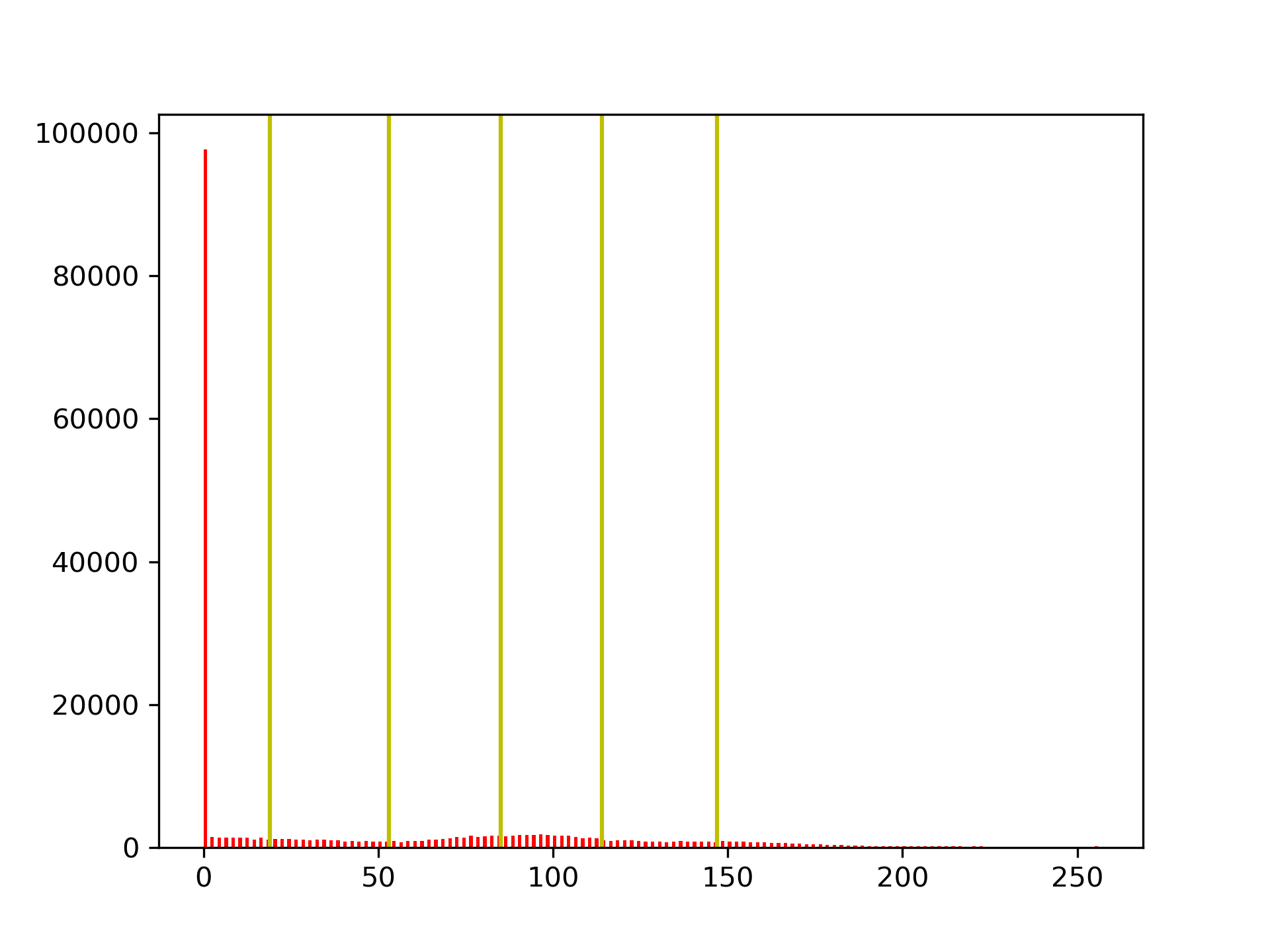}}
    \caption{Original image and thresholded images for slice 42 of the Harvard WBA dataset with number of thresholds: 2, 3, 4 and 5, with their corresponding histograms and thresholds marked with vertical lines.}
    \label{h42}
\end{figure*}

\begin{figure*}
    \centering
    \subfloat[Original]{\includegraphics[scale=0.3]{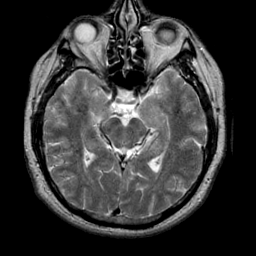}}\;\;
    \subfloat[2 Thresholds]{\includegraphics[scale=0.3]{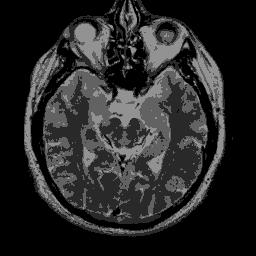}}\;\;
    \subfloat[3 Thresholds]{\includegraphics[scale=0.3]{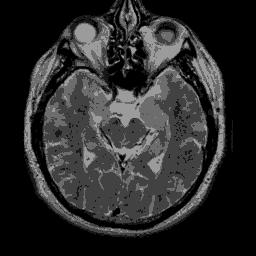}}\;\;
    \subfloat[4 Thresholds]{\includegraphics[scale=0.3]{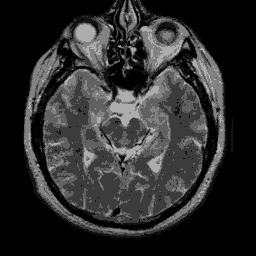}}\;\;
    \subfloat[5 Thresholds]{\includegraphics[scale=0.3]{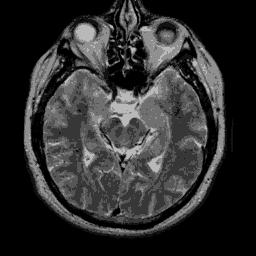}}\\
    \subfloat[Original]{\includegraphics[scale=0.2]{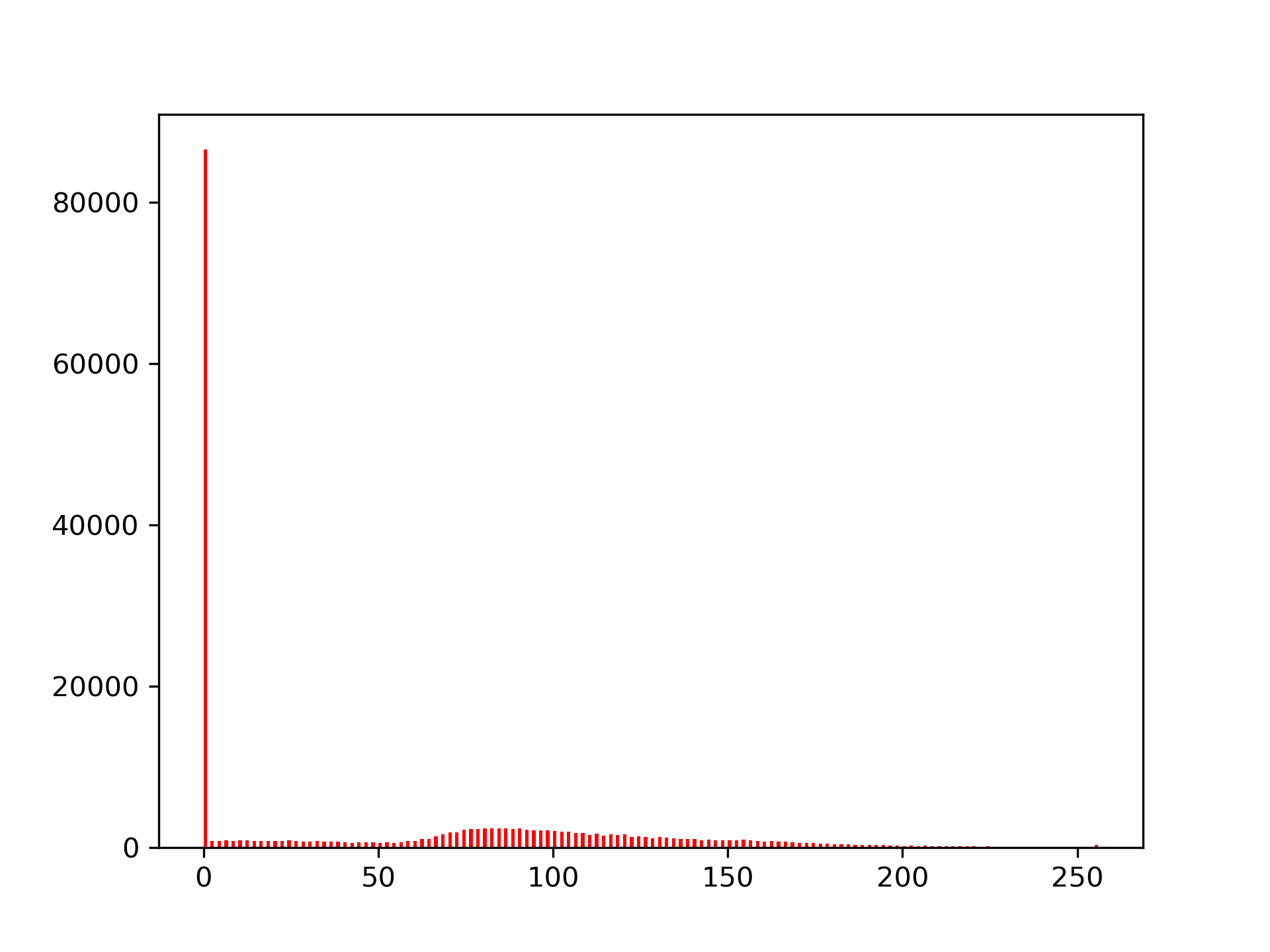}}
    \subfloat[2 Thresholds]{\includegraphics[scale=0.2]{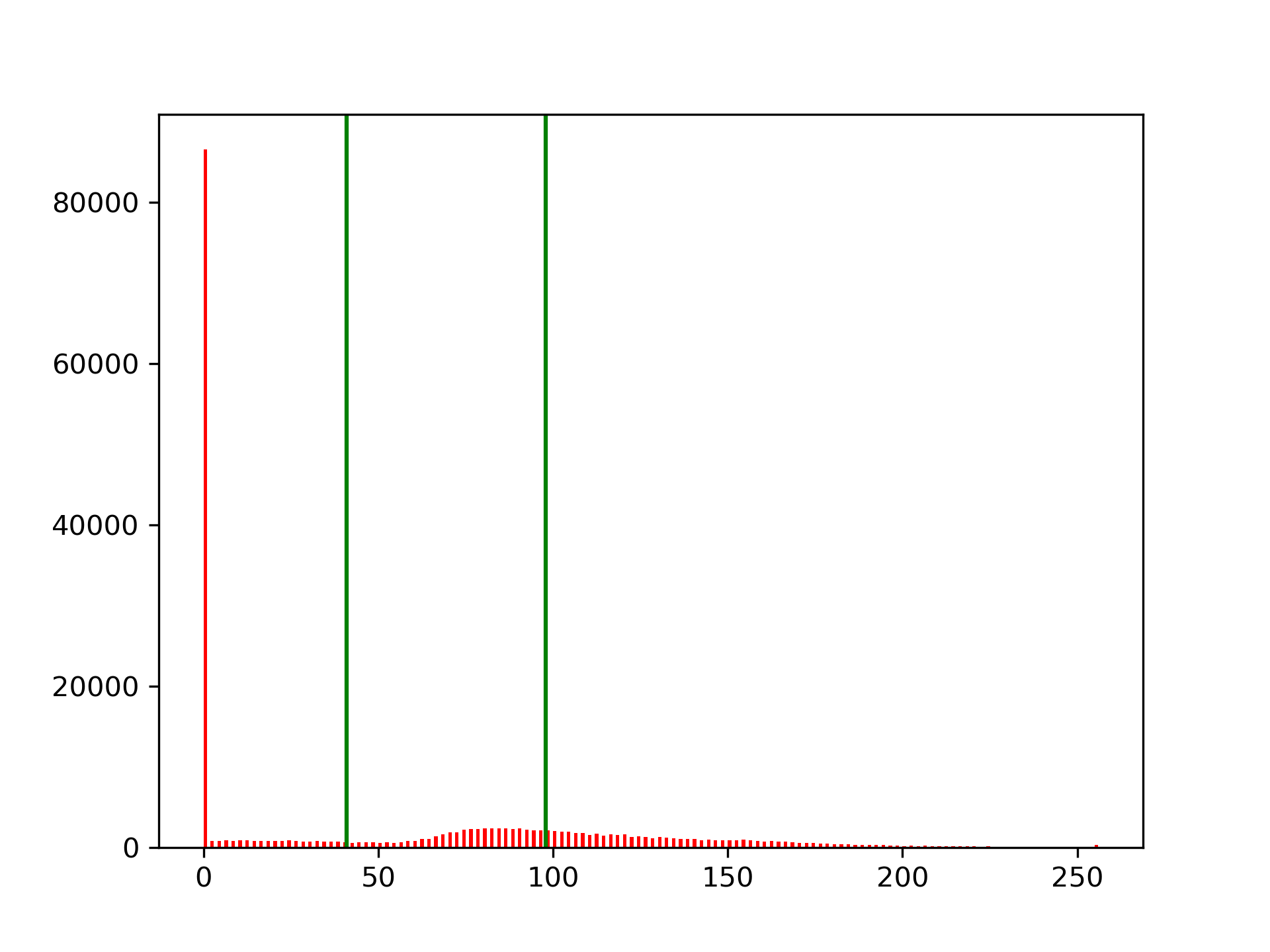}}
    \subfloat[3 Thresholds]{\includegraphics[scale=0.2]{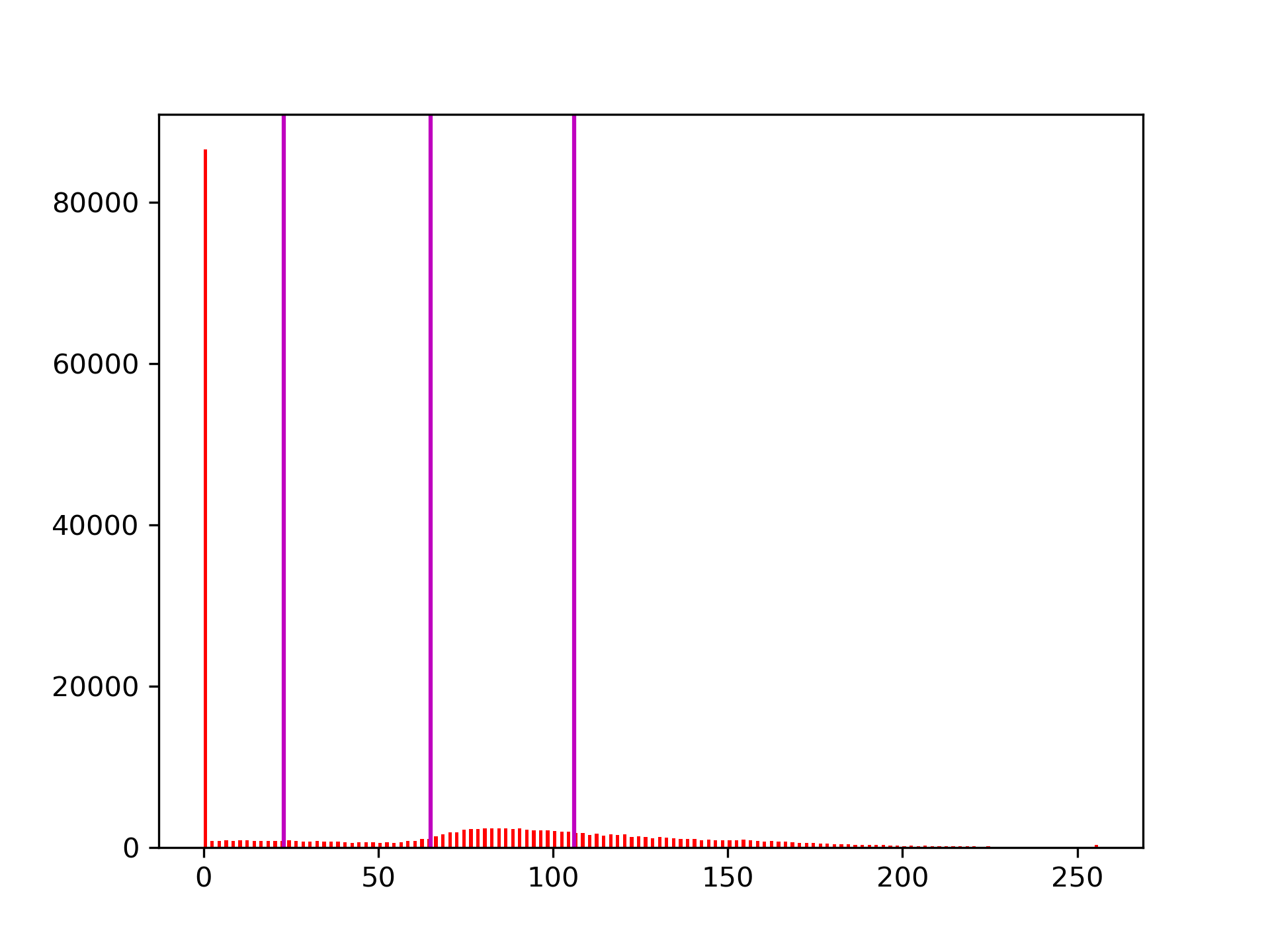}}
    \subfloat[4 Thresholds]{\includegraphics[scale=0.2]{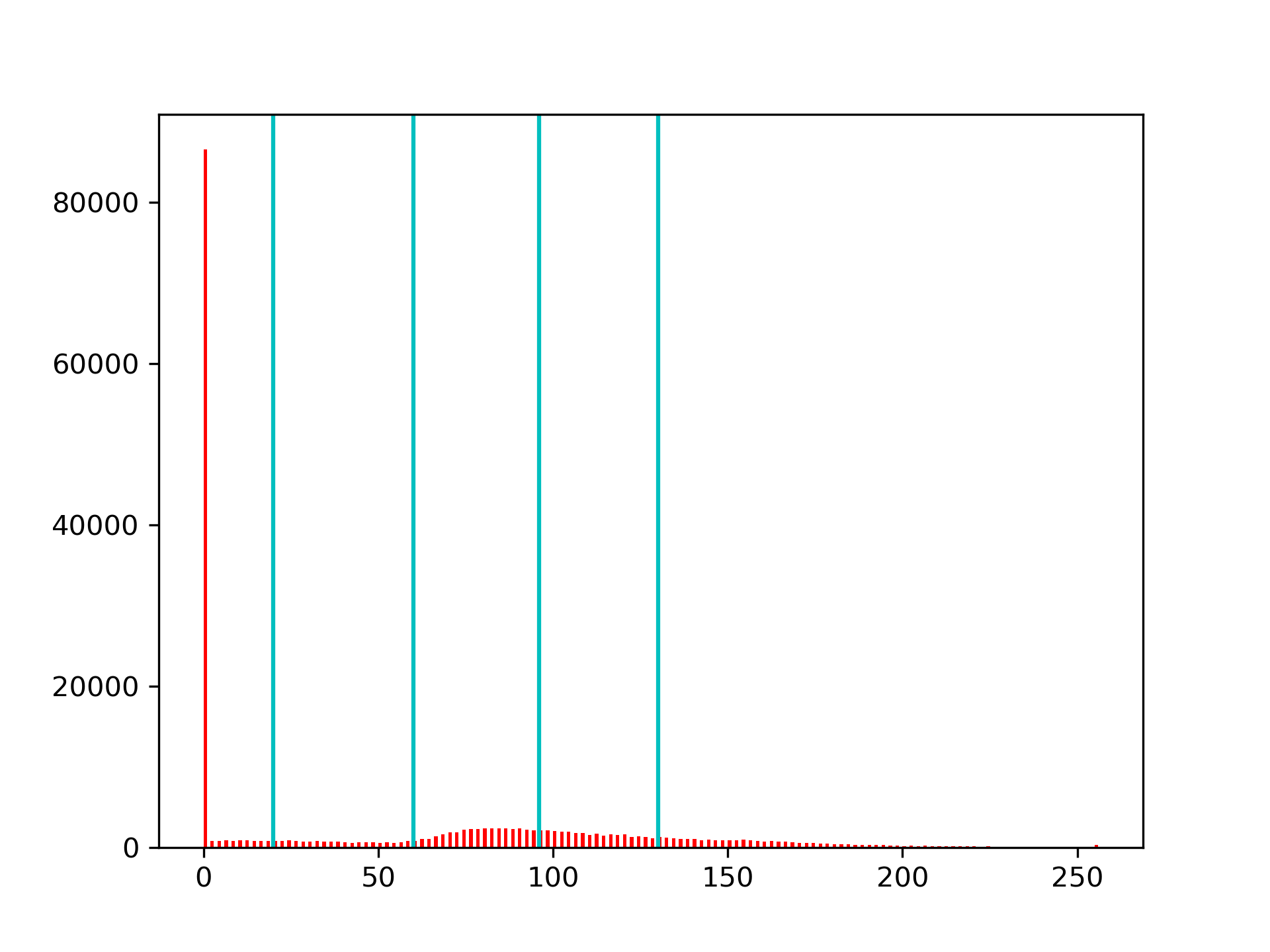}}
    \subfloat[5 Thresholds]{\includegraphics[scale=0.2]{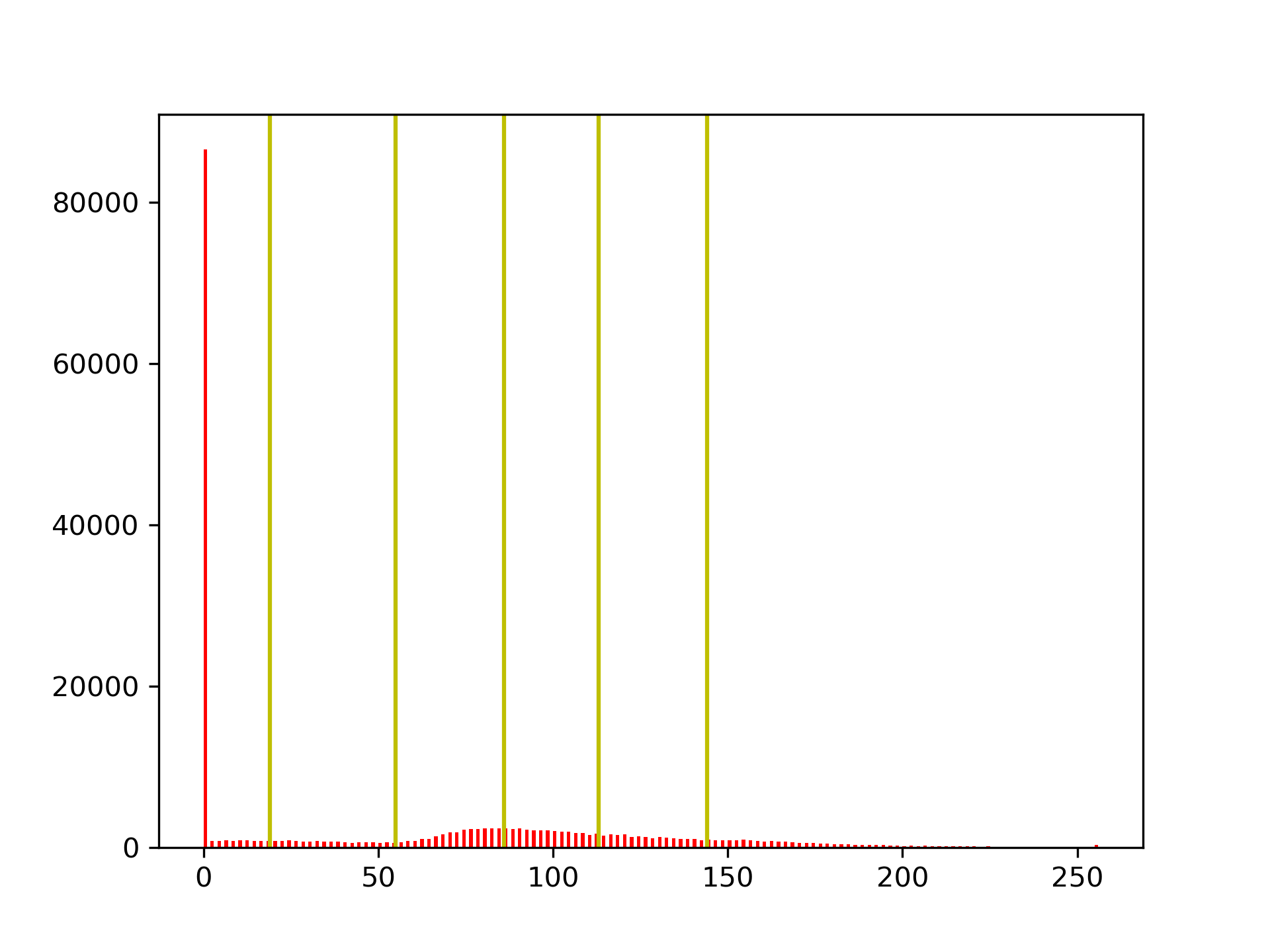}}
    \caption{Original image and thresholded images for slice 62 of the Harvard WBA dataset with number of thresholds: 2, 3, 4 and 5, with their corresponding histograms and thresholds marked with vertical lines.}
    \label{h62}
\end{figure*}

\begin{figure*}
    \centering
    \subfloat[Original]{\includegraphics[scale=0.3]{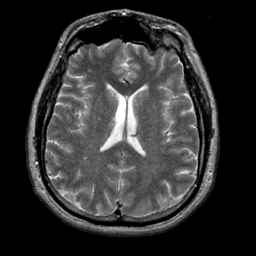}}\;\;
    \subfloat[2 Thresholds]{\includegraphics[scale=0.3]{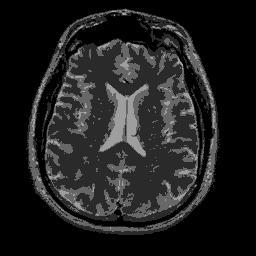}}\;\;
    \subfloat[3 Thresholds]{\includegraphics[scale=0.3]{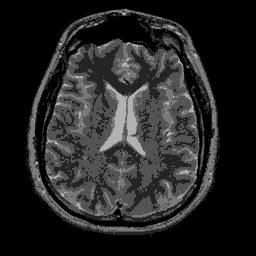}}\;\;
    \subfloat[4 Thresholds]{\includegraphics[scale=0.3]{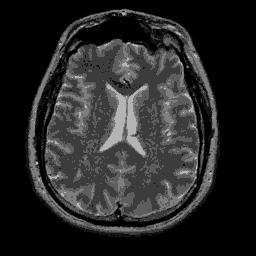}}\;\;
    \subfloat[5 Thresholds]{\includegraphics[scale=0.3]{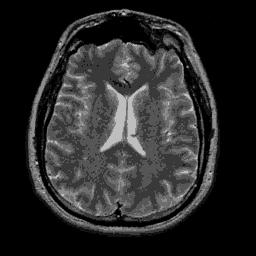}}\\
    \subfloat[Original]{\includegraphics[scale=0.2]{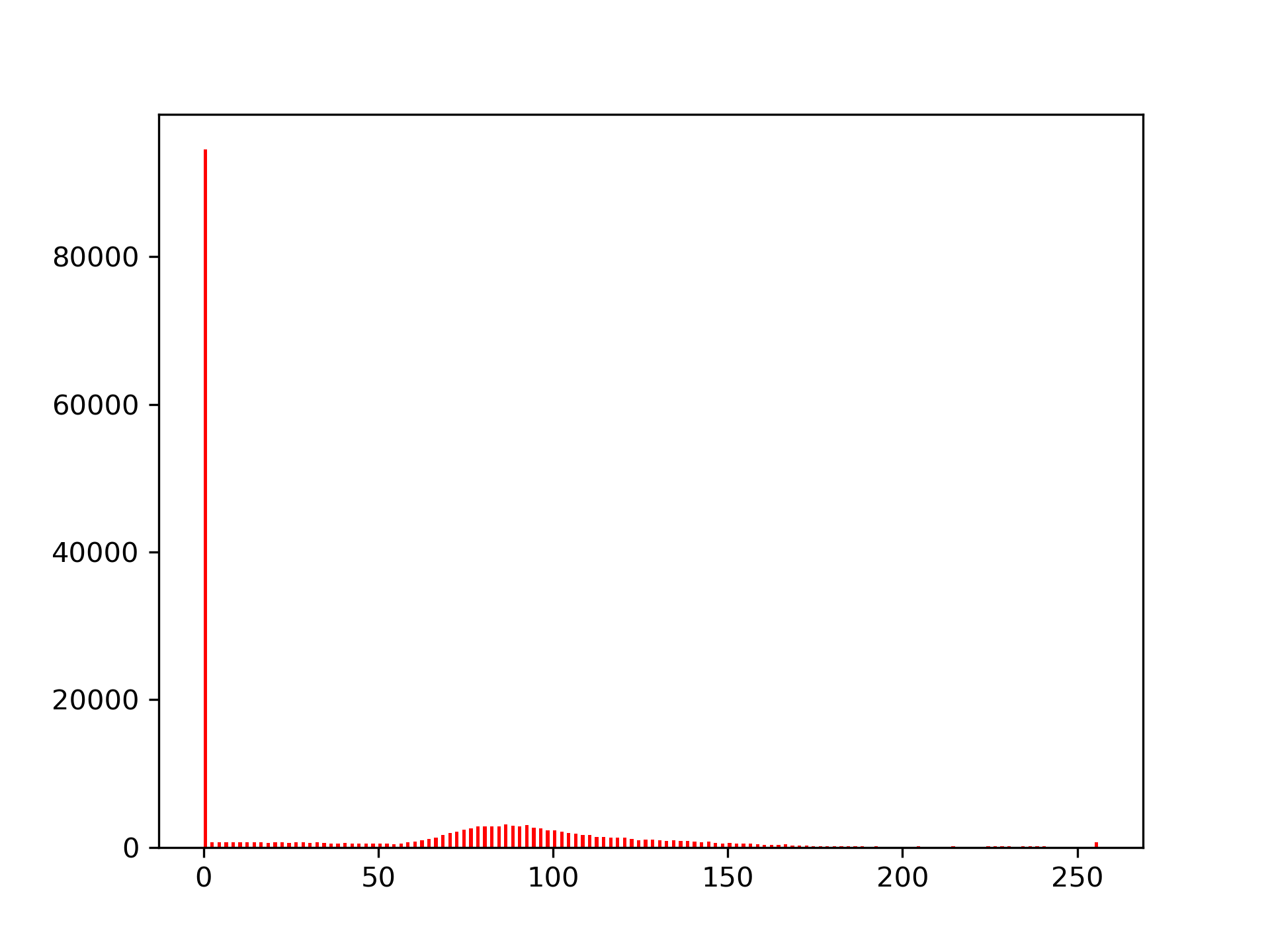}}
    \subfloat[2 Thresholds]{\includegraphics[scale=0.2]{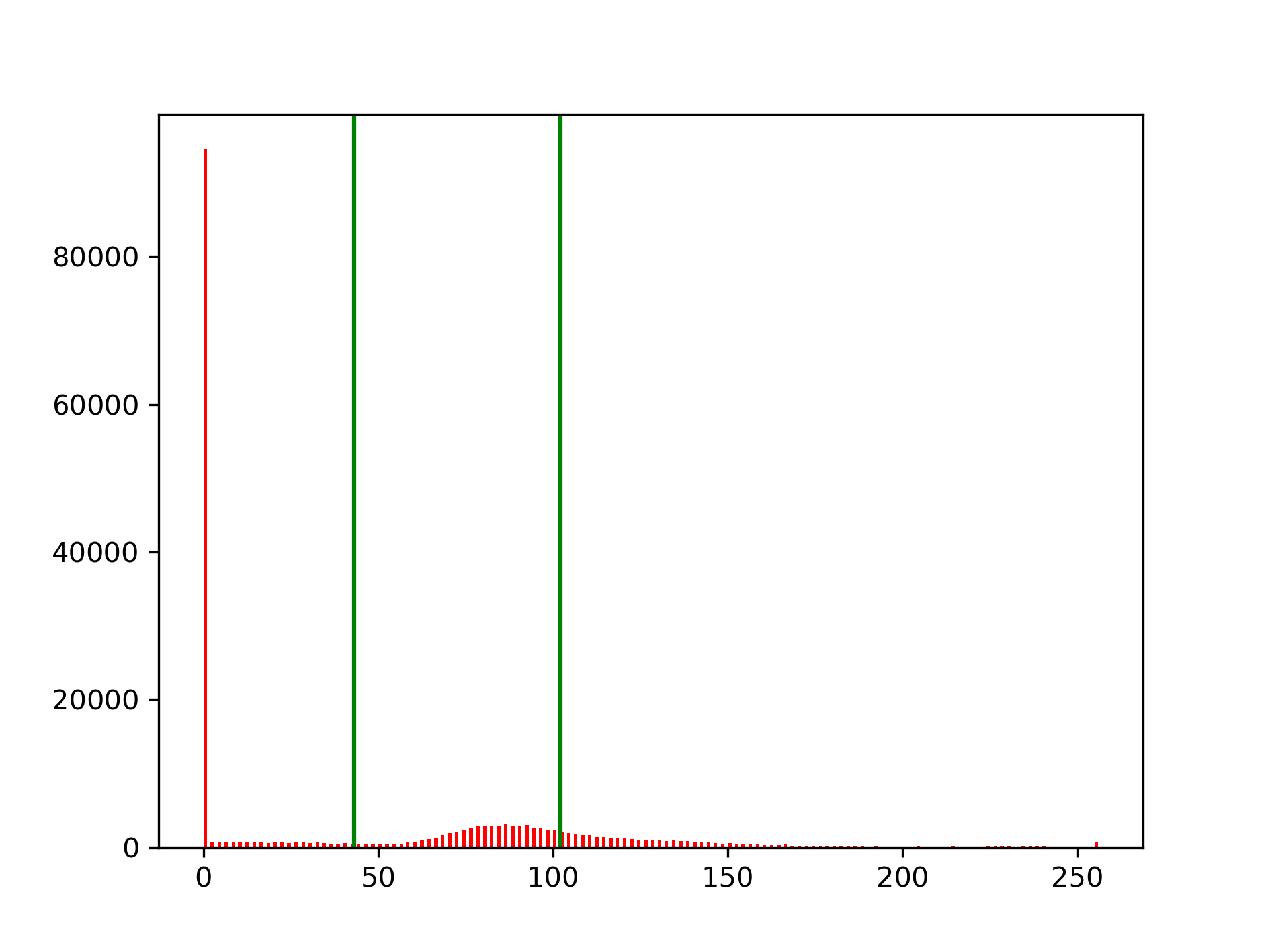}}
    \subfloat[3 Thresholds]{\includegraphics[scale=0.2]{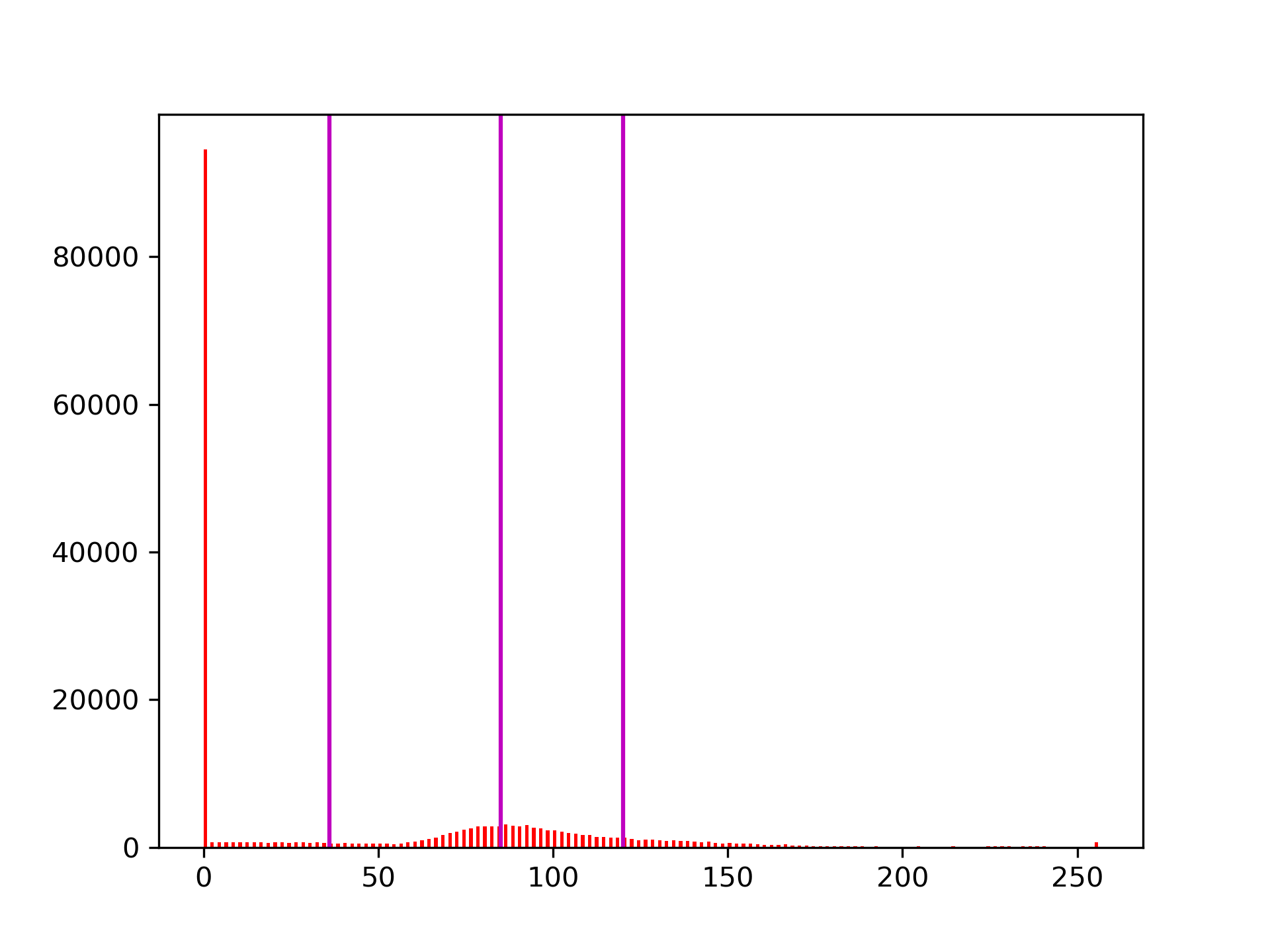}}
    \subfloat[4 Thresholds]{\includegraphics[scale=0.2]{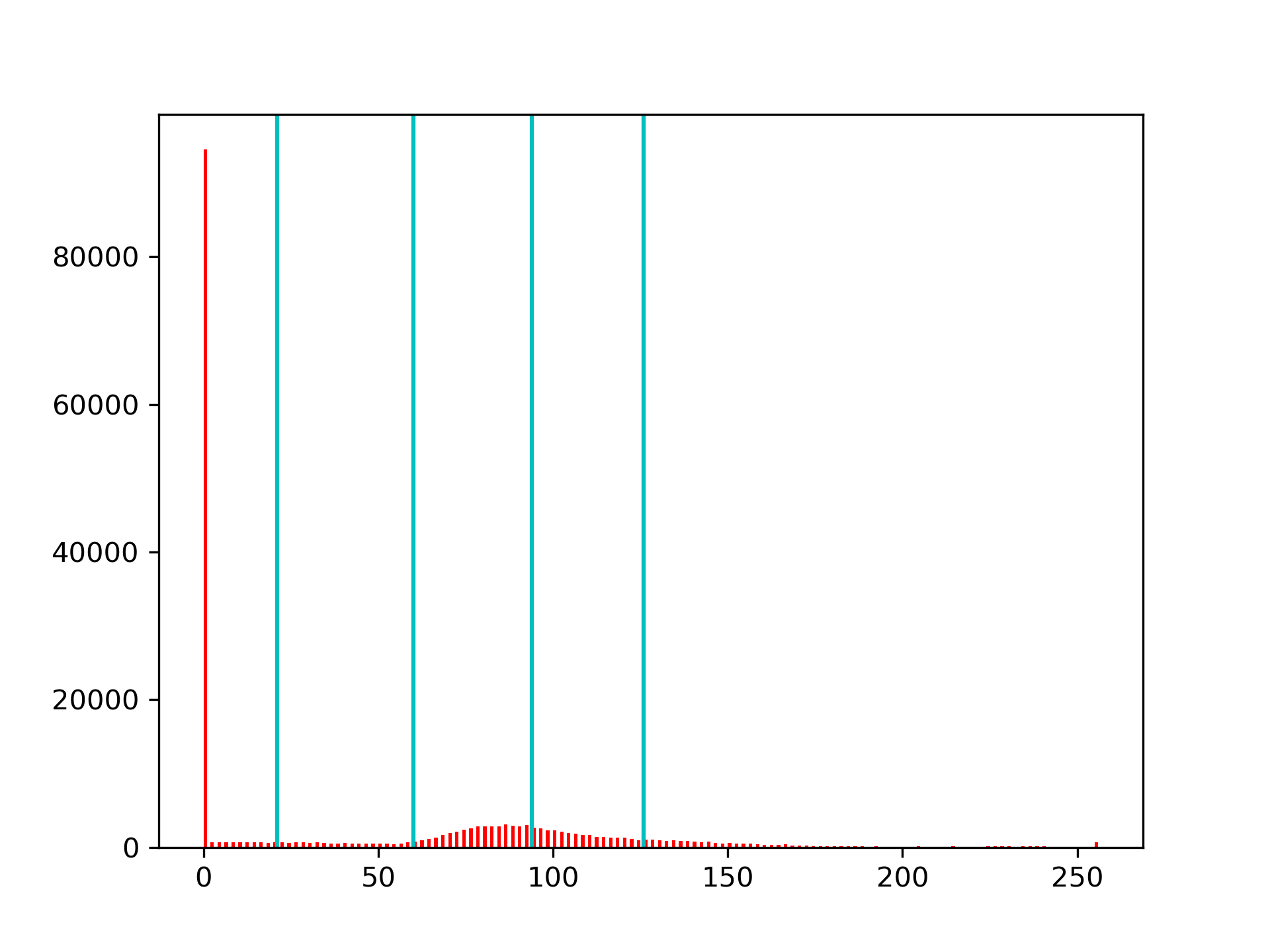}}
    \subfloat[5 Thresholds]{\includegraphics[scale=0.2]{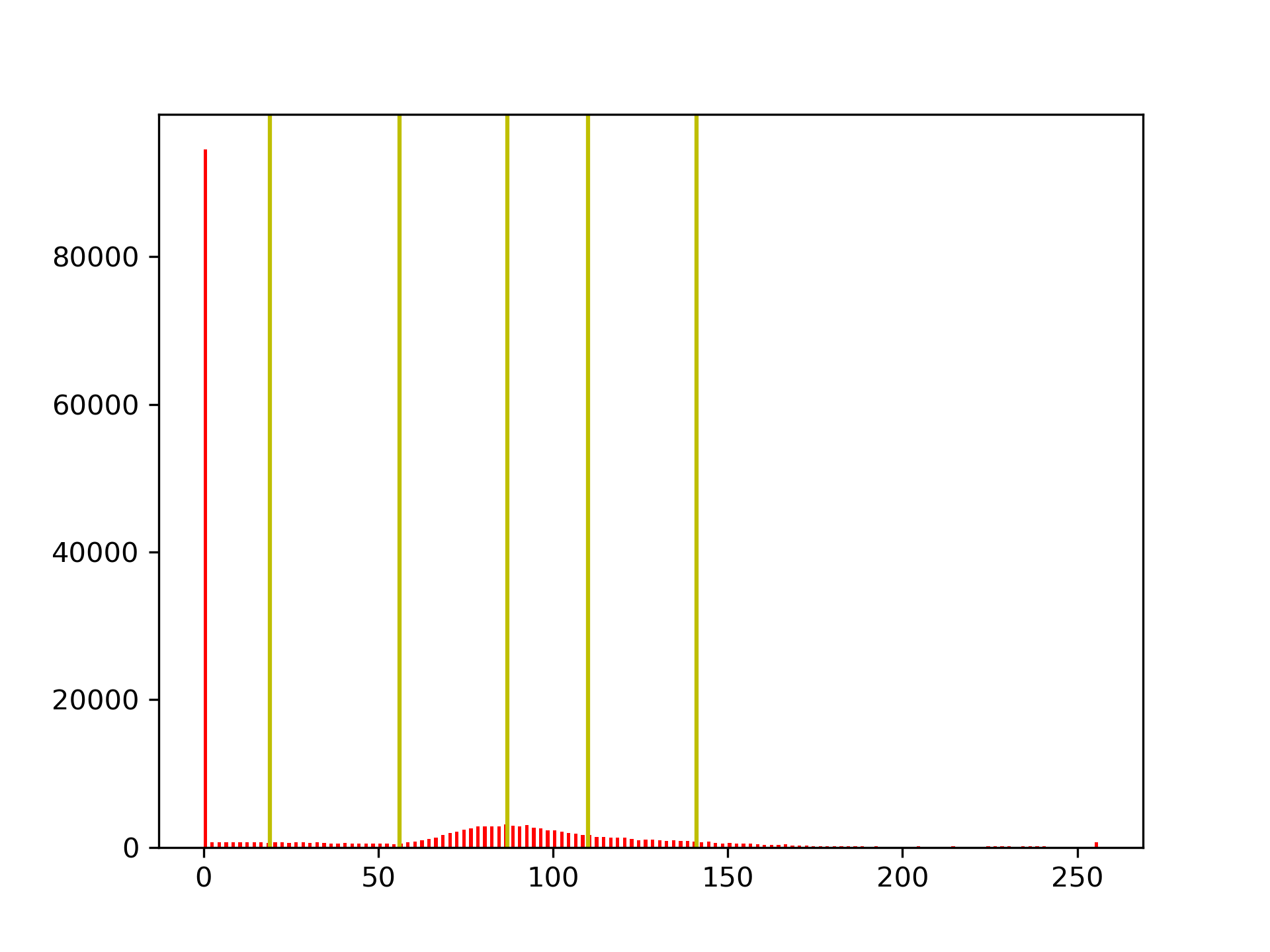}}
    \caption{Original image and thresholded images for slice 82 of the Harvard WBA dataset with number of thresholds: 2, 3, 4 and 5, with their corresponding histograms and thresholds marked with vertical lines.}
    \label{h82}
\end{figure*}

\begin{figure*}
    \centering
    \subfloat[Original]{\includegraphics[scale=0.4]{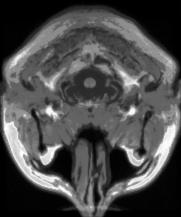}}\;\;
    \subfloat[2 Thresholds]{\includegraphics[scale=0.4]{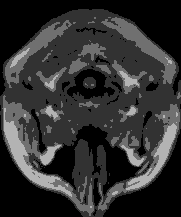}}\;\;
    \subfloat[3 Thresholds]{\includegraphics[scale=0.4]{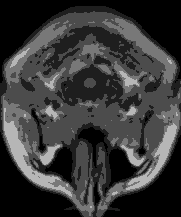}}\;\;
    \subfloat[4 Thresholds]{\includegraphics[scale=0.4]{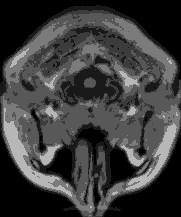}}\;\;
    \subfloat[5 Thresholds]{\includegraphics[scale=0.4]{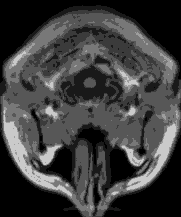}}\\
    \subfloat[Original]{\includegraphics[scale=0.2]{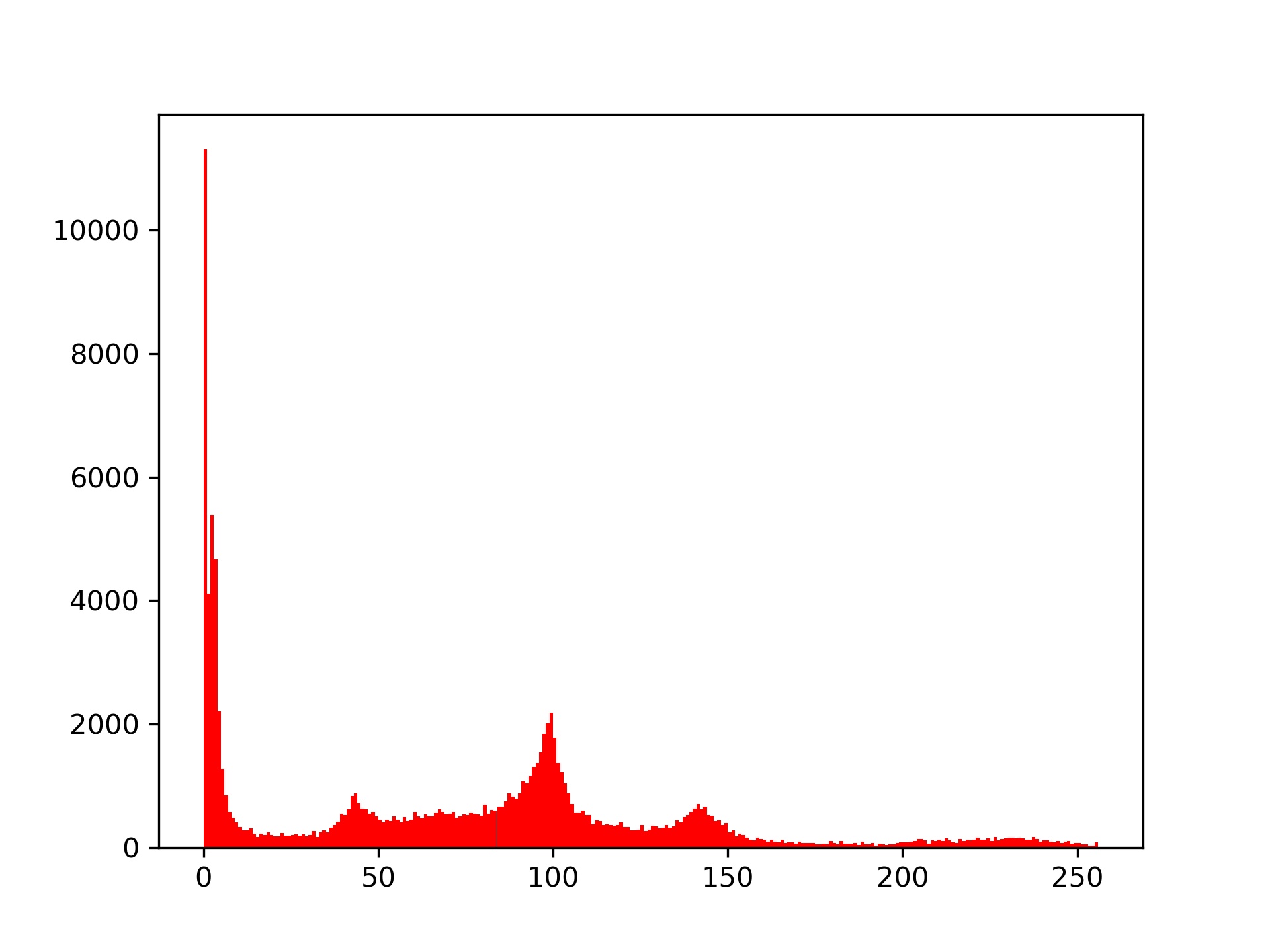}}
    \subfloat[2 Thresholds]{\includegraphics[scale=0.2]{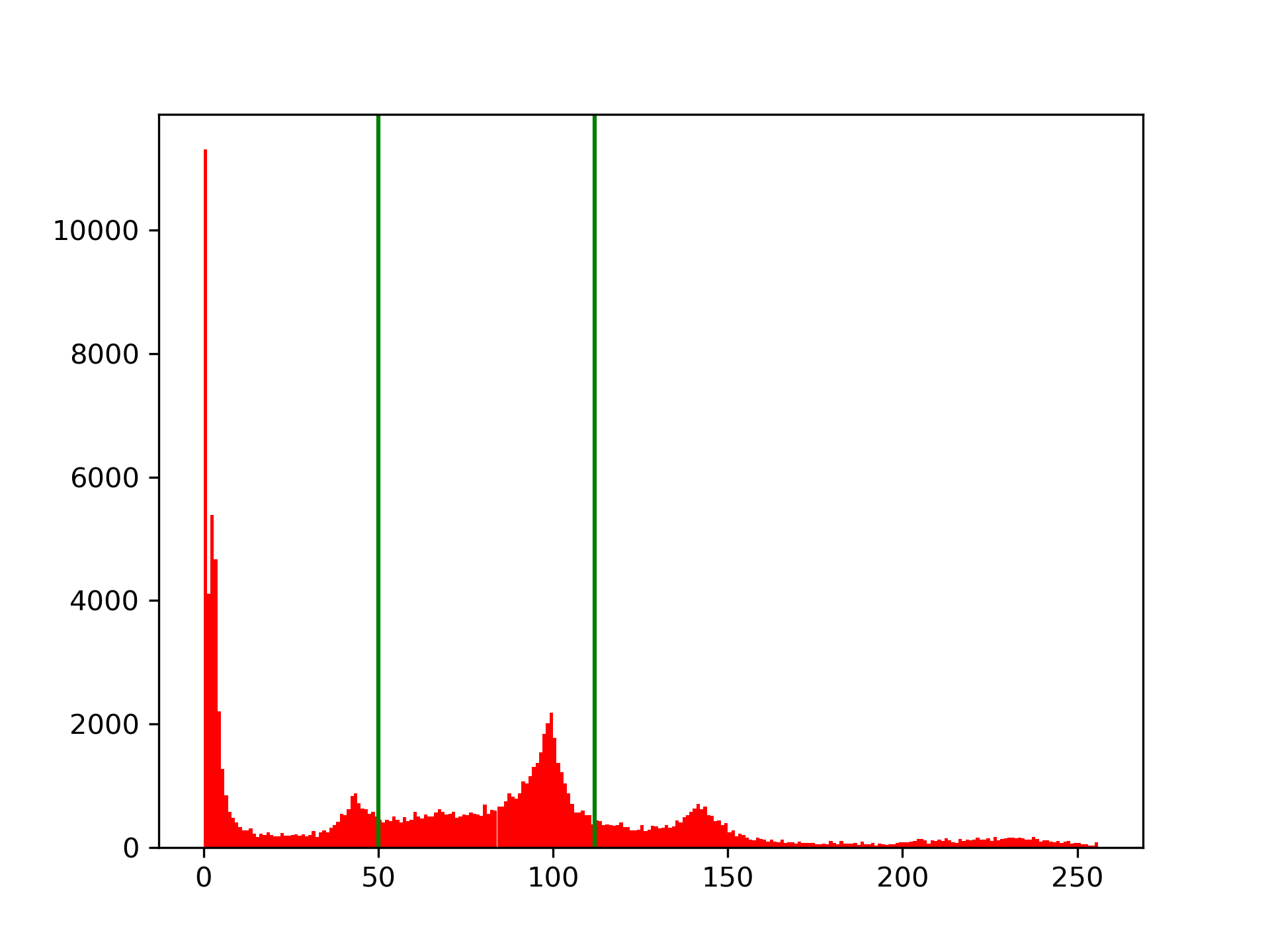}}
    \subfloat[3 Thresholds]{\includegraphics[scale=0.2]{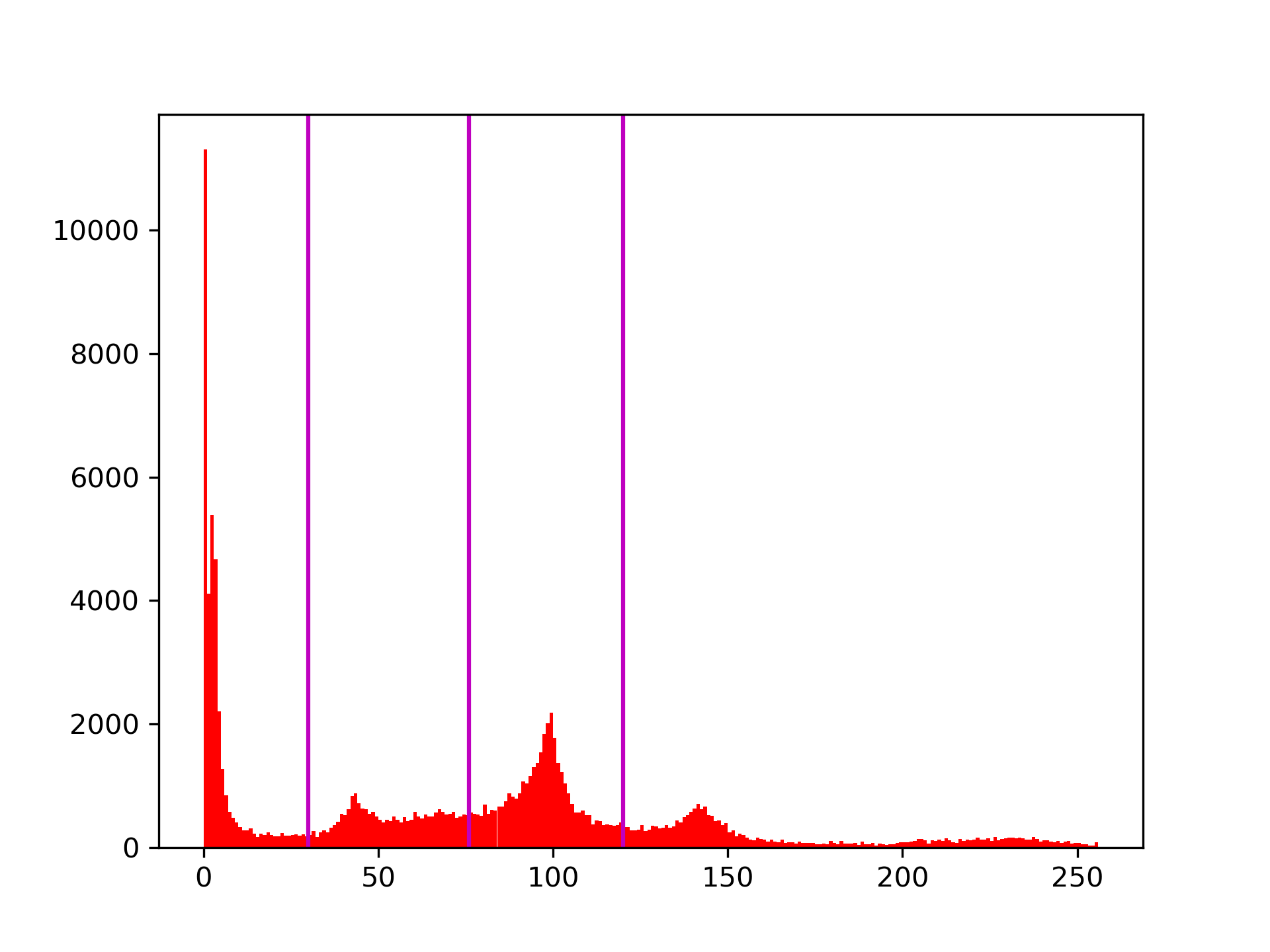}}
    \subfloat[4 Thresholds]{\includegraphics[scale=0.2]{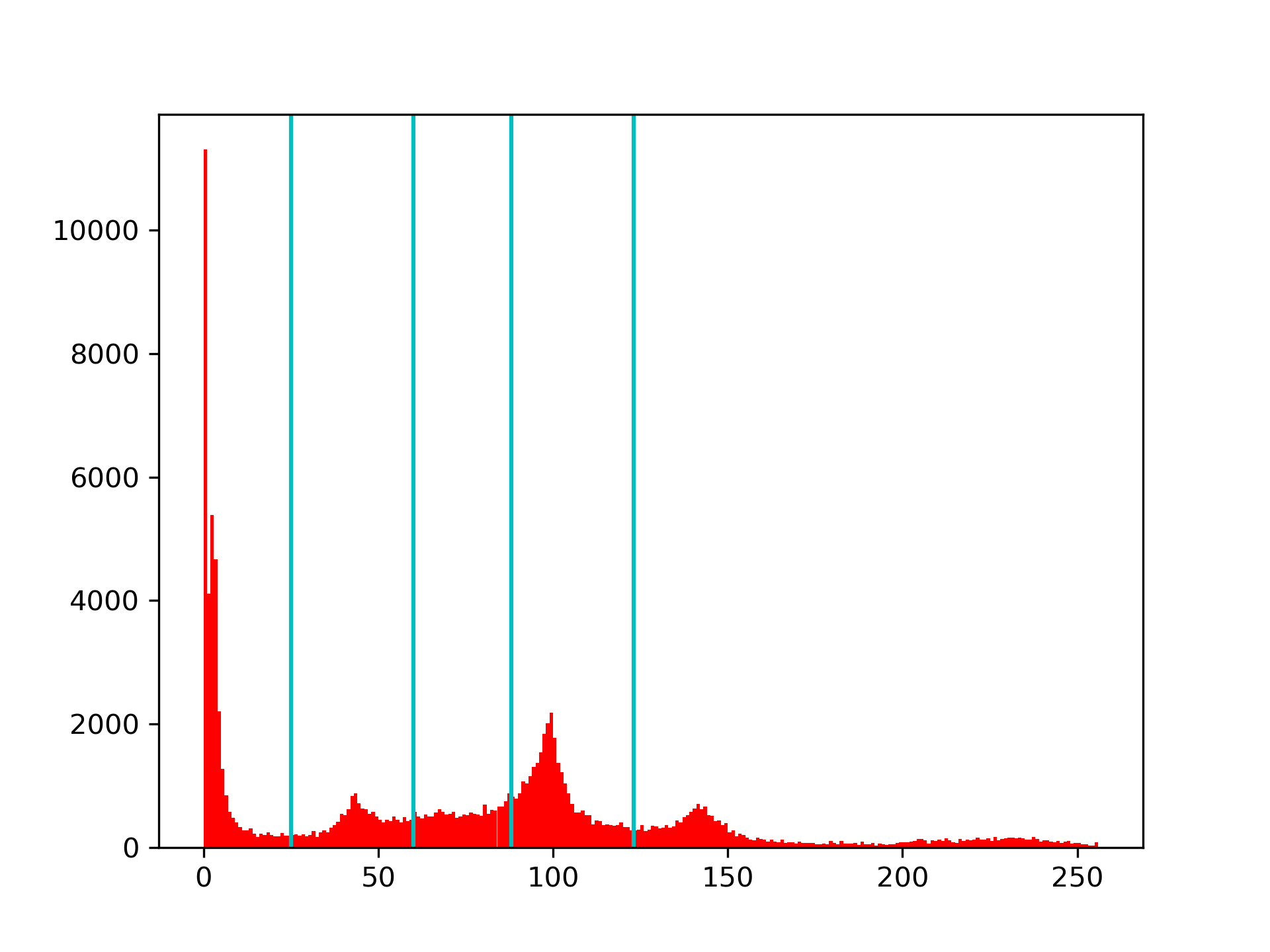}}
    \subfloat[5 Thresholds]{\includegraphics[scale=0.2]{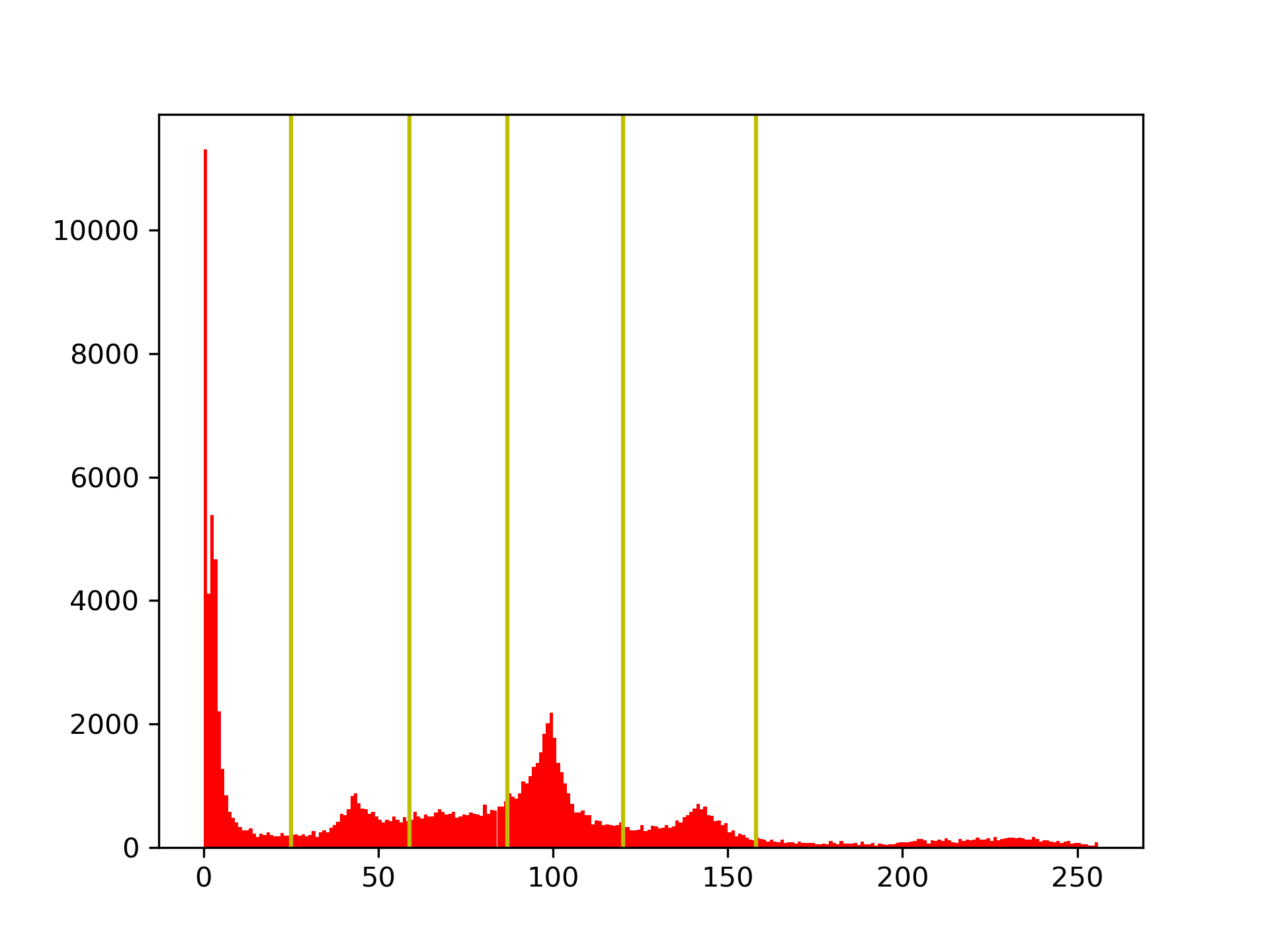}}
    \caption{Original image and thresholded images for Brainweb dataset image Z1 with number of thresholds: 2, 3, 4 and 5, with their corresponding histograms and thresholds marked with vertical lines.}
    \label{mri1}
\end{figure*}

\begin{figure*}
    \centering
    \subfloat[Original]{\includegraphics[scale=0.4]{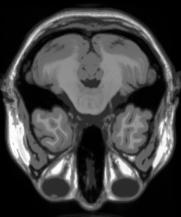}}\;\;
    \subfloat[2 Thresholds]{\includegraphics[scale=0.4]{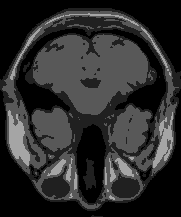}}\;\;
    \subfloat[3 Thresholds]{\includegraphics[scale=0.4]{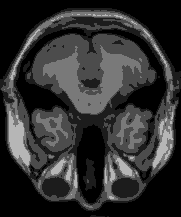}}\;\;
    \subfloat[4 Thresholds]{\includegraphics[scale=0.4]{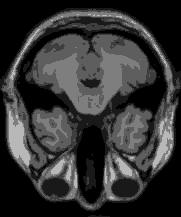}}\;\;
    \subfloat[5 Thresholds]{\includegraphics[scale=0.4]{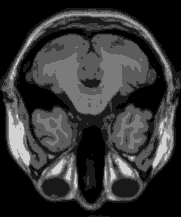}}\\
    \subfloat[Original]{\includegraphics[scale=0.2]{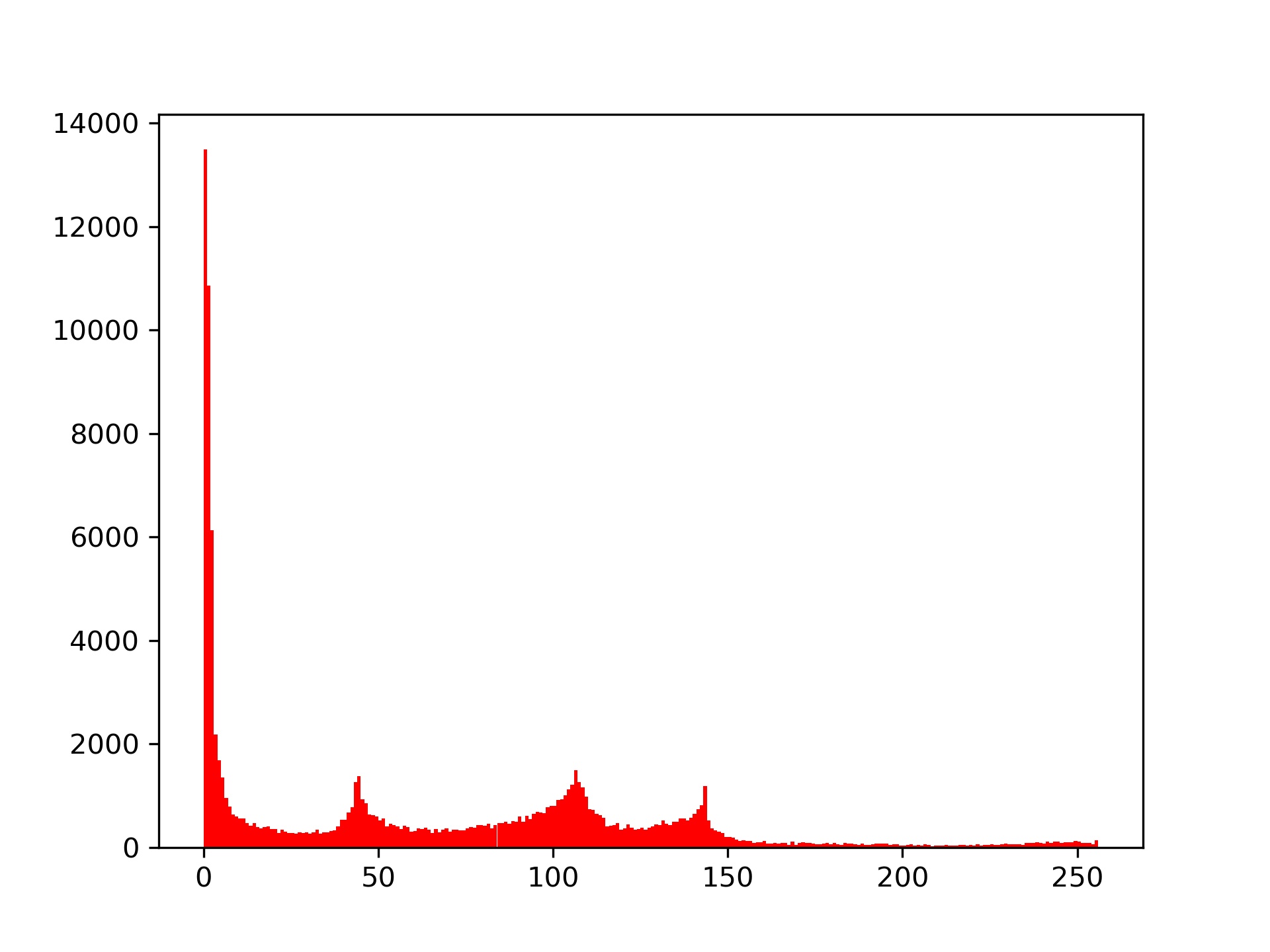}}
    \subfloat[2 Thresholds]{\includegraphics[scale=0.2]{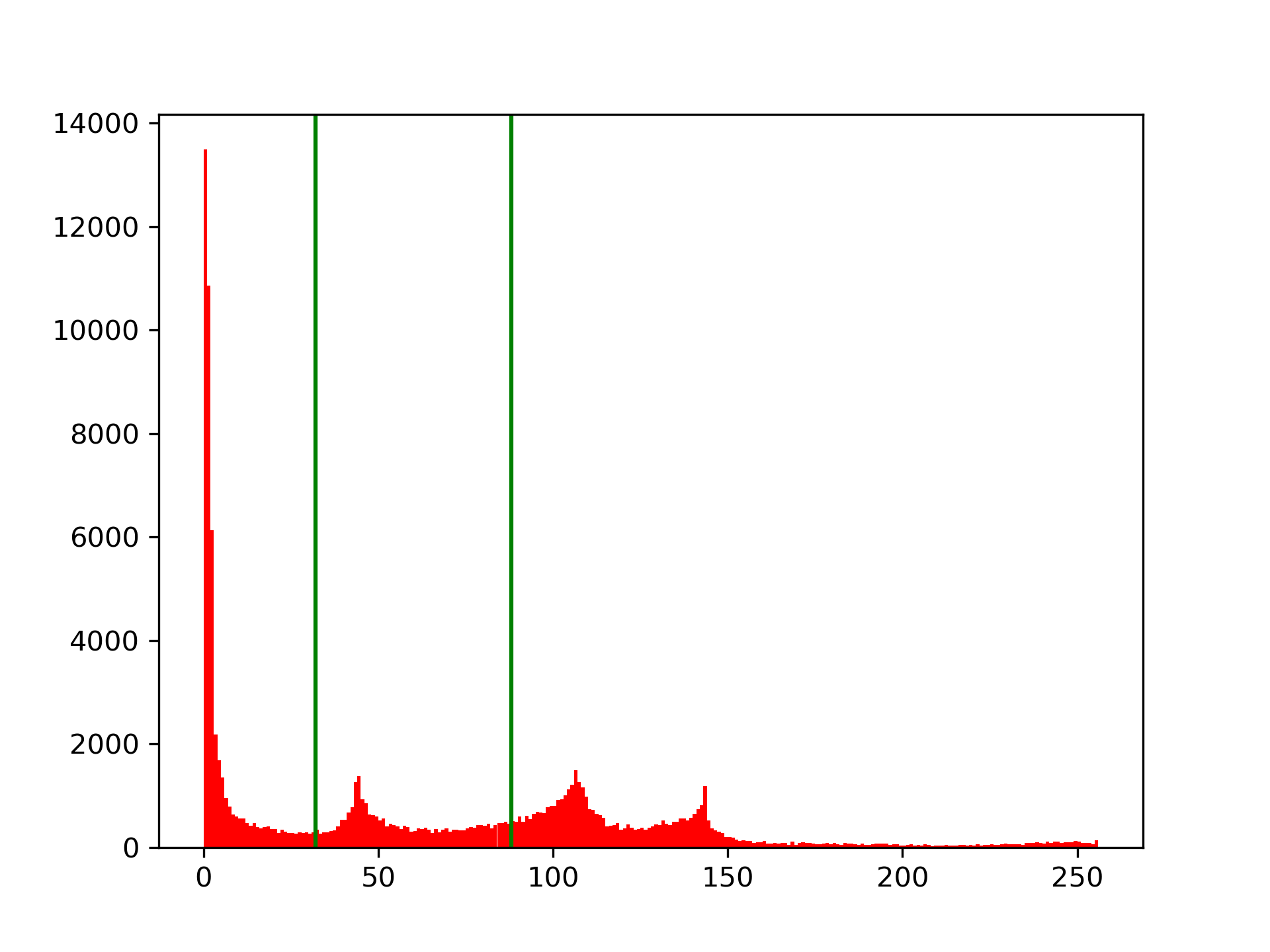}}
    \subfloat[3 Thresholds]{\includegraphics[scale=0.2]{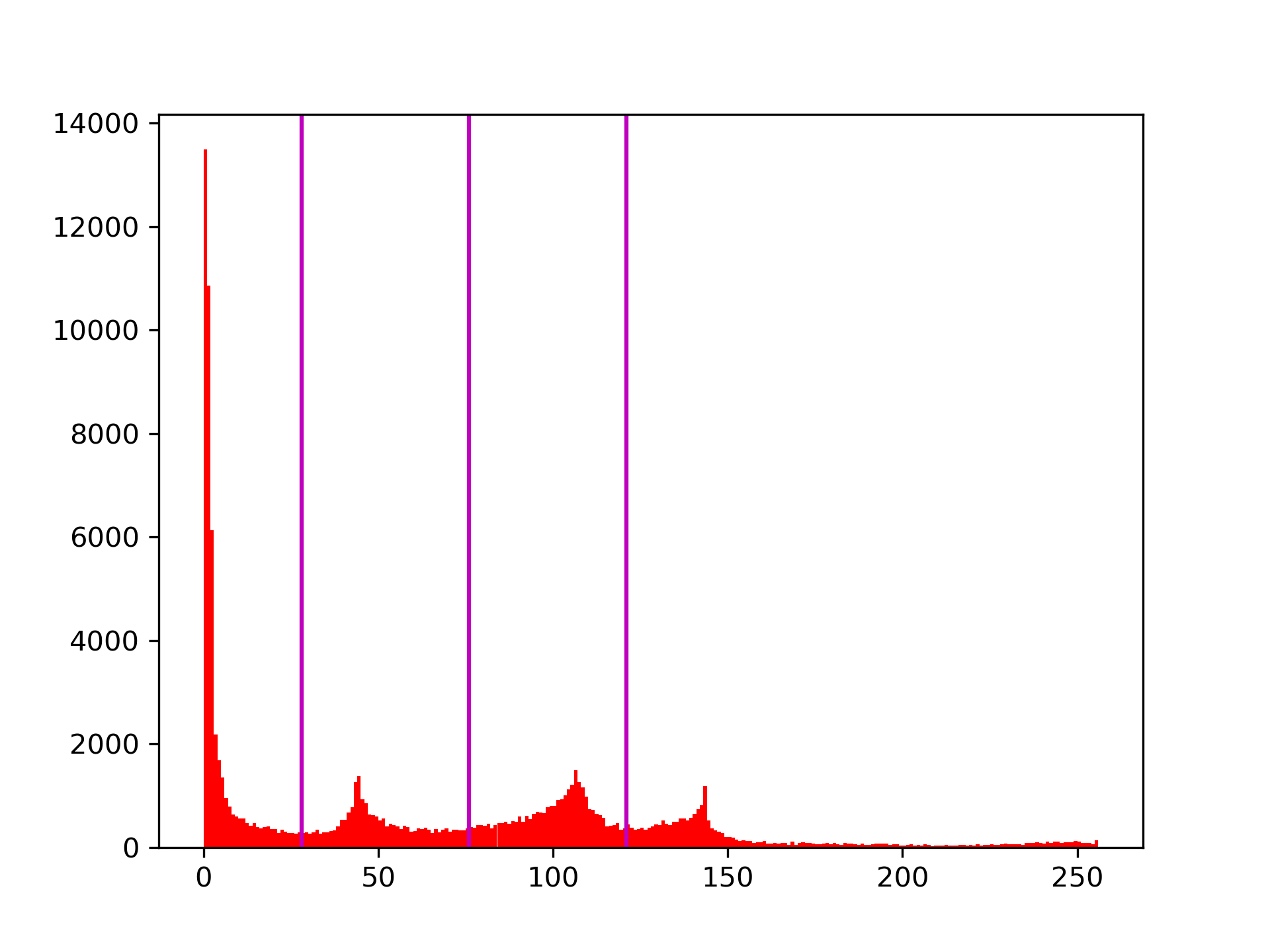}}
    \subfloat[4 Thresholds]{\includegraphics[scale=0.2]{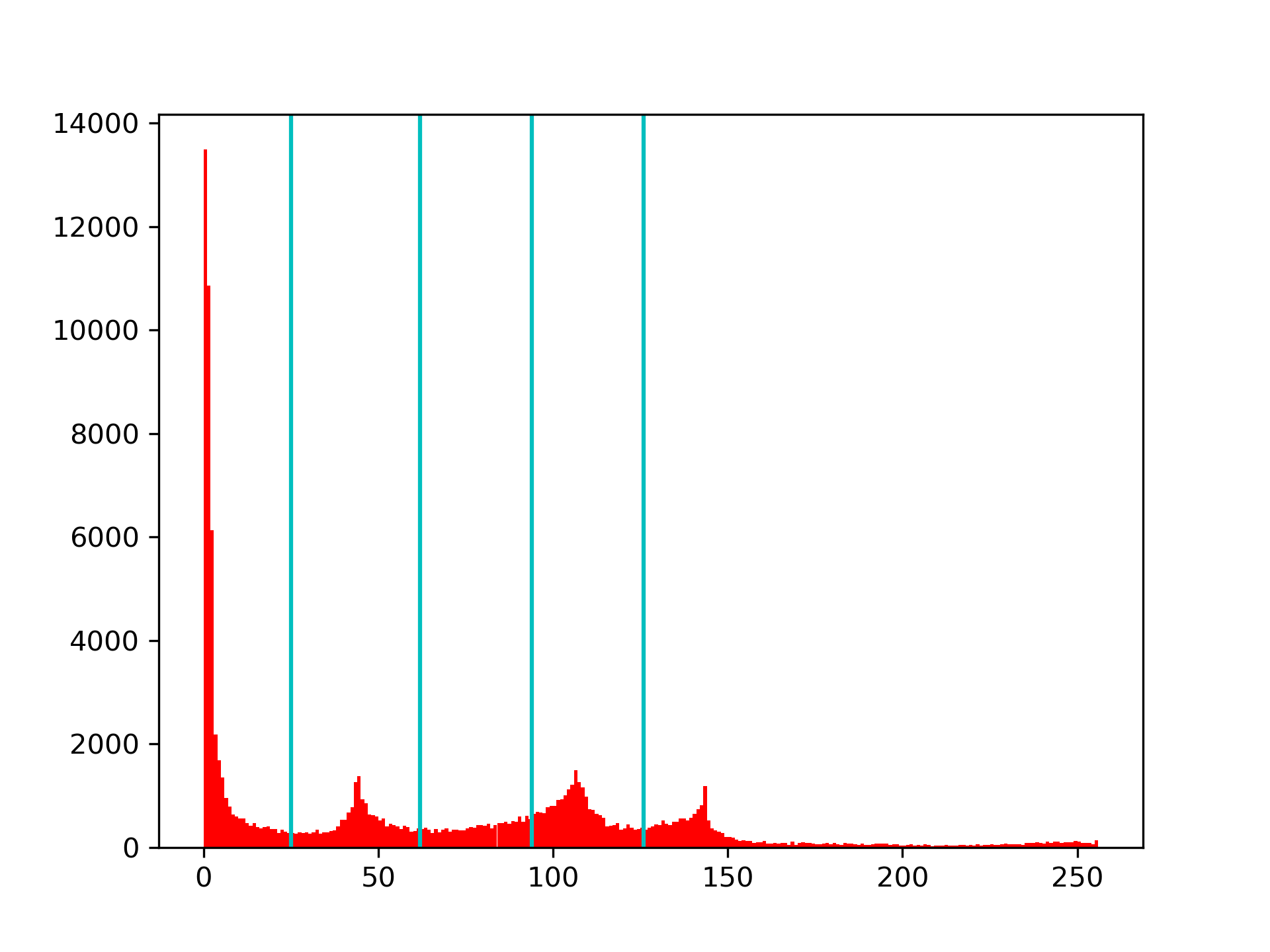}}
    \subfloat[5 Thresholds]{\includegraphics[scale=0.2]{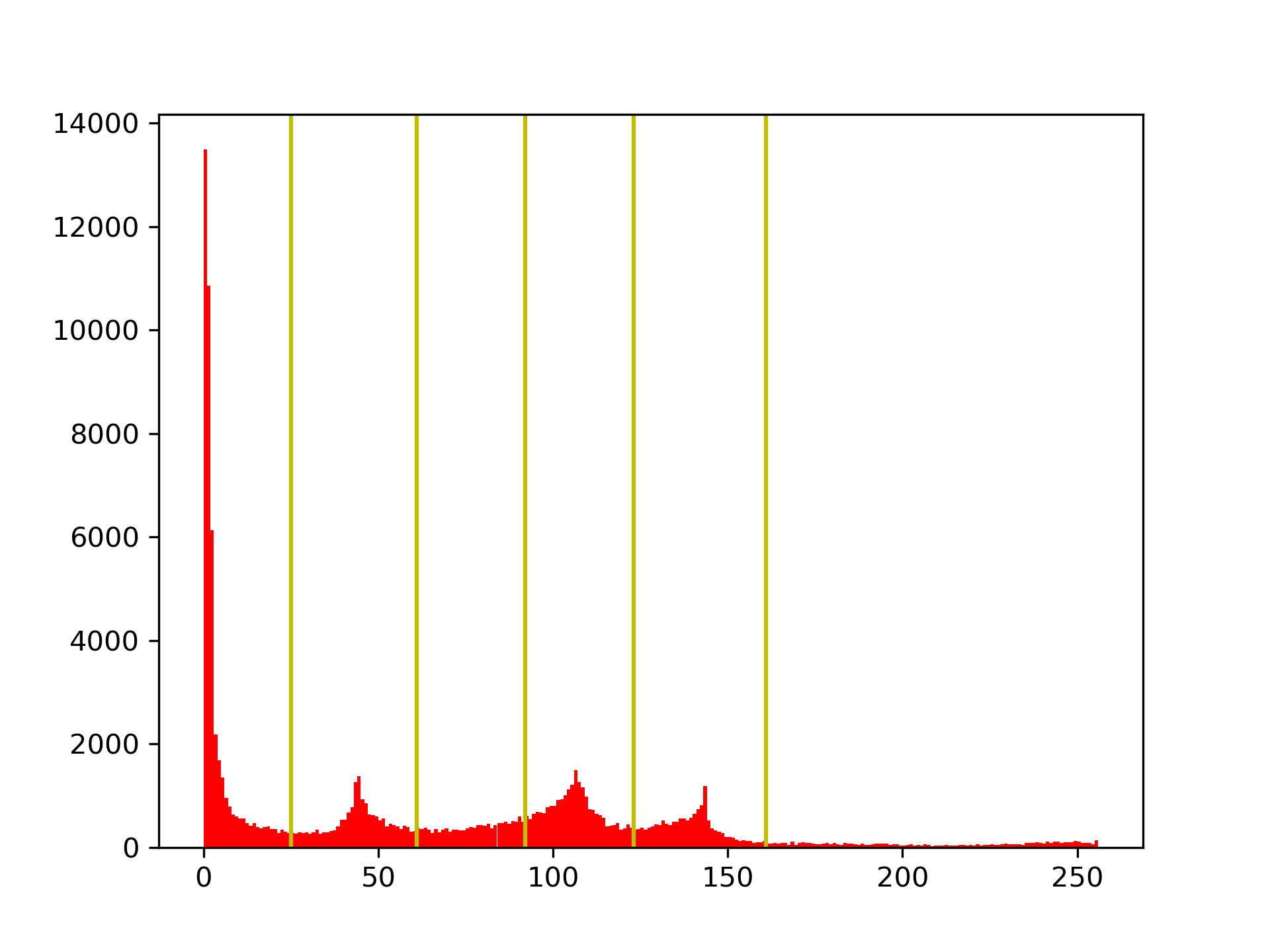}}
    \caption{Original image and thresholded images for Brainweb dataset image Z36 with number of thresholds: 2, 3, 4 and 5, with their corresponding histograms and thresholds marked with vertical lines.}
    \label{mri36}
\end{figure*}

\begin{figure*}
    \centering
    \subfloat[Original]{\includegraphics[scale=0.4]{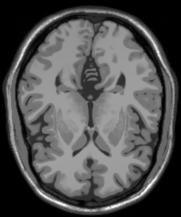}}\;\;
    \subfloat[2 Thresholds]{\includegraphics[scale=0.4]{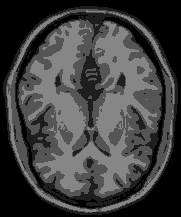}}\;\;
    \subfloat[3 Thresholds]{\includegraphics[scale=0.4]{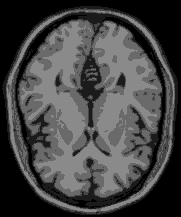}}\;\;
    \subfloat[4 Thresholds]{\includegraphics[scale=0.4]{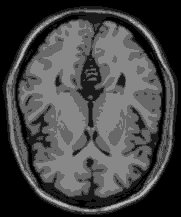}}\;\;
    \subfloat[5 Thresholds]{\includegraphics[scale=0.4]{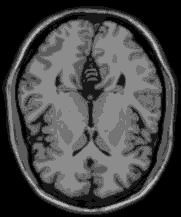}}\\
    \subfloat[Original]{\includegraphics[scale=0.2]{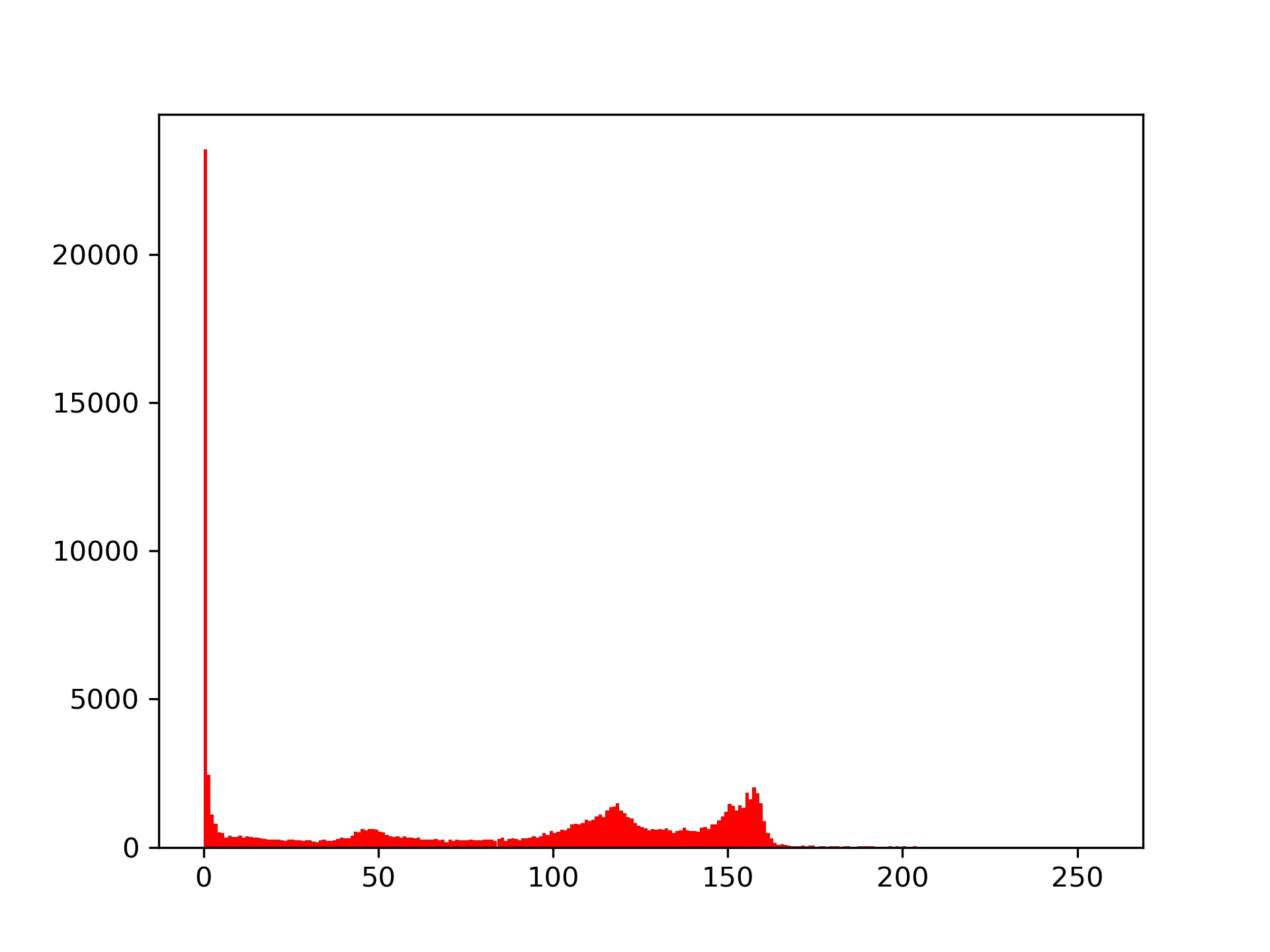}}
    \subfloat[2 Thresholds]{\includegraphics[scale=0.2]{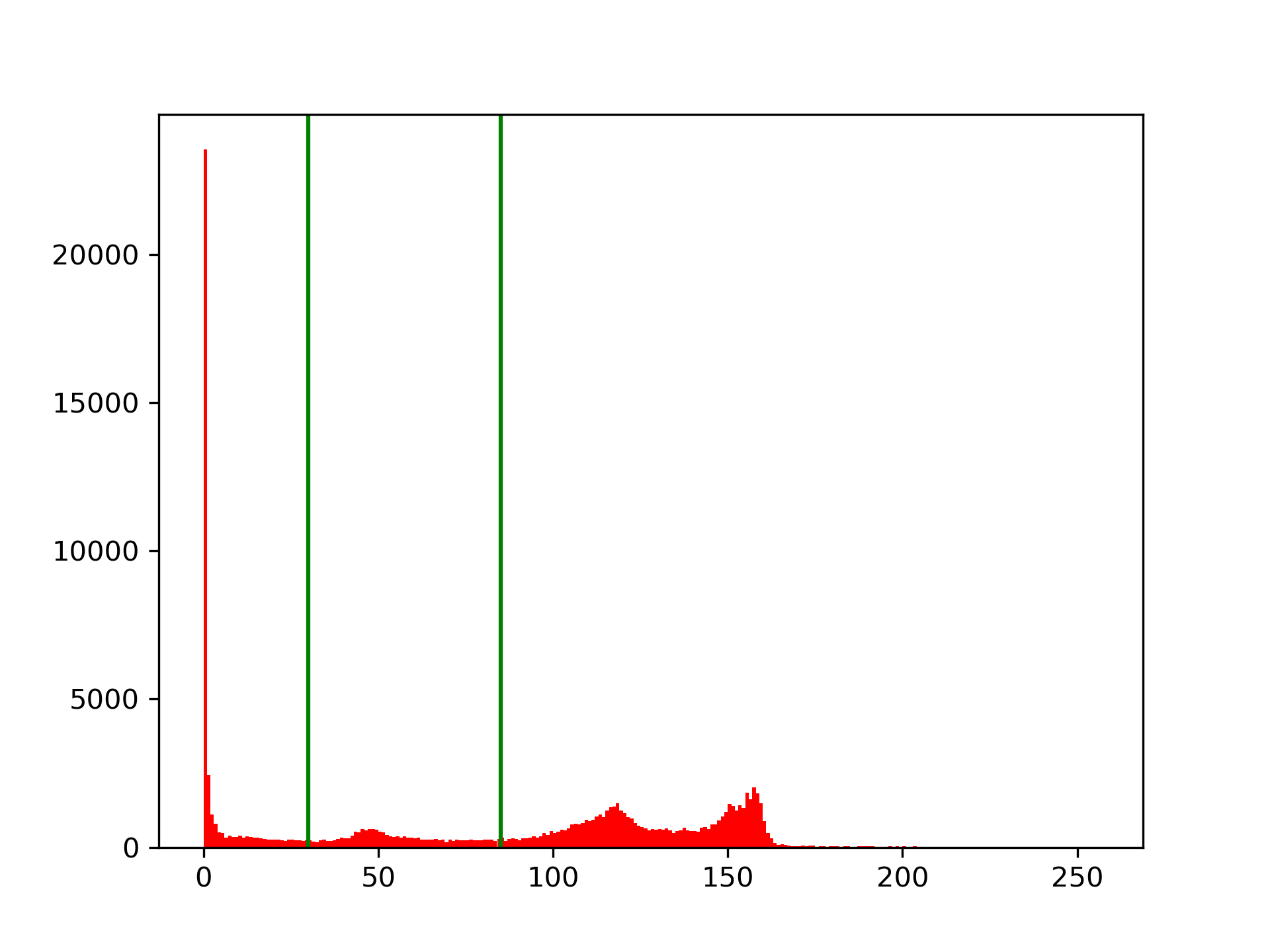}}
    \subfloat[3 Thresholds]{\includegraphics[scale=0.2]{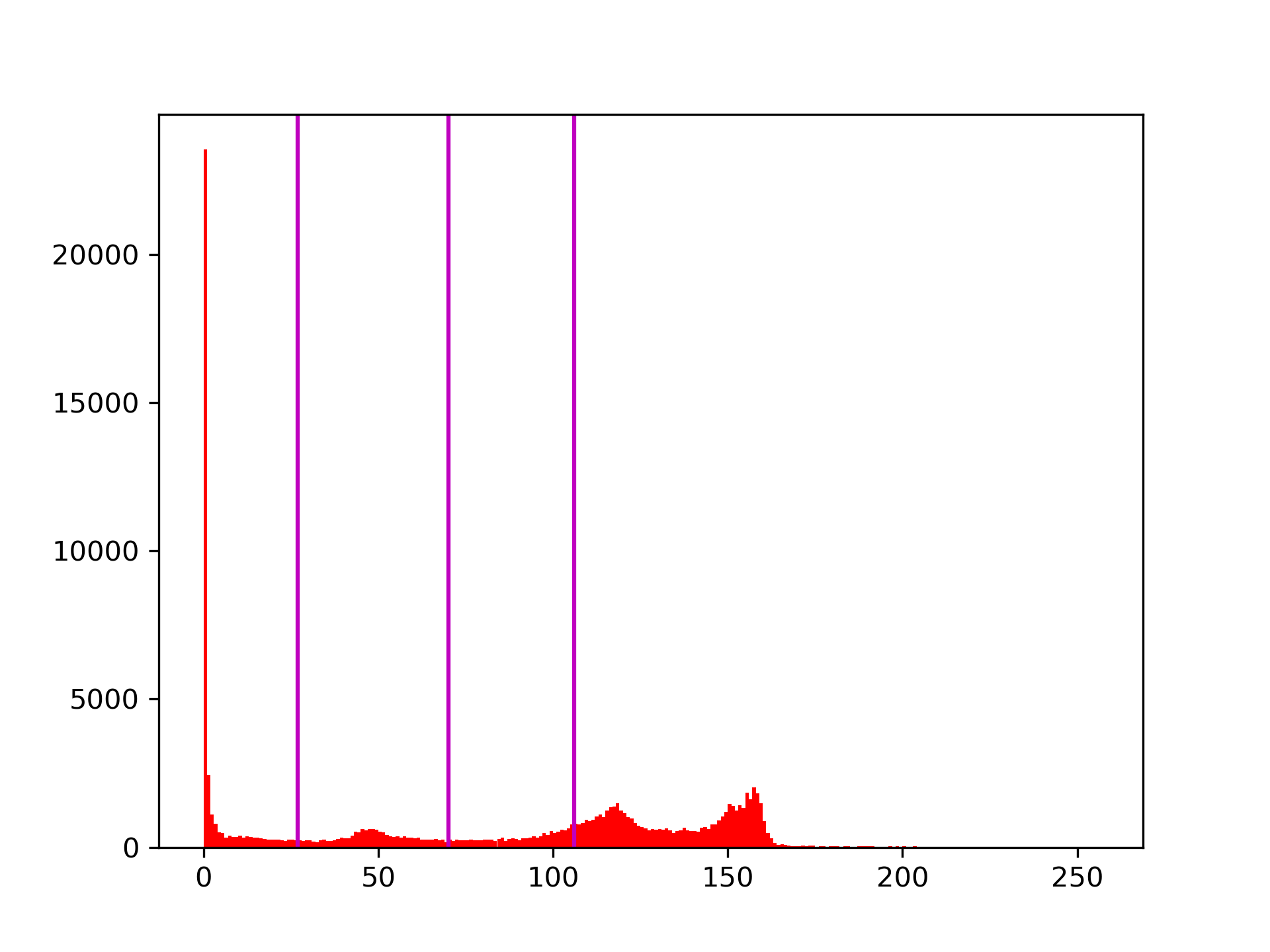}}
    \subfloat[4 Thresholds]{\includegraphics[scale=0.2]{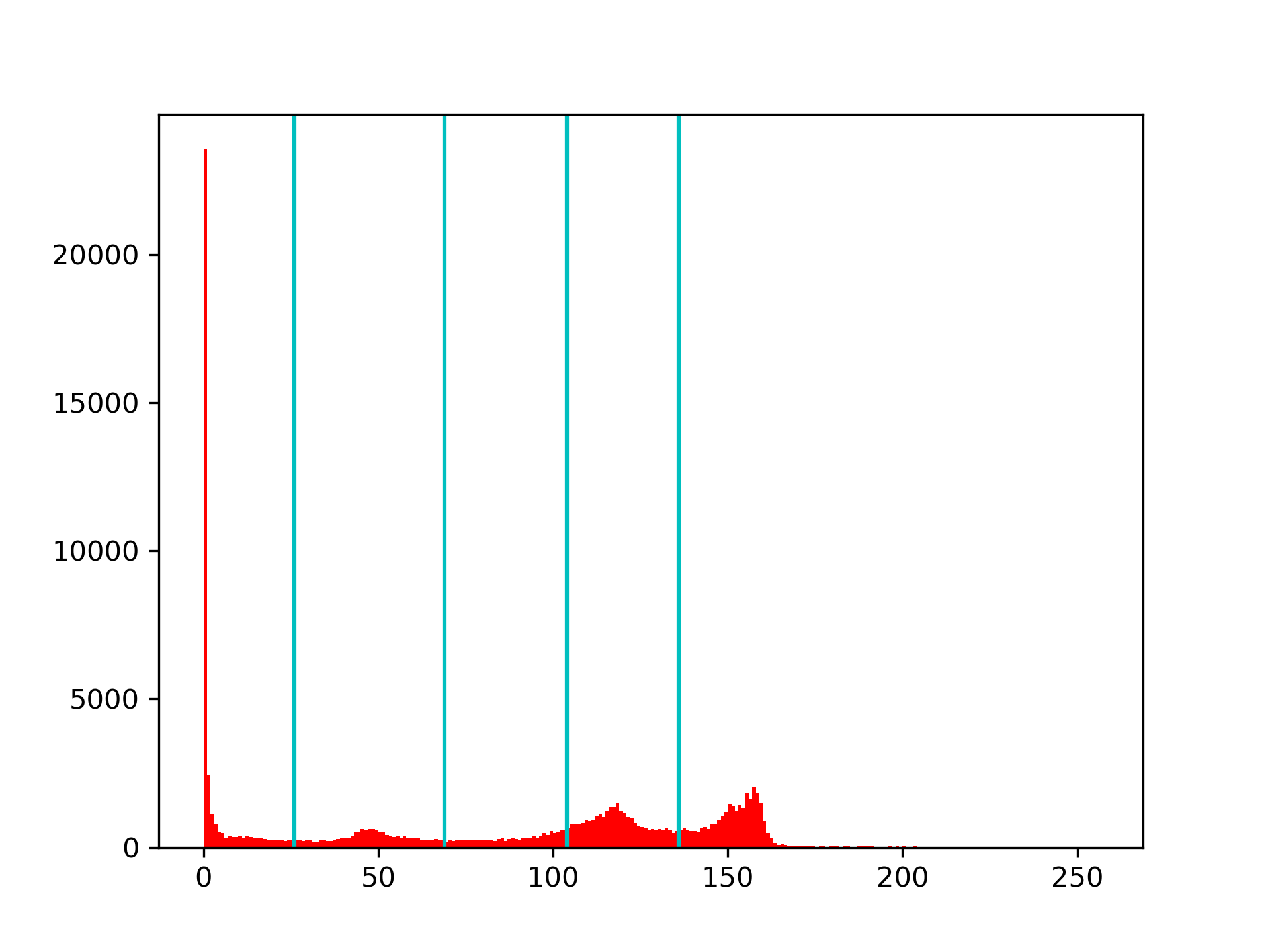}}
    \subfloat[5 Thresholds]{\includegraphics[scale=0.2]{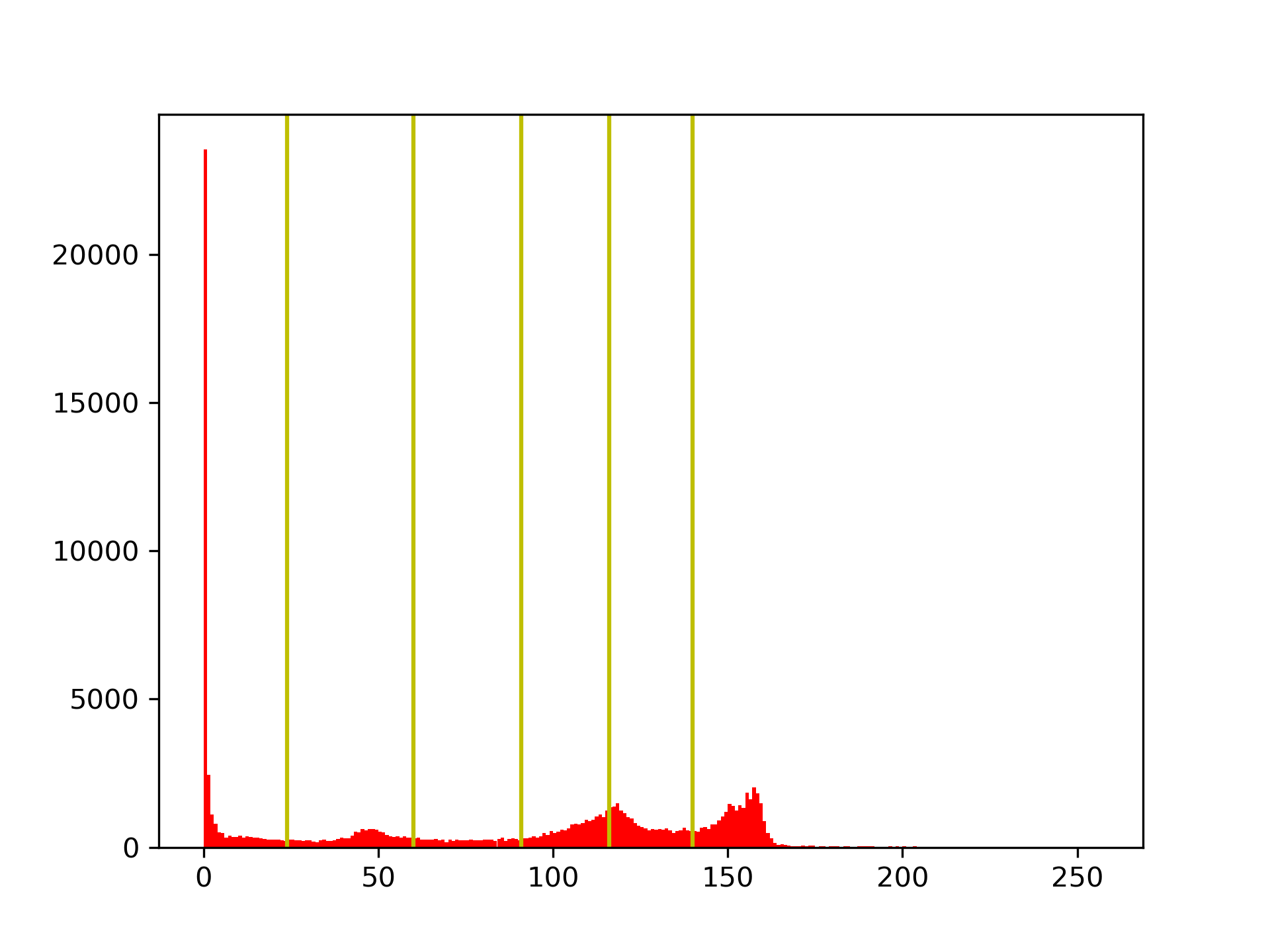}}
    \caption{Original image and thresholded images for Brainweb dataset image Z72 with number of thresholds: 2, 3, 4 and 5, with their corresponding histograms and thresholds marked with vertical lines.}
    \label{mri72}
\end{figure*}

\begin{figure*}
    \centering
    \subfloat[Original]{\includegraphics[scale=0.4]{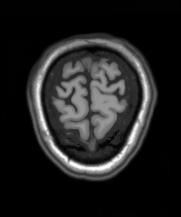}}\;\;
    \subfloat[2 Thresholds]{\includegraphics[scale=0.4]{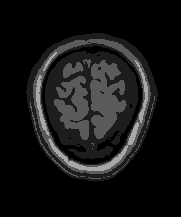}}\;\;
    \subfloat[3 Thresholds]{\includegraphics[scale=0.4]{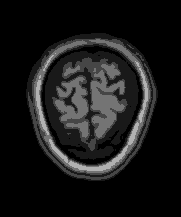}}\;\;
    \subfloat[4 Thresholds]{\includegraphics[scale=0.4]{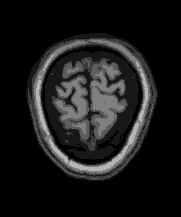}}\;\;
    \subfloat[5 Thresholds]{\includegraphics[scale=0.4]{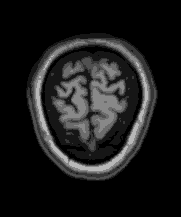}}\\
    \subfloat[Original]{\includegraphics[scale=0.2]{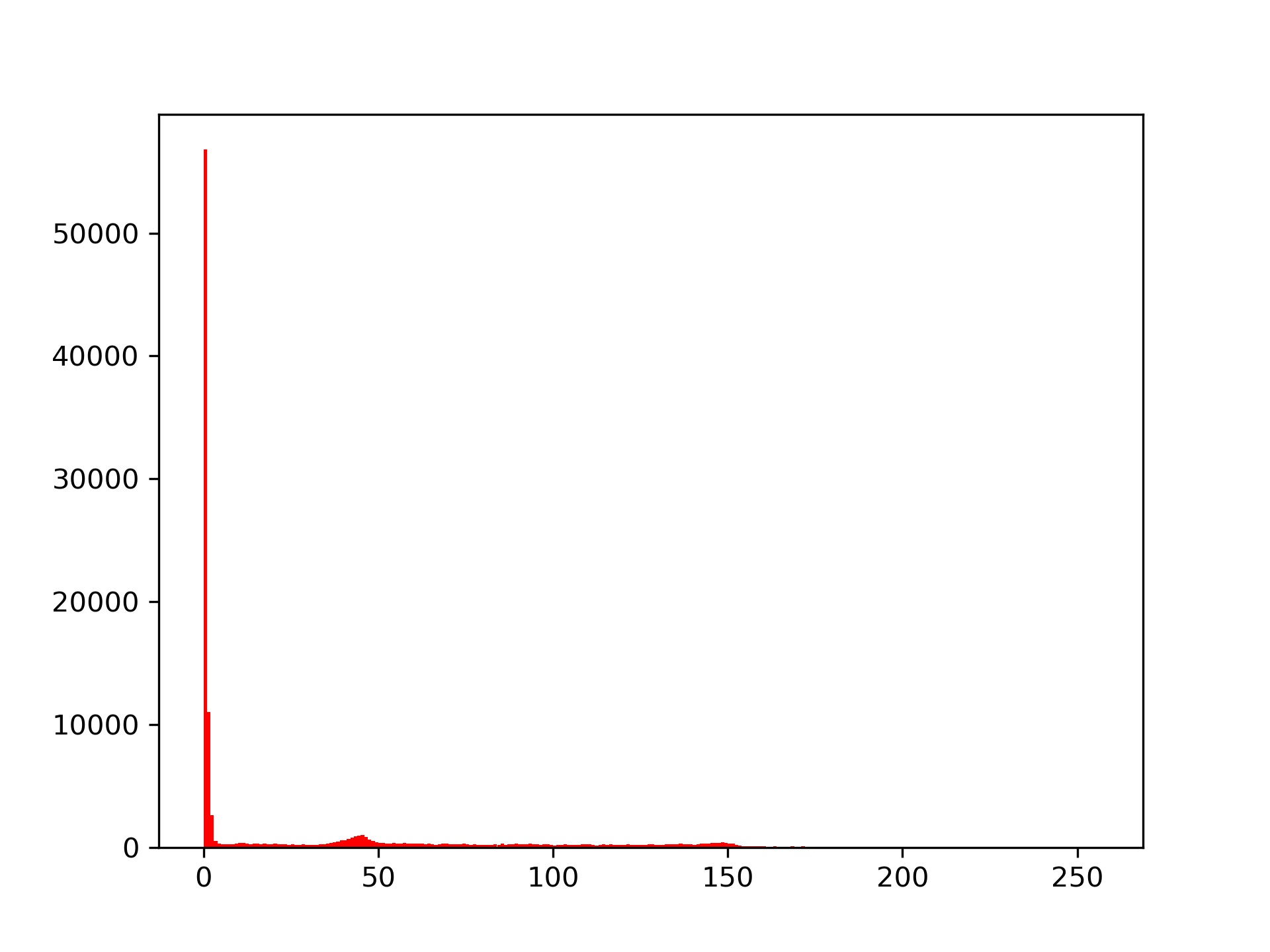}}
    \subfloat[2 Thresholds]{\includegraphics[scale=0.2]{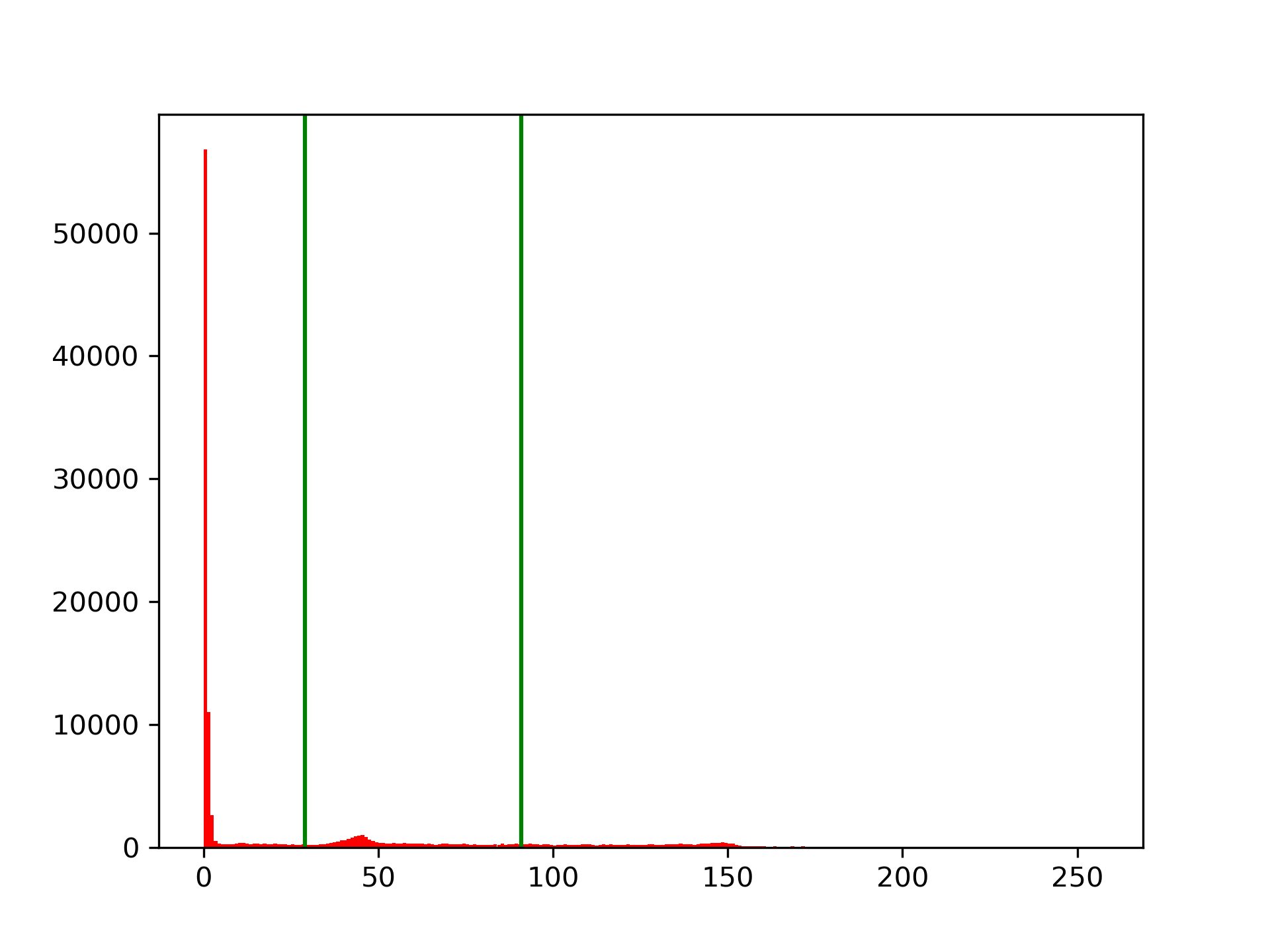}}
    \subfloat[3 Thresholds]{\includegraphics[scale=0.2]{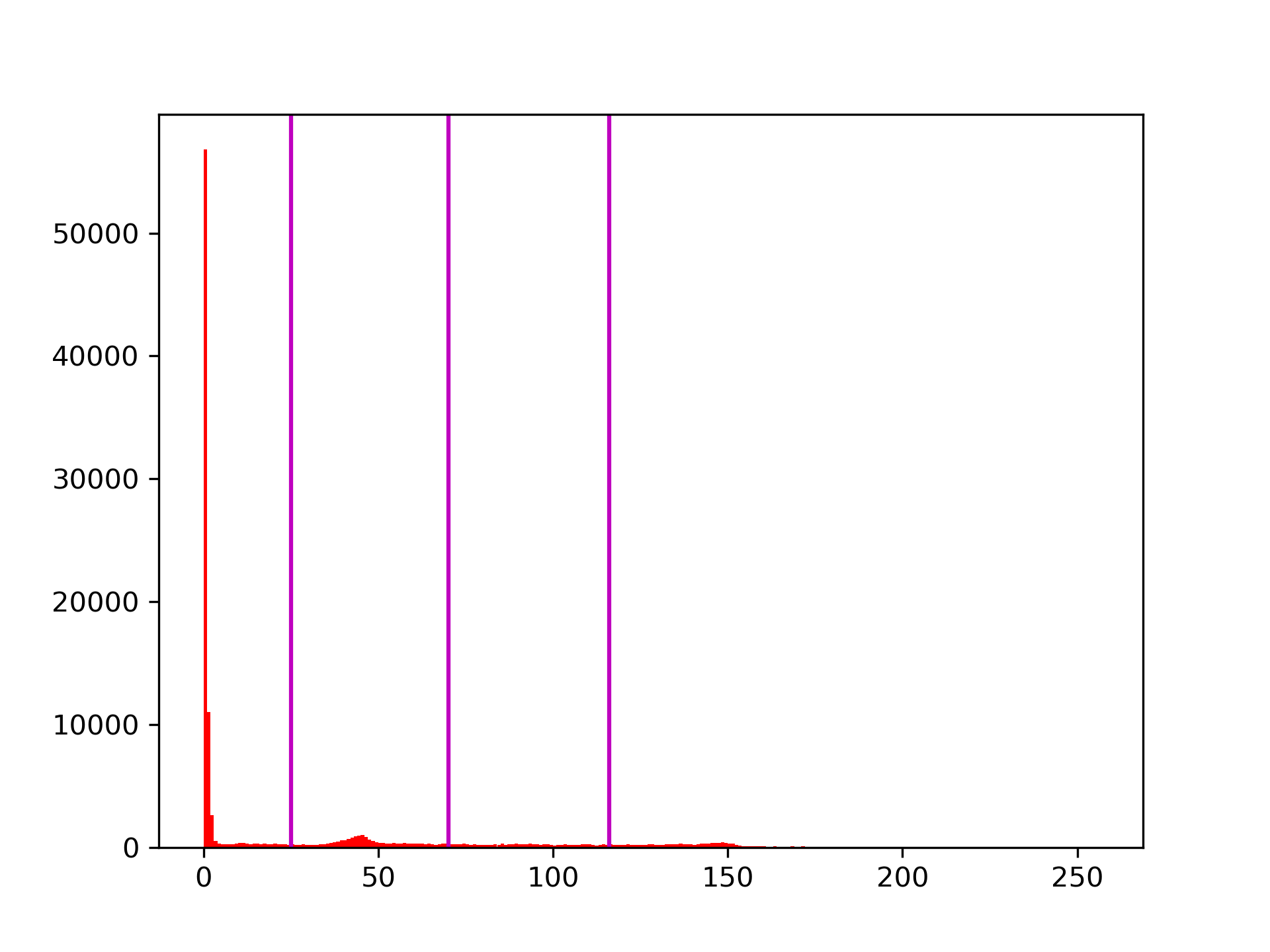}}
    \subfloat[4 Thresholds]{\includegraphics[scale=0.2]{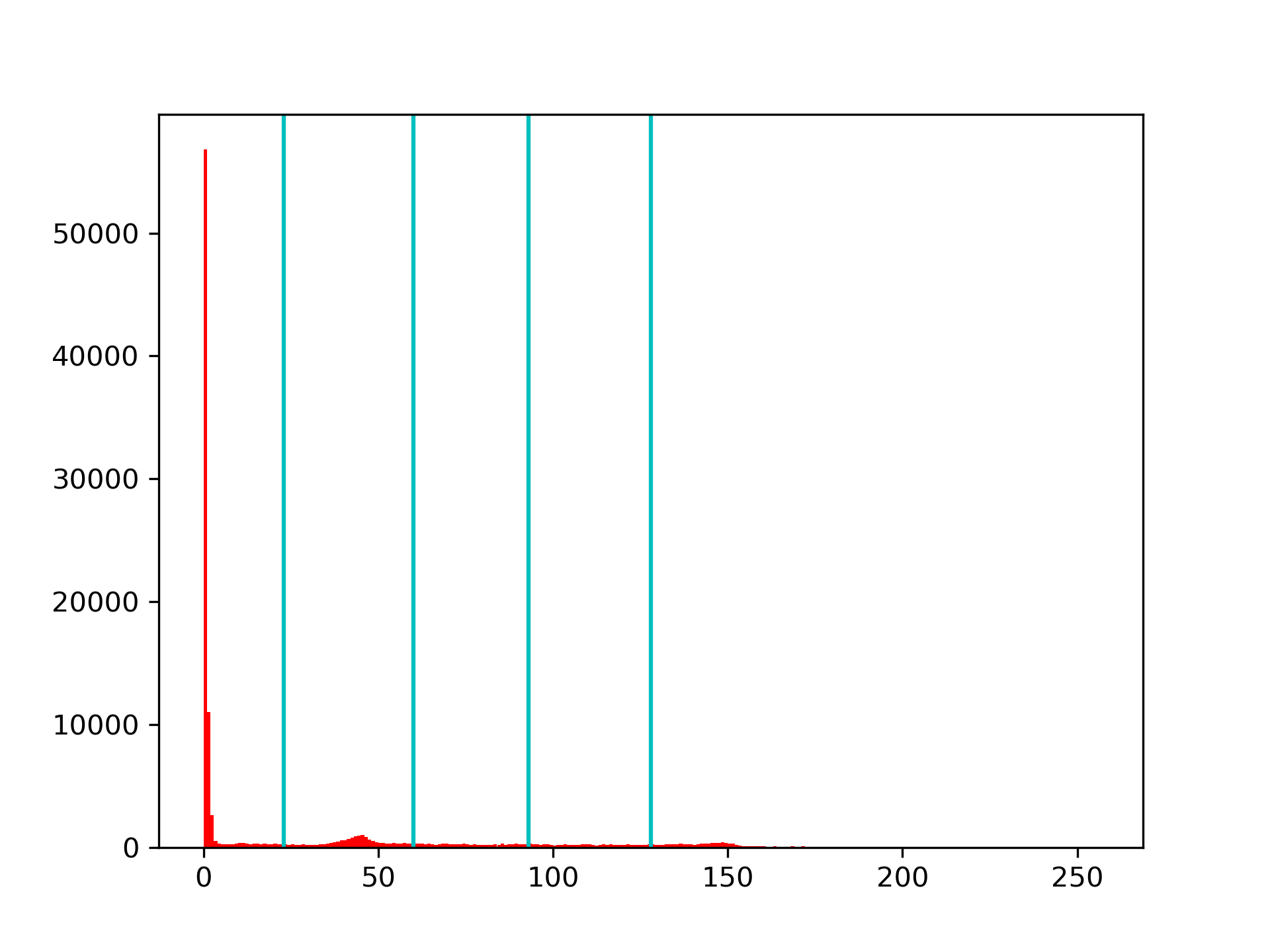}}
    \subfloat[5 Thresholds]{\includegraphics[scale=0.2]{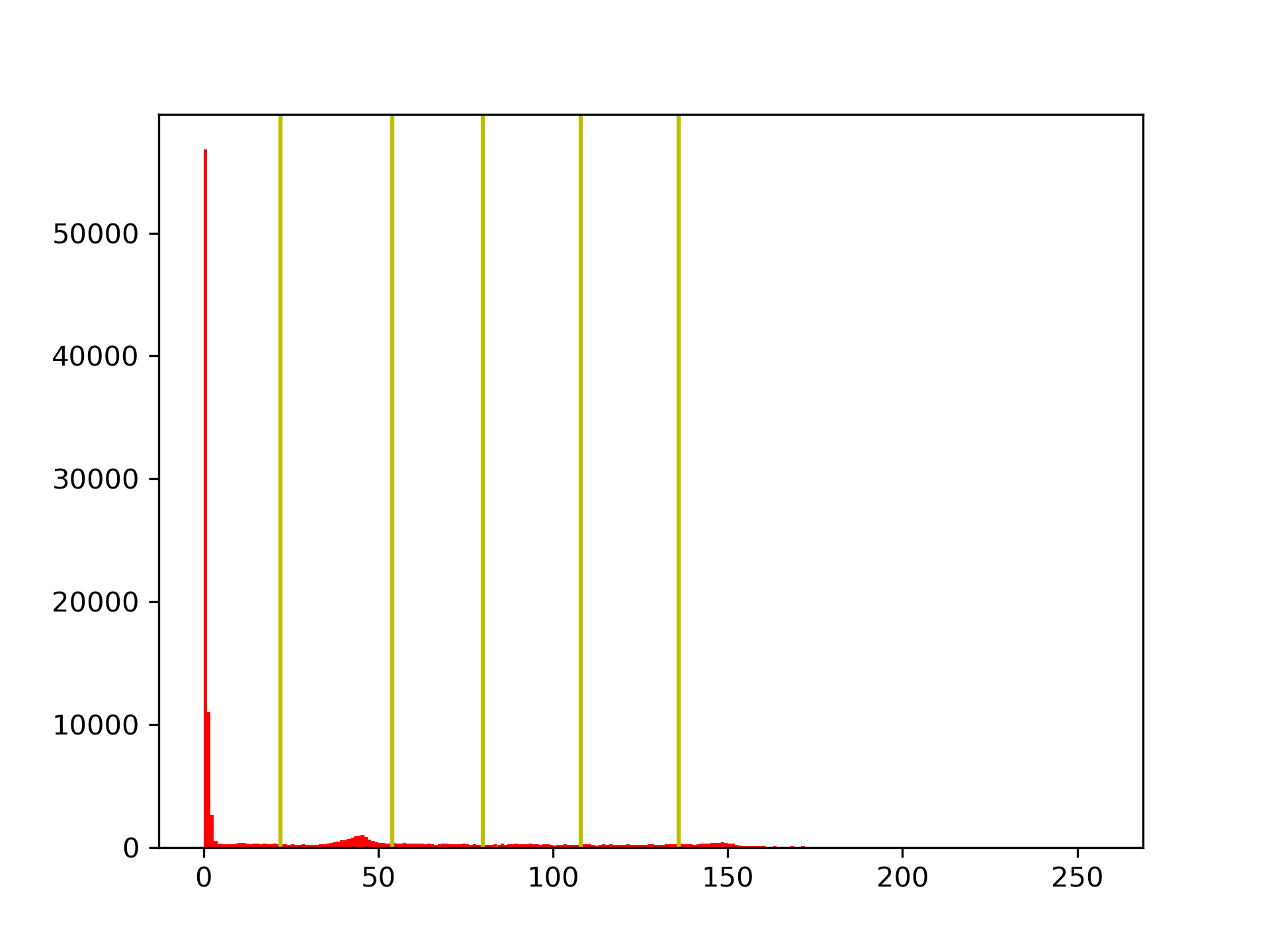}}
    \caption{Original image and thresholded images for Brainweb dataset image Z144 with number of thresholds: 2, 3, 4 and 5, with their corresponding histograms and thresholds marked with vertical lines.}
    \label{mri144}
\end{figure*}

Experiments have been performed using 2, 3, 4 and 5 thresholds for the multi-level thresholding problem. \autoref{res_harv} and \autoref{res_mri} show the results of the proposed model on the two publicly available datasets used in this study. For each of the images, the high values for the six evaluation metrics justify the fact that the proposed segmentation framework is reliable. The values of the quantitative parameters differ from each other for each of the images and it is also observed that as the number of thresholds is increased for a particular image, the values of the metrics are seen to increase. This is because on increasing the number of thresholds, we get an image that is segmented into more classes and hence, the result. The FSIM, SSIM and PSNR metrics are generally used to gauge the quality of the segmented image by drawing a comparison with the original image. HPSI measures the visual similarity between two images with respect to a human observer. UIQI is used in comparison by modelling the image distortion which comprises loss of correlation, luminance and contrast. QILV is used to measure the non-stationarity of an image by centring on the basic image structure. Figures \ref{h22}, \ref{h42}, \ref{h62} and \ref{h82} show some qualitative results from the experiments on the Harvard WBA dataset and Figures \ref{mri1}, \ref{mri36}, \ref{mri72} and \ref{mri144} show some qualitative results from the Brainweb dataset using number of thresholds as 2, 3, 4 and 5 respectively.  In each of the figures, we have placed the initial image, as well as the thresholded images by varying the number of thresholds. We have also included a graph marking the point of each of the thresholds for each of our thresholded images.

\begin{table*}[]
\centering
\caption{Quantitative results as obtained on using various chaotic maps while initializing the feature vectors in HHO}
\label{chaotic_map_table}
\begin{tabular}{|c|c|c|c|c|c|c|c|}
\hline
\textbf{Chaotic   map} & \textbf{No. of Thresholds} & \textbf{PSNR} & \textbf{SSIM} & \textbf{FSIM} & \textbf{UIQI} & \textbf{QILV} & \textbf{HPSI} \\ \hline
\multirow{4}{*}{No map}   & 2 & 23.0206 & 0.8162 & 0.7949 & 0.4371 & 0.7426 & 0.6863 \\ \cline{2-8} 
                            & 3 & 24.9477 & 0.8658 & 0.8603 & 0.5442 & 0.8440 & 0.7867 \\ \cline{2-8} 
                            & 4 & 26.4417 & 0.8960 & 0.8961 & 0.6215 & 0.8935 & 0.8389 \\ \cline{2-8} 
                            & 5 & 27.5454 & 0.9158 & 0.9205 & 0.6779 & 0.9223 & 0.8711 \\ \hline
\multirow{4}{*}{Sine}       & 2 & 23.0206 & 0.8194 & 0.7959 & 0.4396 & 0.7426 & 0.6893 \\ \cline{2-8} 
                            & 3 & 24.8599 & 0.8681 & 0.8619 & 0.5478 & 0.8396 & 0.7900 \\ \cline{2-8} 
                            & 4 & 26.5405 & 0.8983 & 0.8983 & 0.6250 & 0.8947 & 0.8423 \\ \cline{2-8} 
                            & 5 & 27.5543 & 0.9146 & 0.9220 & 0.6744 & 0.9213 & 0.8746 \\ \hline
\multirow{4}{*}{Singer}     & 2 & 23.0206 & 0.8162 & 0.7949 & 0.4371 & 0.7426 & 0.6852 \\ \cline{2-8} 
                            & 3 & 24.9173 & 0.8658 & 0.8603 & 0.5442 & 0.8418 & 0.7867 \\ \cline{2-8} 
                            & 4 & 26.3635 & 0.8972 & 0.8992 & 0.6242 & 0.8921 & 0.8440 \\ \cline{2-8} 
                            & 5 & 27.6836 & 0.9159 & 0.9211 & 0.6777 & 0.9237 & 0.8737 \\ \hline
\multirow{4}{*}{Sinusoidal} & 2 & 22.8736 & 0.8167 & 0.7945 & 0.4366 & 0.7381 & 0.6863 \\ \cline{2-8} 
                            & 3 & 24.9147 & 0.8657 & 0.8584 & 0.5419 & 0.8434 & 0.7813 \\ \cline{2-8} 
                            & 4 & 26.3635 & 0.8988 & 0.8992 & 0.6277 & 0.8898 & 0.8440 \\ \cline{2-8} 
                            & 5 & 27.6072 & 0.9192 & 0.9226 & 0.6844 & 0.9216 & 0.8766 \\ \hline
\multirow{4}{*}{Chebyshev}  & 2 & 23.0206 & 0.8162 & 0.7949 & 0.4371 & 0.7426 & 0.6863 \\ \cline{2-8} 
                            & 3 & 24.9477 & 0.8636 & 0.8599 & 0.5395 & 0.8440 & 0.7875 \\ \cline{2-8} 
                            & 4 & 26.4528 & 0.8978 & 0.8980 & 0.6241 & 0.8925 & 0.8422 \\ \cline{2-8} 
                            & 5 & 27.6964 & 0.9175 & 0.9235 & 0.6805 & 0.9227 & 0.8785 \\ \hline
\multirow{4}{*}{Tent}       & 2 & 22.9281 & 0.8174 & 0.7956 & 0.4382 & 0.7412 & 0.6904 \\ \cline{2-8} 
                            & 3 & 24.9173 & 0.8658 & 0.8603 & 0.5442 & 0.8418 & 0.7867 \\ \cline{2-8} 
                            & 4 & 26.4464 & 0.8997 & 0.8975 & 0.6280 & 0.8942 & 0.8413 \\ \cline{2-8} 
                            & 5 & 27.6484 & 0.9175 & 0.9235 & 0.6803 & 0.9216 & 0.8785 \\ \hline
\multirow{4}{*}{Logistic}   & 2 & 23.0206 & 0.8174 & 0.7949 & 0.4382 & 0.7426 & 0.6852 \\ \cline{2-8} 
                            & 3 & 24.8212 & 0.8651 & 0.8602 & 0.5426 & 0.8408 & 0.7843 \\ \cline{2-8} 
                            & 4 & 26.3912 & 0.8972 & 0.8992 & 0.6242 & 0.8936 & 0.8440 \\ \cline{2-8} 
                            & 5 & 27.7382 & 0.9197 & 0.9230 & 0.6852 & 0.9239 & 0.8764 \\ \hline
\multirow{4}{*}{Iterative}  & 2 & 22.9959 & 0.8202 & 0.7959 & 0.4415 & 0.7415 & 0.6893 \\ \cline{2-8} 
                            & 3 & 24.9477 & 0.8663 & 0.8615 & 0.5446 & 0.8440 & 0.7924 \\ \cline{2-8} 
                            & 4 & 26.3417 & 0.8966 & 0.8993 & 0.6238 & 0.8935 & 0.8443 \\ \cline{2-8} 
                            & 5 & 27.6072 & 0.9192 & 0.9226 & 0.6844 & 0.9218 & 0.8766 \\ \hline
\multirow{4}{*}{Gauss}      & 2 & 22.9585 & 0.8137 & 0.7945 & 0.4343 & 0.7424 & 0.6863 \\ \cline{2-8} 
                            & 3 & 25.0224 & 0.8667 & 0.8600 & 0.5443 & 0.8445 & 0.7886 \\ \cline{2-8} 
                            & 4 & 26.2731 & 0.8939 & 0.8969 & 0.6181 & 0.8882 & 0.8396 \\ \cline{2-8} 
                            & 5 & 27.5186 & 0.9159 & 0.9208 & 0.6777 & 0.9217 & 0.8743 \\ \hline
\end{tabular}
\end{table*}

\autoref{chaotic_map_table} thus demonstrates that amongst the various maps given, the one that gives the best results is the Logistic map. The values are provided in the table for each of the maps, for each threshold, and are computed by averaging each of the evaluation parameters, for all the images used in this paper. As expected, the chaotic maps boost the performance by increasing the diversity of the solutions. It is seen that the metrics obtained for most of the maps are very close to each other and differ marginally. Since the performance delivered by the Logistic chaotic map is the best, we use this particular map while doing further experiments.

\begin{table*}[]
\centering
\caption{Comparison in performance of the proposed framework based on the objective function for fitness assignment to the Harris hawks population.}
\label{comp_loss_fn}
\begin{tabular}{|c|c|c|c|c|c|c|c|}
\hline
\textbf{Objective   Function} &
  \textbf{No. of thresholds} &
  \textbf{PSNR} &
  \textbf{SSIM} &
  \textbf{FSIM} &
  \textbf{UIQI} &
  \textbf{QILV} &
  \textbf{HPSI} \\ \hline
\multirow{4}{*}{Cross Entropy}               & 2          & 18.8024          & 0.7158          & 0.7553          & 0.2720          & 0.6109          & 0.5665          \\ \cline{2-8} 
                                             & 3          & 20.4756          & 0.7703          & 0.8151          & 0.3389          & 0.7542          & 0.7001          \\ \cline{2-8} 
                                             & 4          & 22.2979          & 0.8242          & 0.8577          & 0.4099          & 0.8139          & 0.7705          \\ \cline{2-8} 
                                             & 5          & 23.6517          & 0.8555          & 0.8866          & 0.4676          & 0.8545          & 0.8167          \\ \hline
\multirow{4}{*}{Mean Squared Error}          & 2          & 17.0006          & 0.5524          & 0.6512          & 0.2234          & 0.7231          & 0.3524          \\ \cline{2-8} 
                                             & 3          & 18.3757          & 0.6121          & 0.7019          & 0.2903          & 0.7999          & 0.4120          \\ \cline{2-8} 
                                             & 4          & 19.6511          & 0.6574          & 0.7420          & 0.3509          & 0.8653          & 0.4622          \\ \cline{2-8} 
                                             & 5          & 20.8165          & 0.6976          & 0.7764          & 0.4031          & 0.9059          & 0.5076          \\ \hline
\multirow{4}{*}{Proposed Objective Function} & 2          & 18.6361          & 0.7077          & 0.7621          & 0.2709          & 0.7008          & 0.5810          \\ \cline{2-8} 
                                             & 3          & 20.5767          & 0.7823          & 0.8162          & 0.3451          & 0.8215          & 0.7136          \\ \cline{2-8} 
                                             & 4          & 23.5121          & 0.8132          & 0.8740          & 0.4212          & 0.8456          & 0.7816          \\ \cline{2-8} 
                                             & 5          & 24.7360          & 0.8410          & 0.8842          & 0.4531          & 0.9093          & 0.8142          \\ \hline
\multirow{4}{*}{\textbf{Proposed Hybrid}} &
  \textbf{2} &
  \textbf{19.6240} &
  \textbf{0.7176} &
  \textbf{0.7652} &
  \textbf{0.2711} &
  \textbf{0.7098} &
  \textbf{0.5818} \\ \cline{2-8} 
                                             & \textbf{3} & \textbf{21.8767} & \textbf{0.7913} & \textbf{0.8368} & \textbf{0.3549} & \textbf{0.8258} & \textbf{0.7285} \\ \cline{2-8} 
                                             & \textbf{4} & \textbf{23.5343} & \textbf{0.8251} & \textbf{0.8741} & \textbf{0.4237} & \textbf{0.8725} & \textbf{0.7857} \\ \cline{2-8} 
                                             & \textbf{5} & \textbf{24.7686} & \textbf{0.8608} & \textbf{0.8959} & \textbf{0.4670} & \textbf{0.9192} & \textbf{0.8228} \\ \hline
\end{tabular}
\end{table*}

\autoref{comp_loss_fn} outlines the quantitative results obtained by different objective functions used for fitness assignment in the HHO algorithm. Evidently, the proposed hybrid objective function performs better than the rest, where the weights are set as: $\alpha = 0.35$ and $\beta = 0.65$ experimentally. The hybrid objective function augments the values of all the metrics, as is evident from the results. There is a clear gap between the values returned by this function and those returned by the other objective functions.

\begin{figure*}
    \centering
    \subfloat[Harvard PSNR]{\includegraphics[scale=0.3]{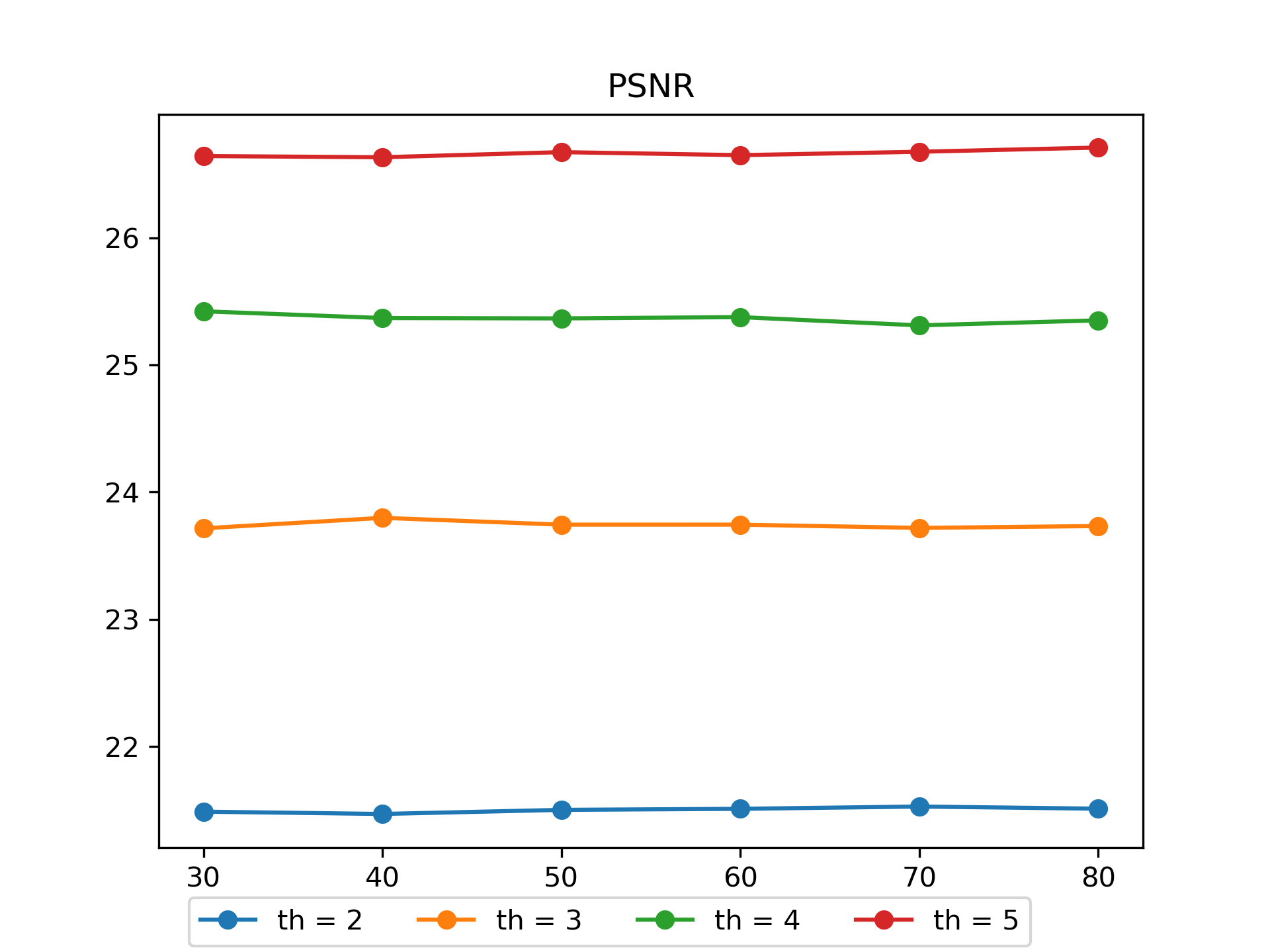}}\;\;
    \subfloat[Harvard SSIM]{\includegraphics[scale=0.3]{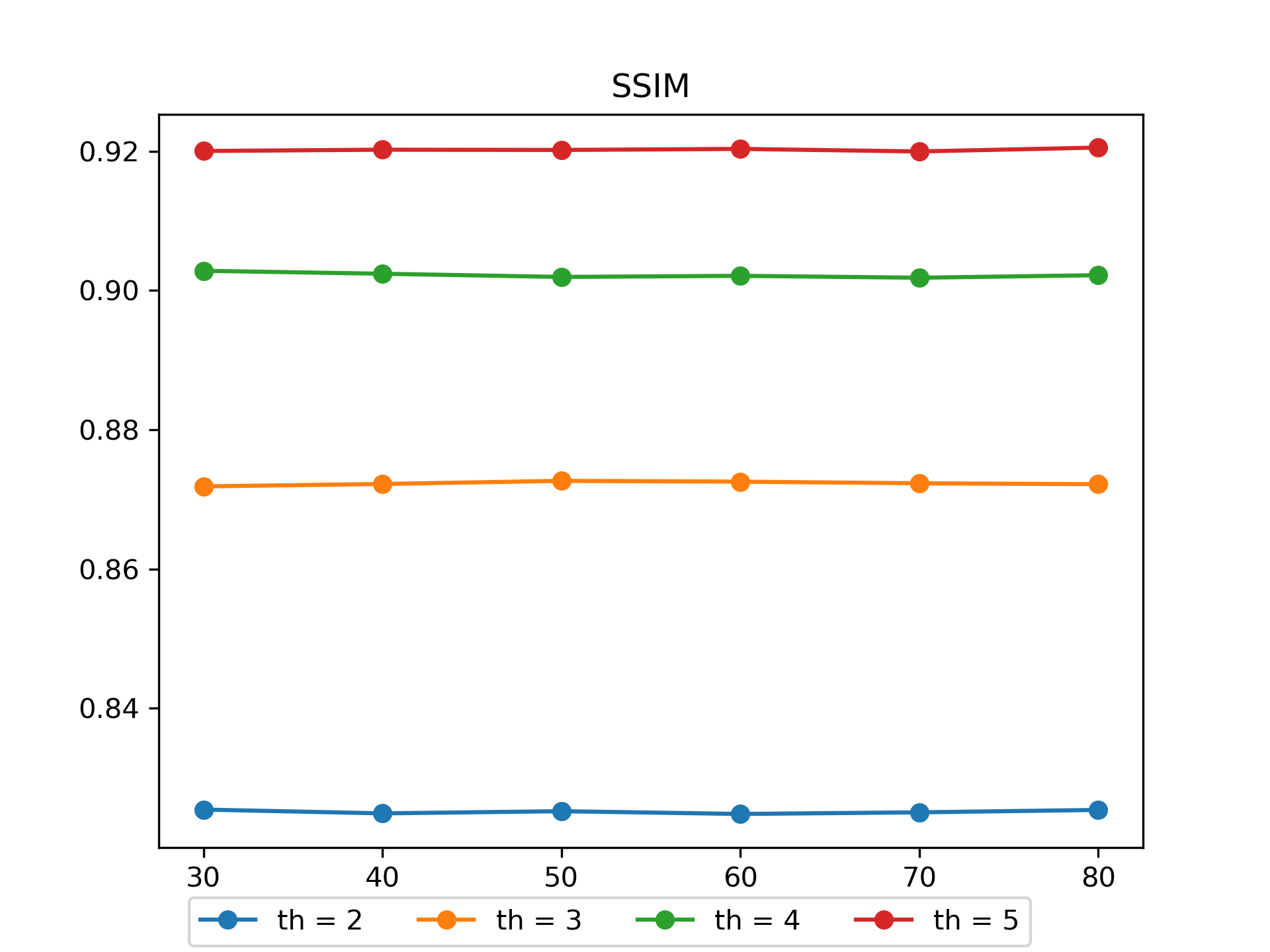}}\;\;
    \subfloat[Harvard FSIM]{\includegraphics[scale=0.3]{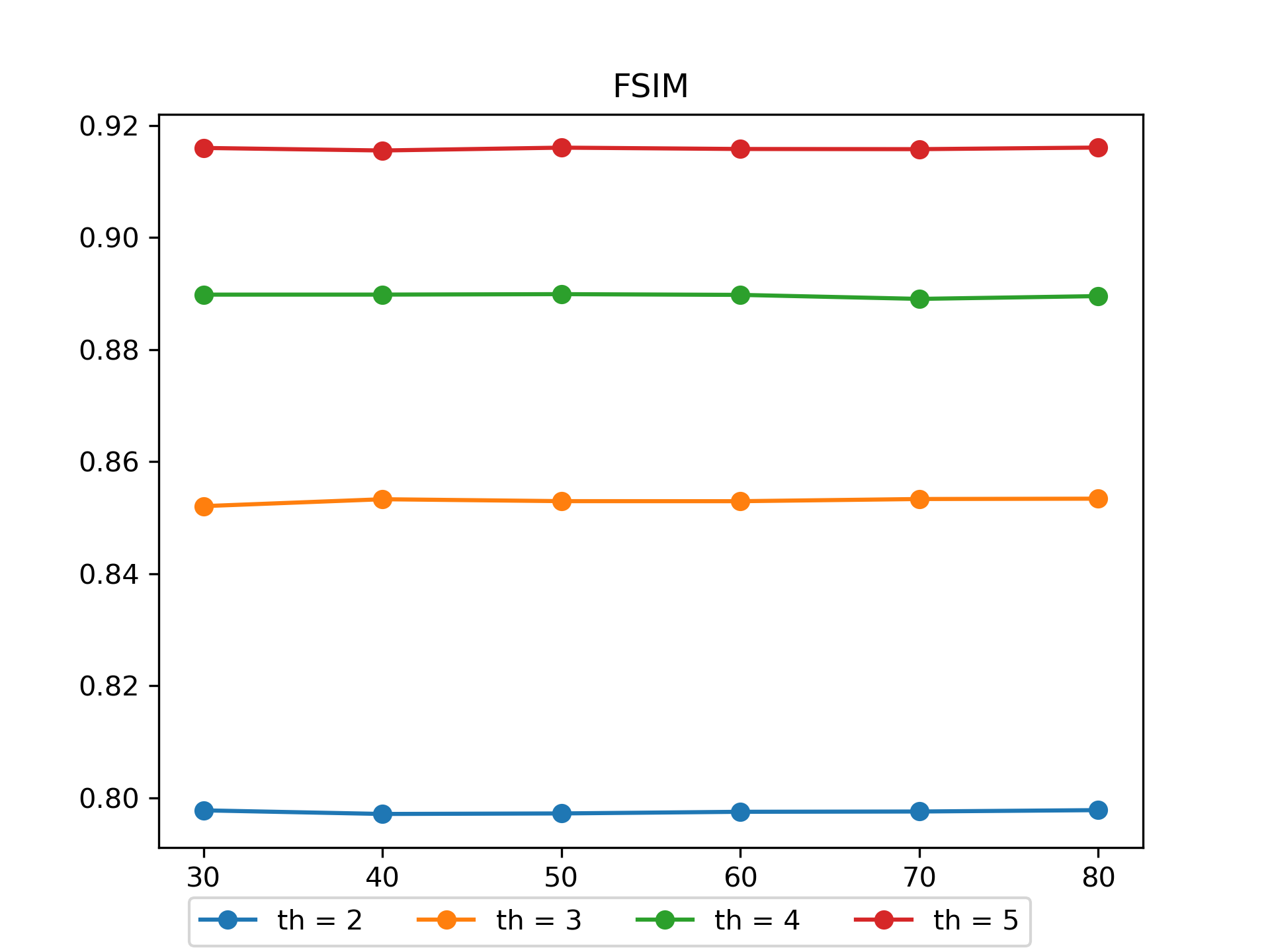}}\\
    \subfloat[Harvard UIQI]{\includegraphics[scale=0.3]{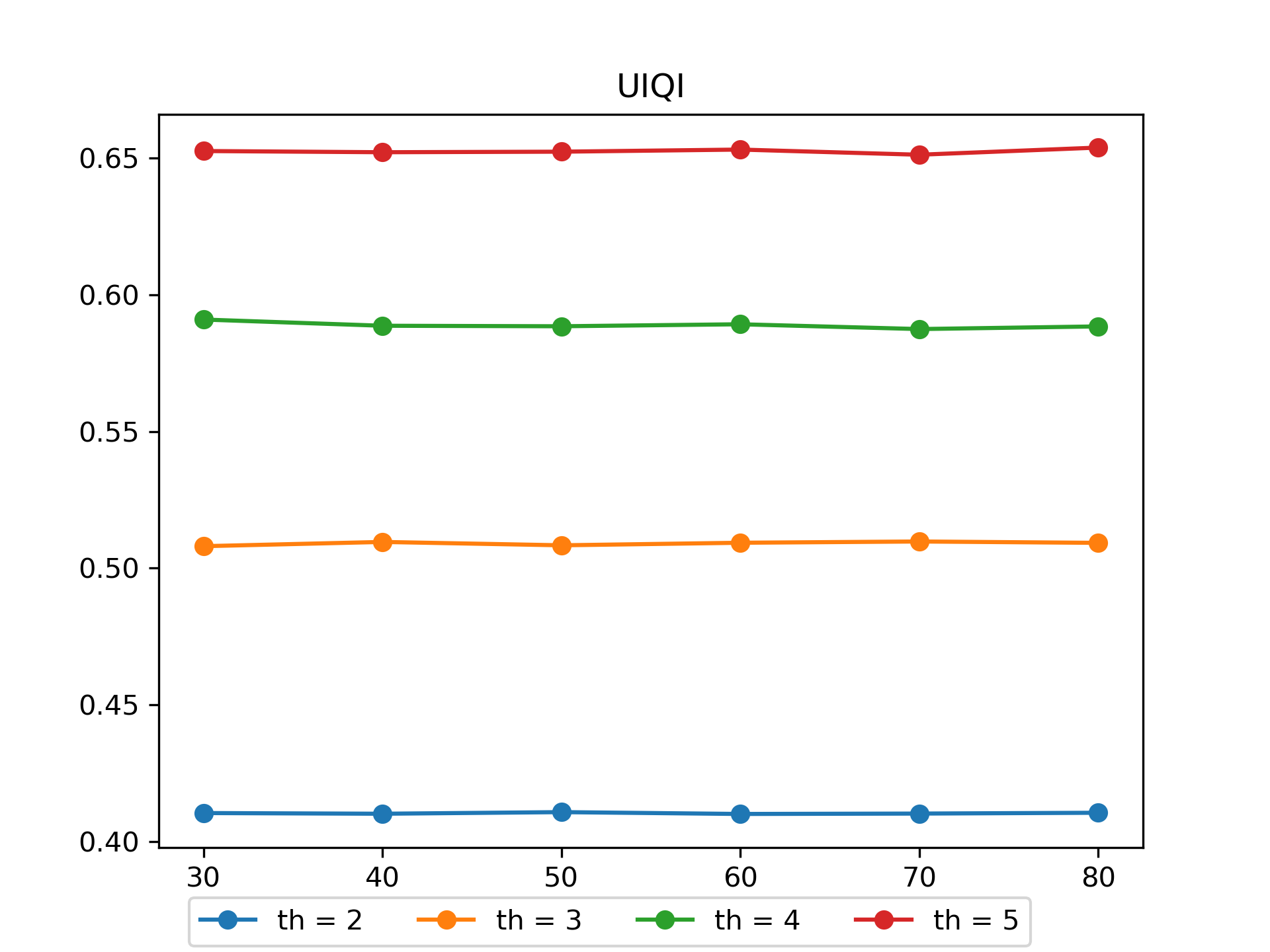}}\;\;
    \subfloat[Harvard QILV]{\includegraphics[scale=0.3]{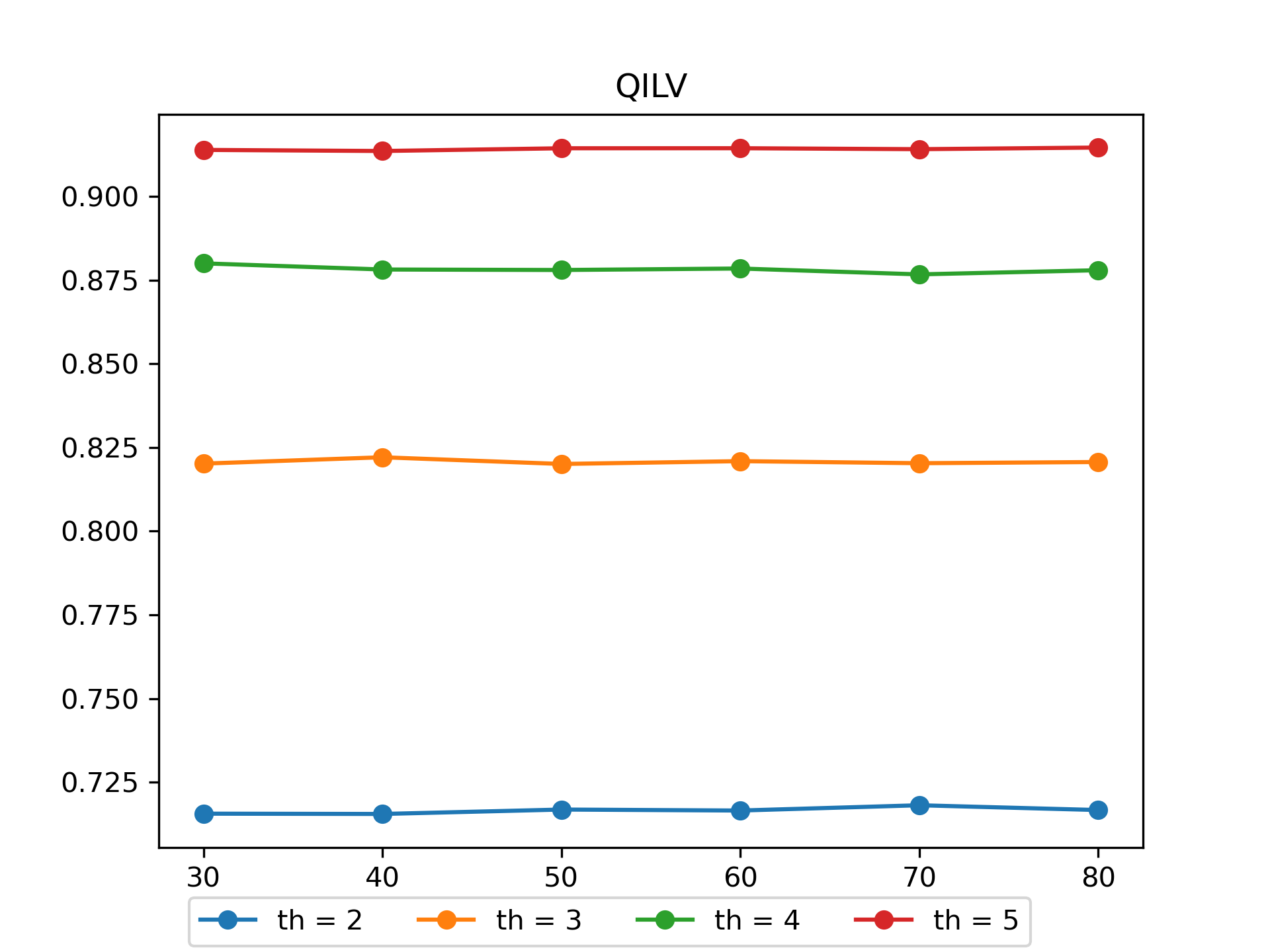}}\;\;
    \subfloat[Harvard HPSI]{\includegraphics[scale=0.3]{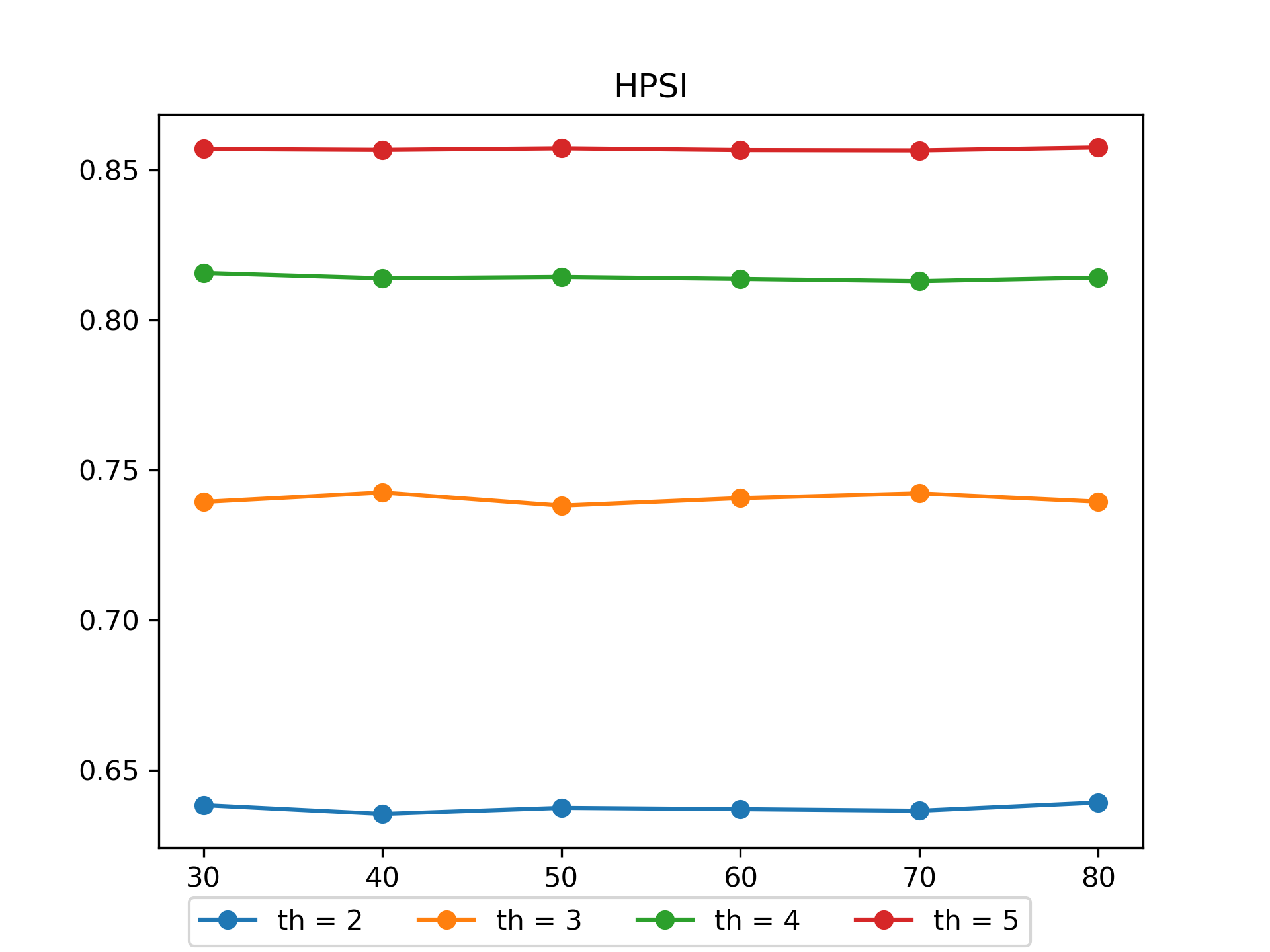}}\\
    \subfloat[Brainweb PSNR]{\includegraphics[scale=0.3]{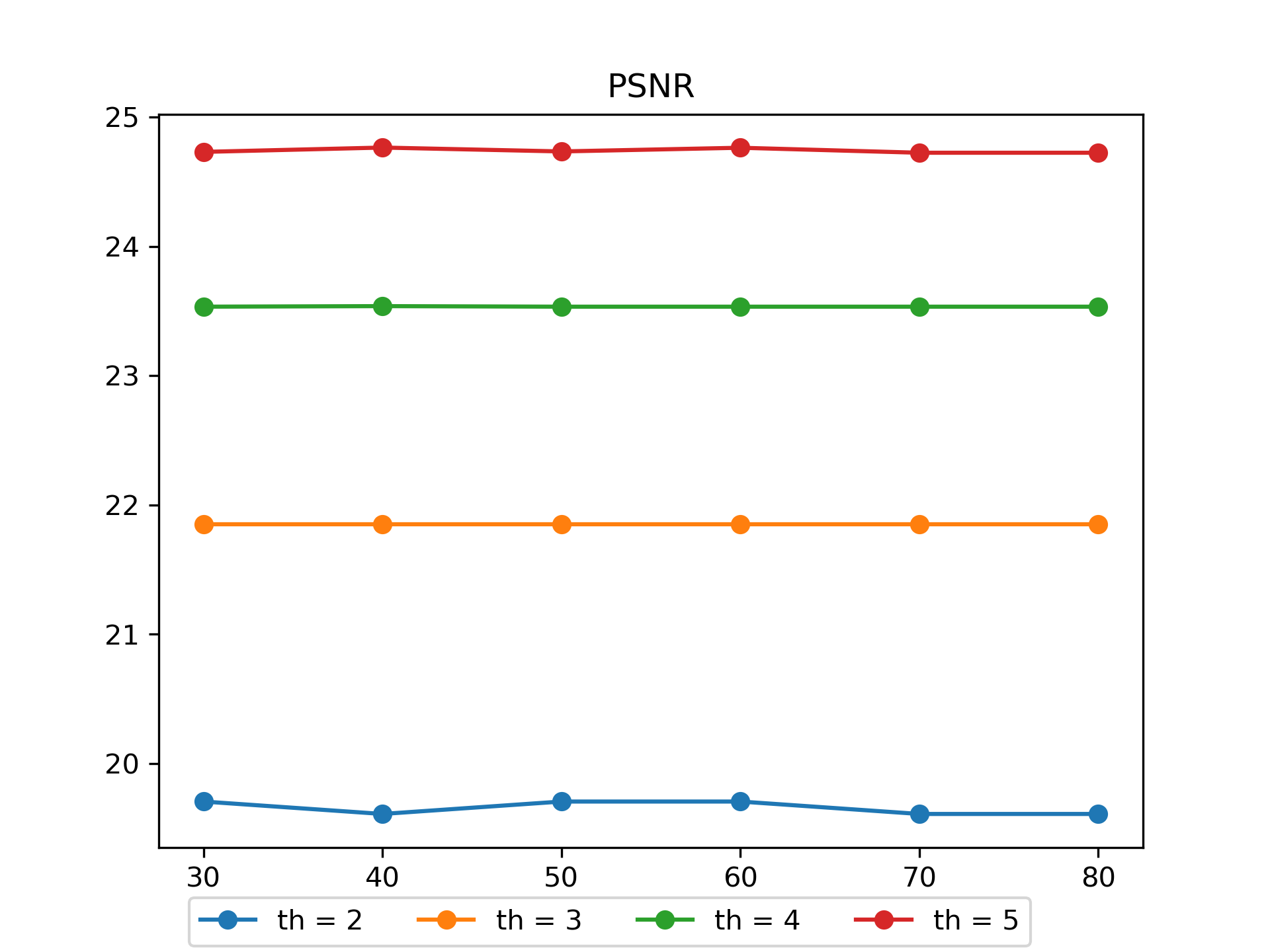}}\;\;
    \subfloat[Brainweb SSIM]{\includegraphics[scale=0.3]{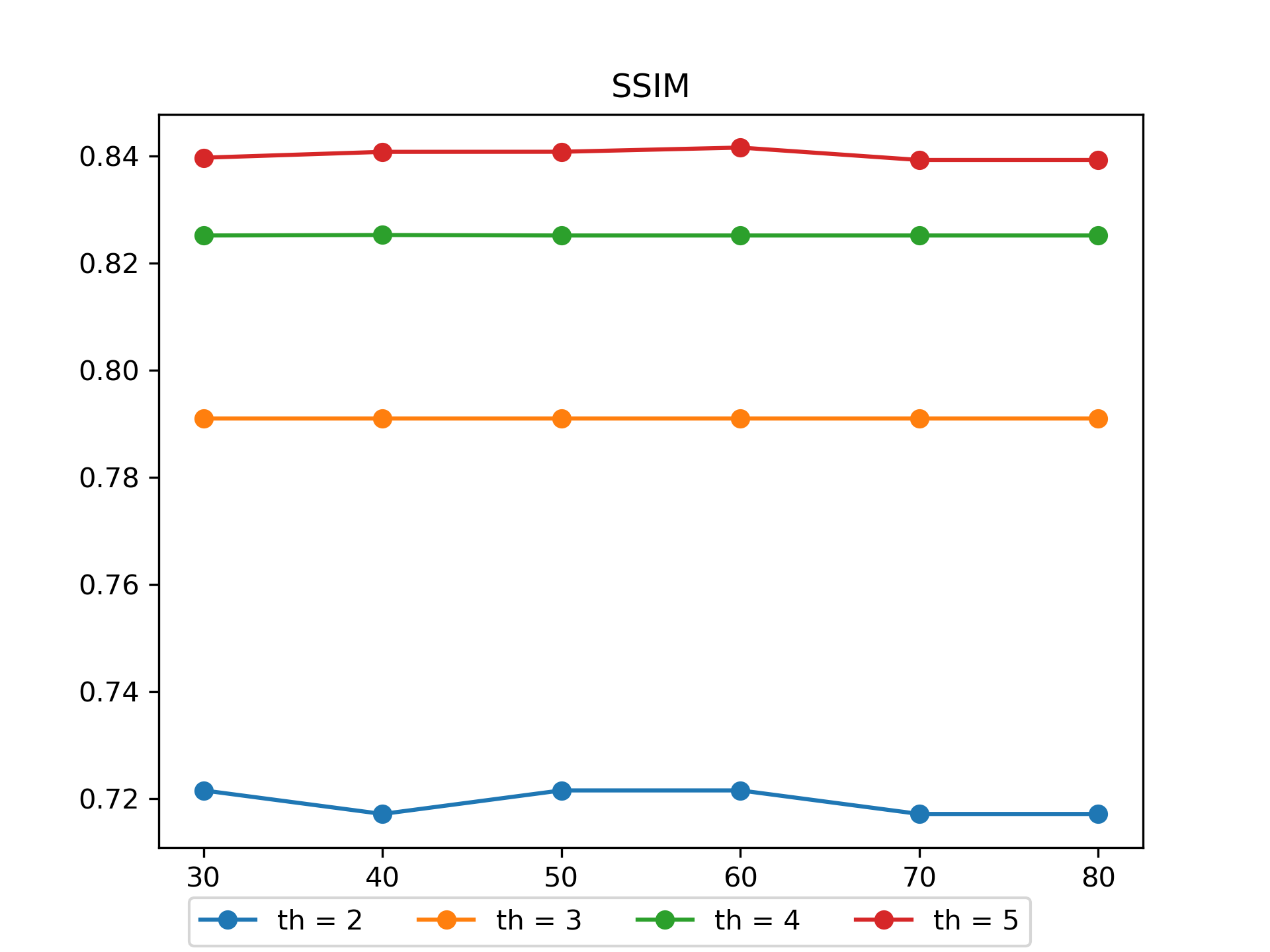}}\;\;
    \subfloat[Brainweb FSIM]{\includegraphics[scale=0.3]{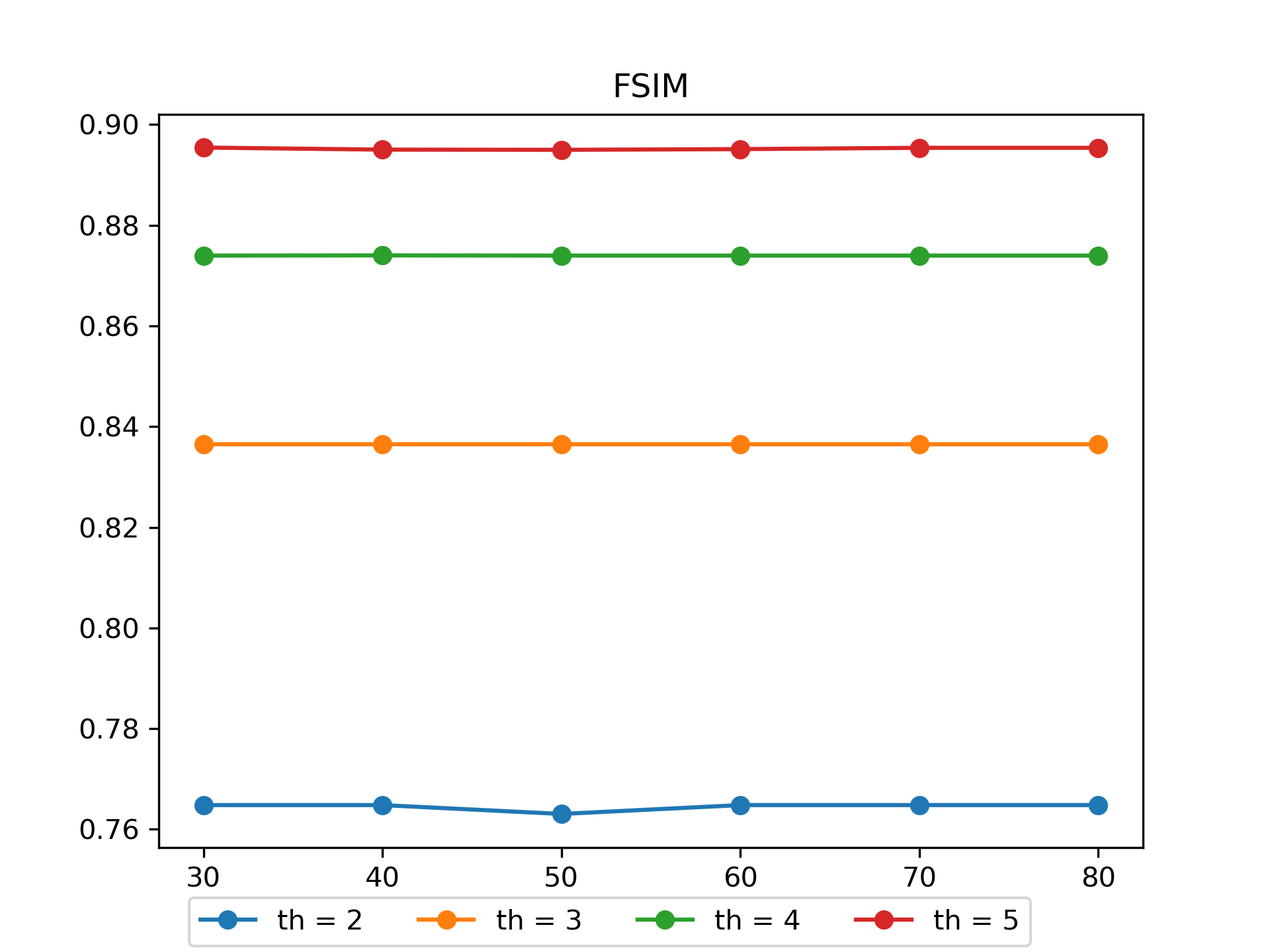}}\\
    \subfloat[Brainweb UIQI]{\includegraphics[scale=0.3]{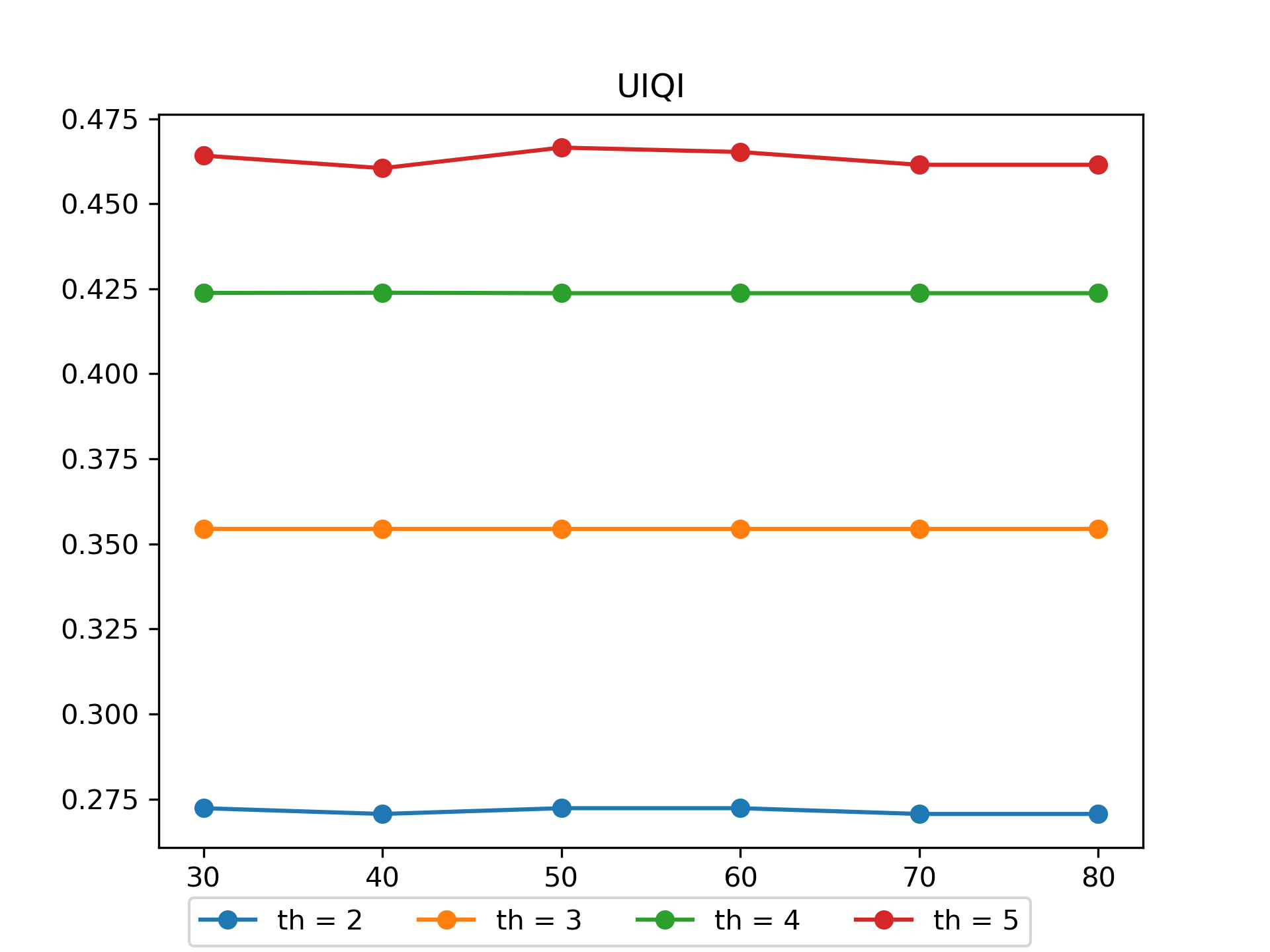}}\;\;
    \subfloat[Brainweb QILV]{\includegraphics[scale=0.3]{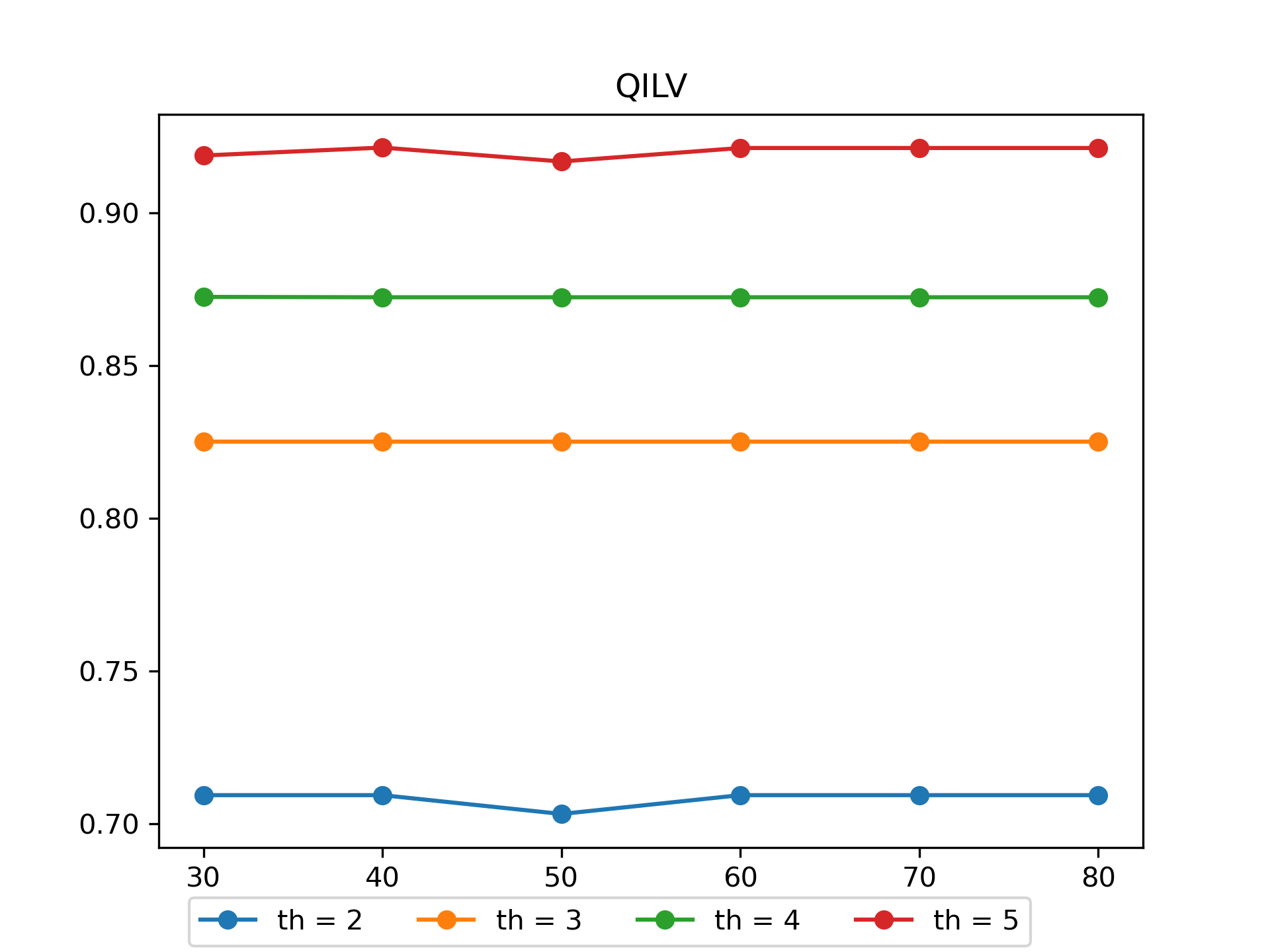}}\;\;
    \subfloat[Brainweb HPSI]{\includegraphics[scale=0.3]{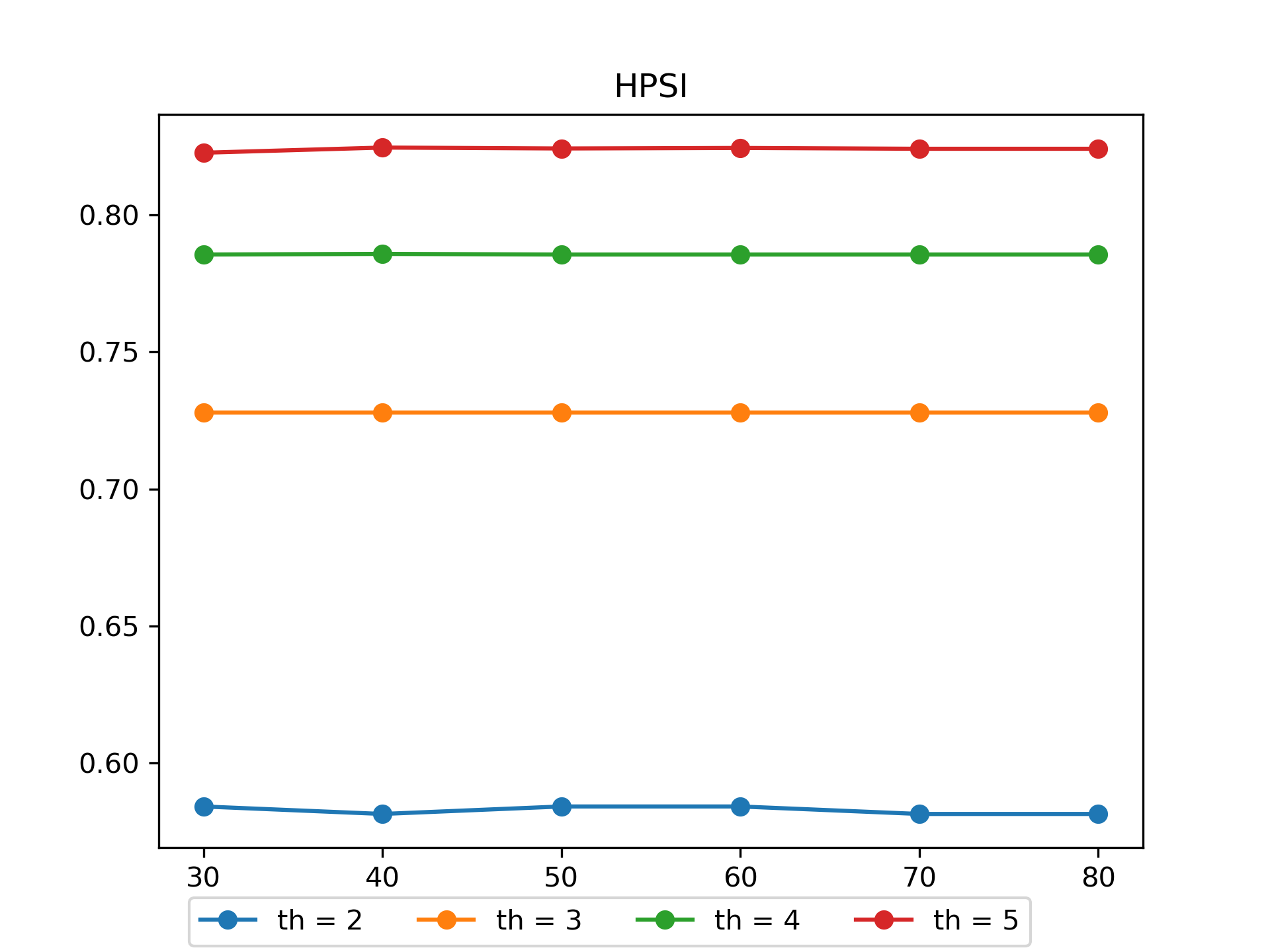}}
    \caption{Variation of performance with change in population size in HHO (altruism is set to 4 in each case): (a)-(f) in the Harvard WBA dataset; (g)-(l) in the Brainweb dataset.}
    \label{pop}
\end{figure*}

\begin{figure*}
    \centering
    \subfloat[Harvard PSNR]{\includegraphics[scale=0.3]{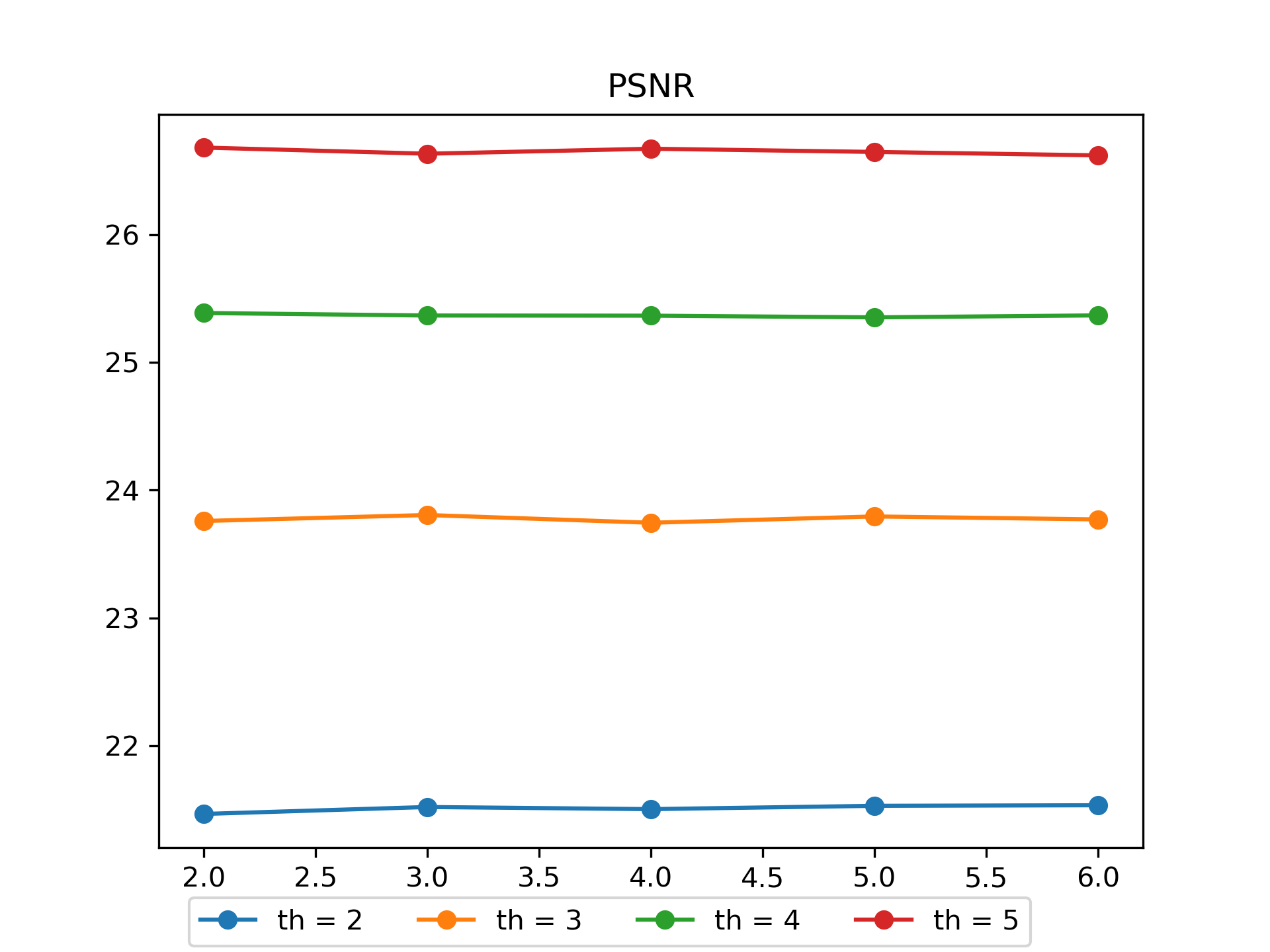}}\;\;
    \subfloat[Harvard SSIM]{\includegraphics[scale=0.3]{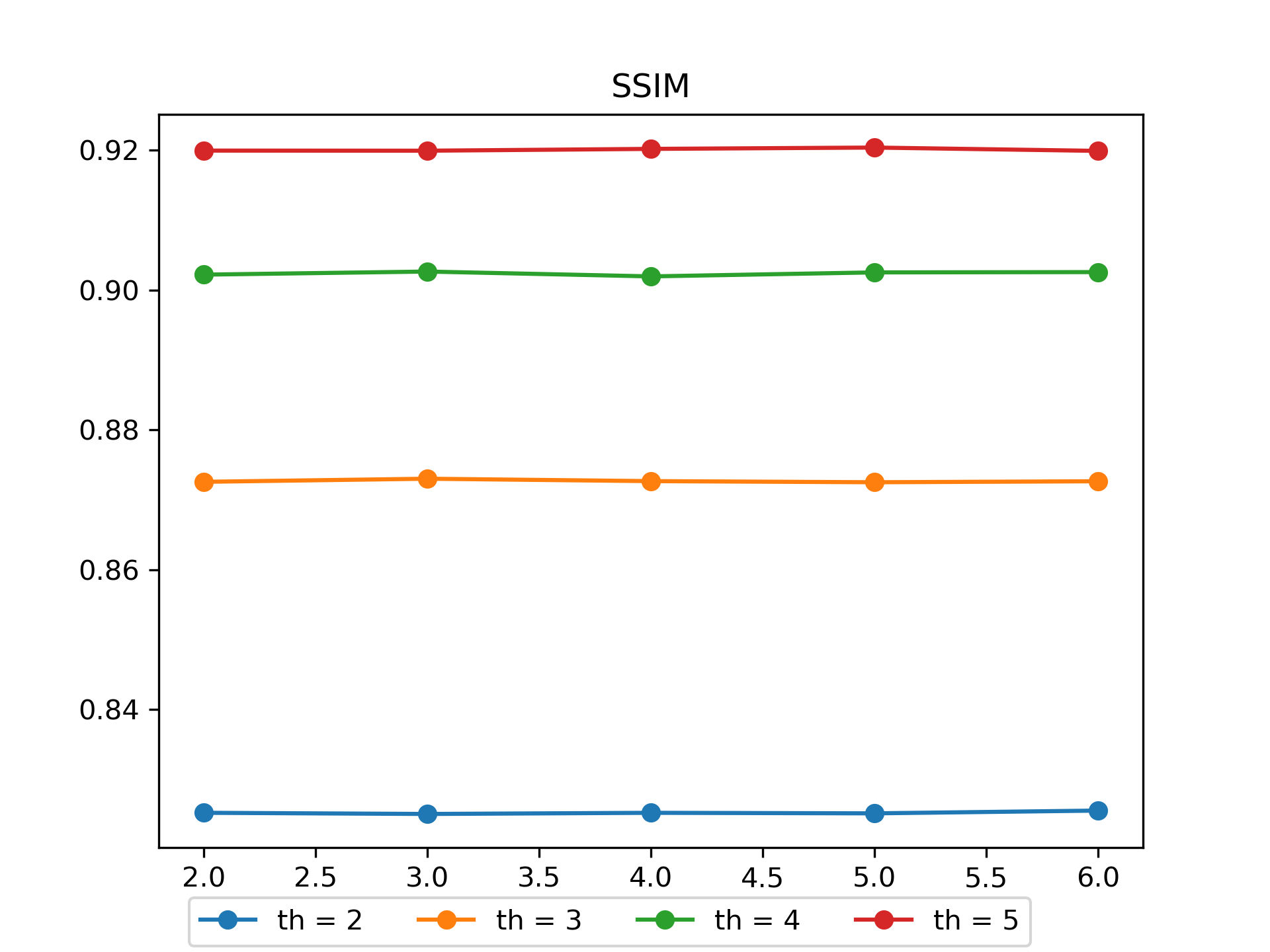}}\;\;
    \subfloat[Harvard FSIM]{\includegraphics[scale=0.3]{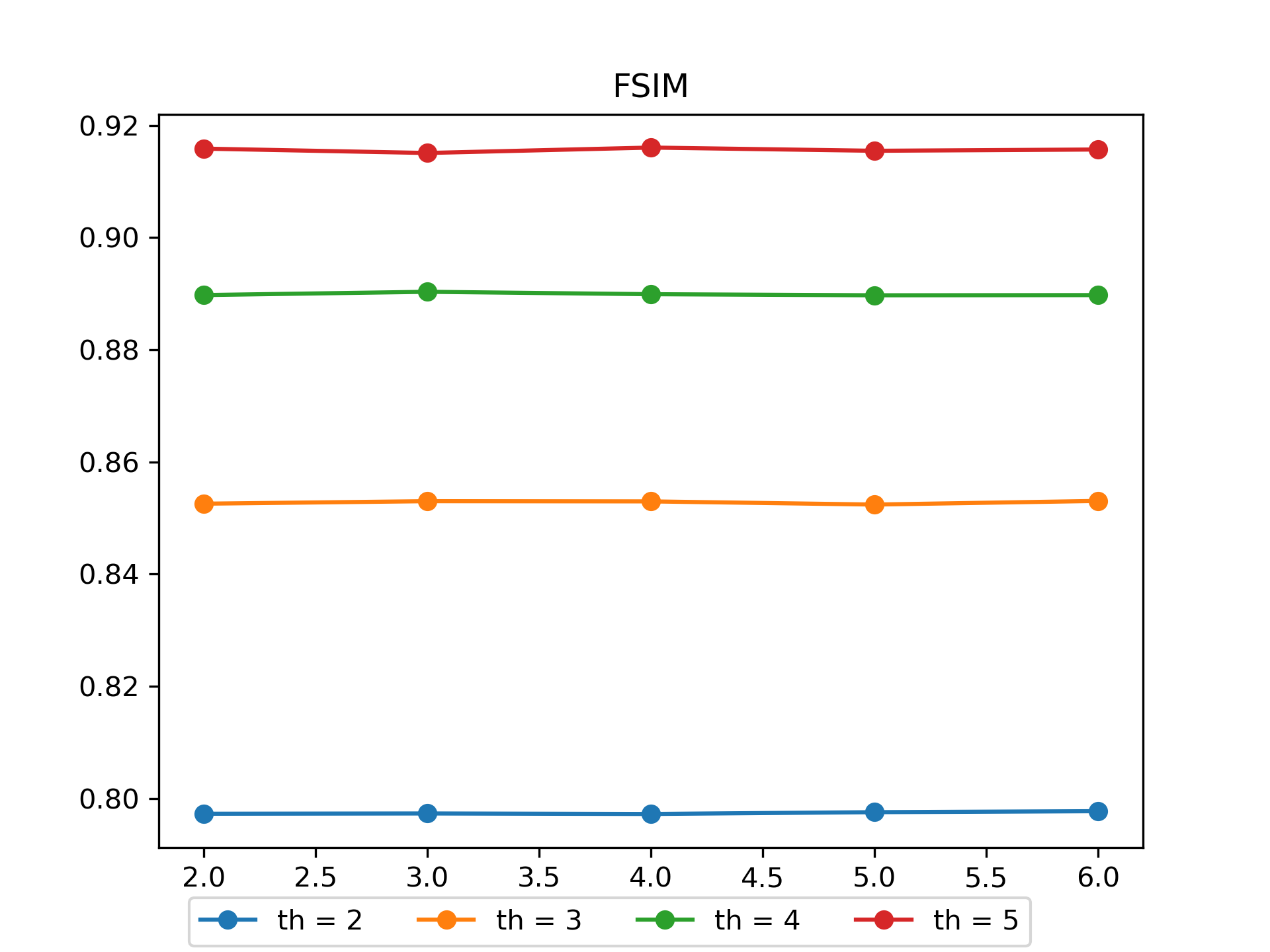}}\\
    \subfloat[Harvard UIQI]{\includegraphics[scale=0.3]{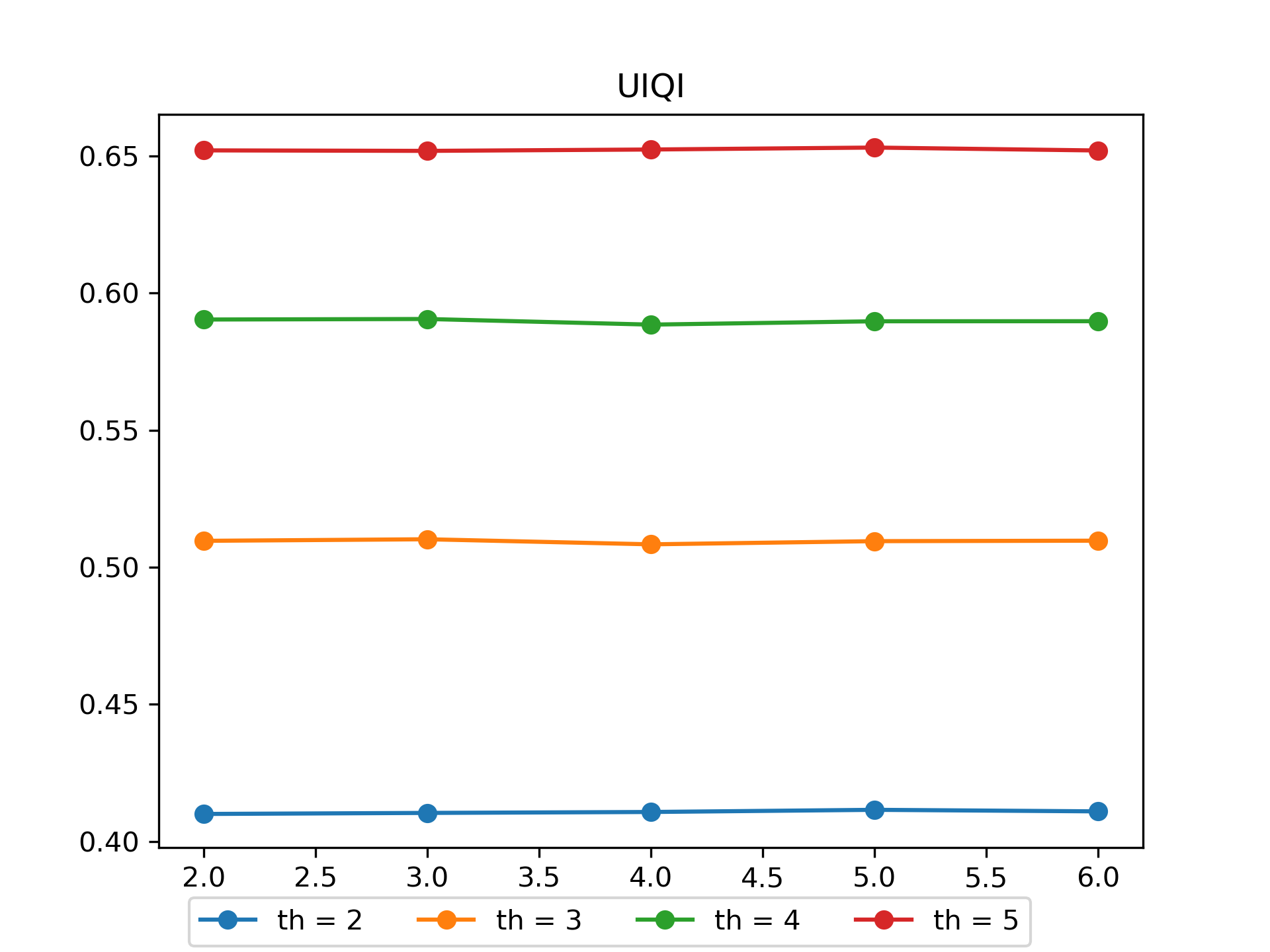}}\;\;
    \subfloat[Harvard QILV]{\includegraphics[scale=0.3]{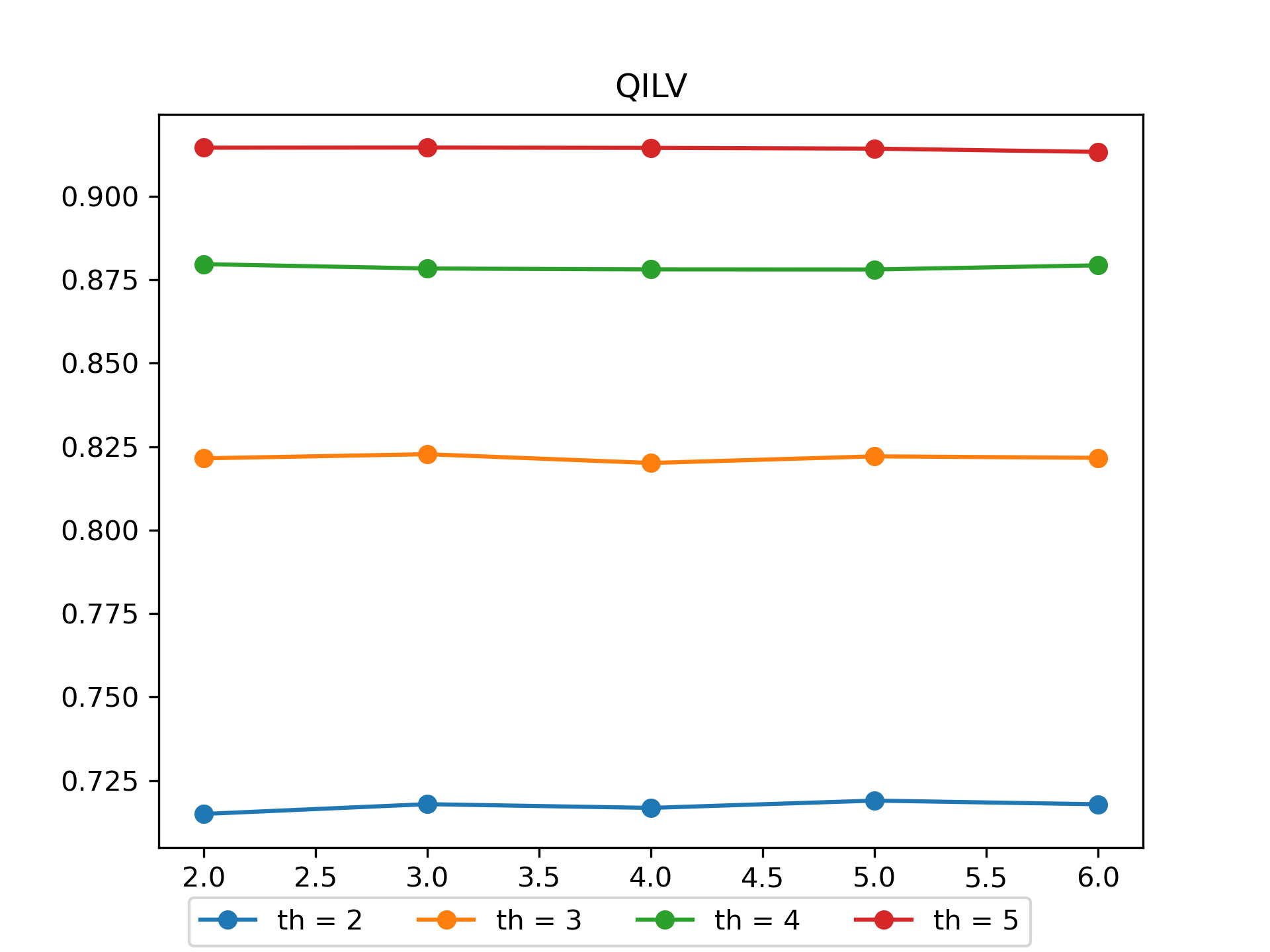}}\;\;
    \subfloat[Harvard HPSI]{\includegraphics[scale=0.3]{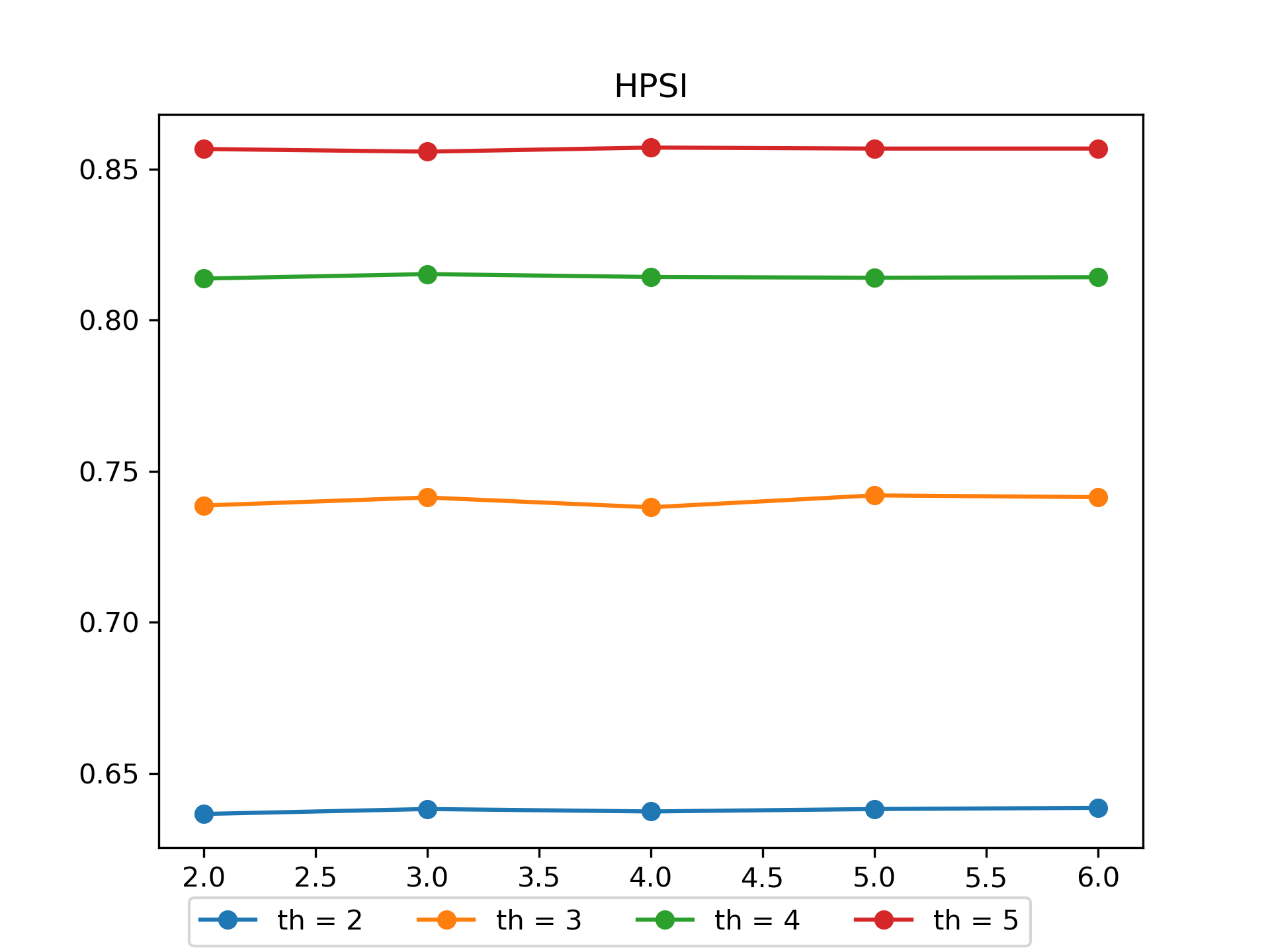}}\\
    \subfloat[Brainweb PSNR]{\includegraphics[scale=0.3]{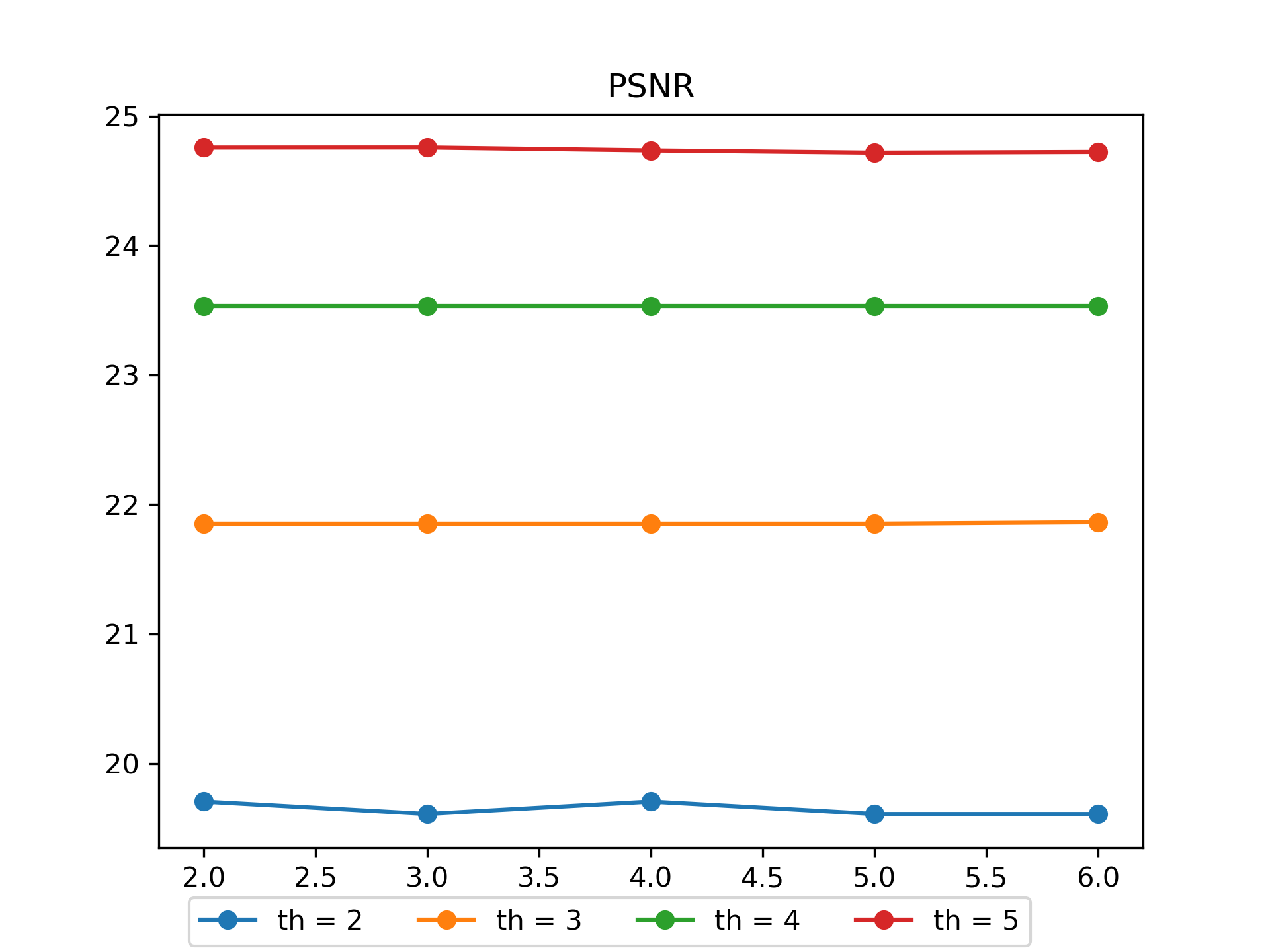}}\;\;
    \subfloat[Brainweb SSIM]{\includegraphics[scale=0.3]{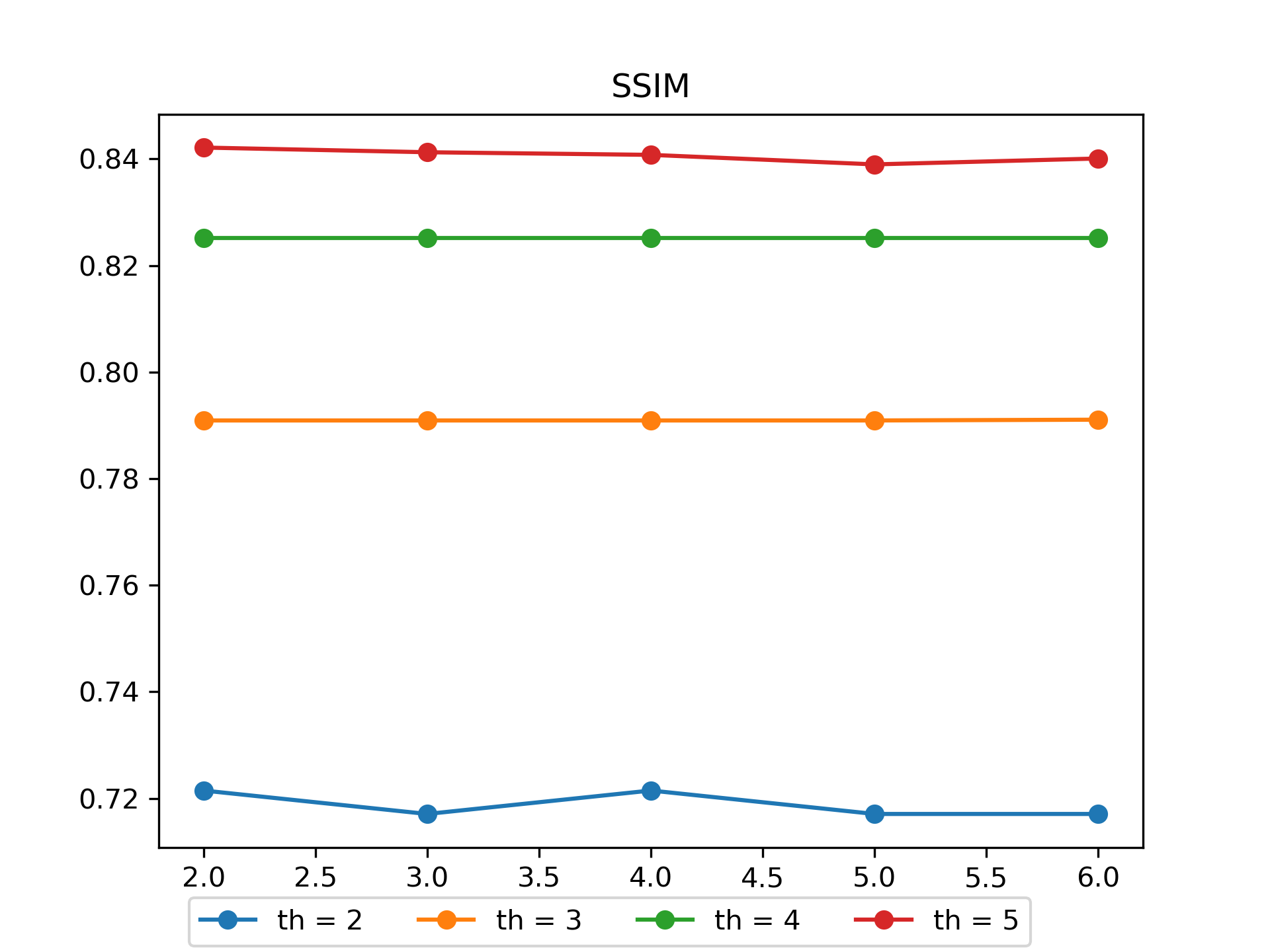}}\;\;
    \subfloat[Brainweb FSIM]{\includegraphics[scale=0.3]{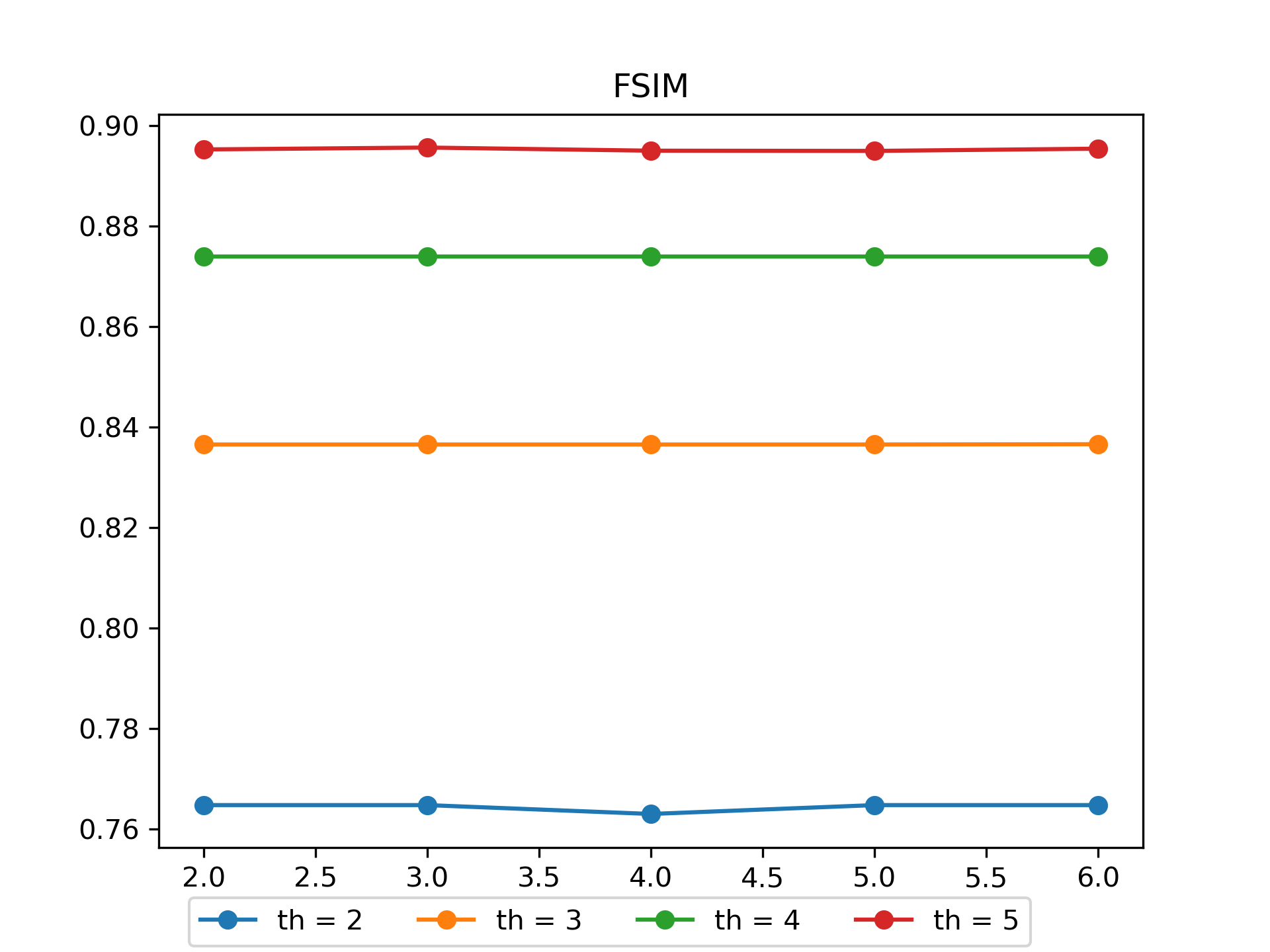}}\\
    \subfloat[Brainweb UIQI]{\includegraphics[scale=0.3]{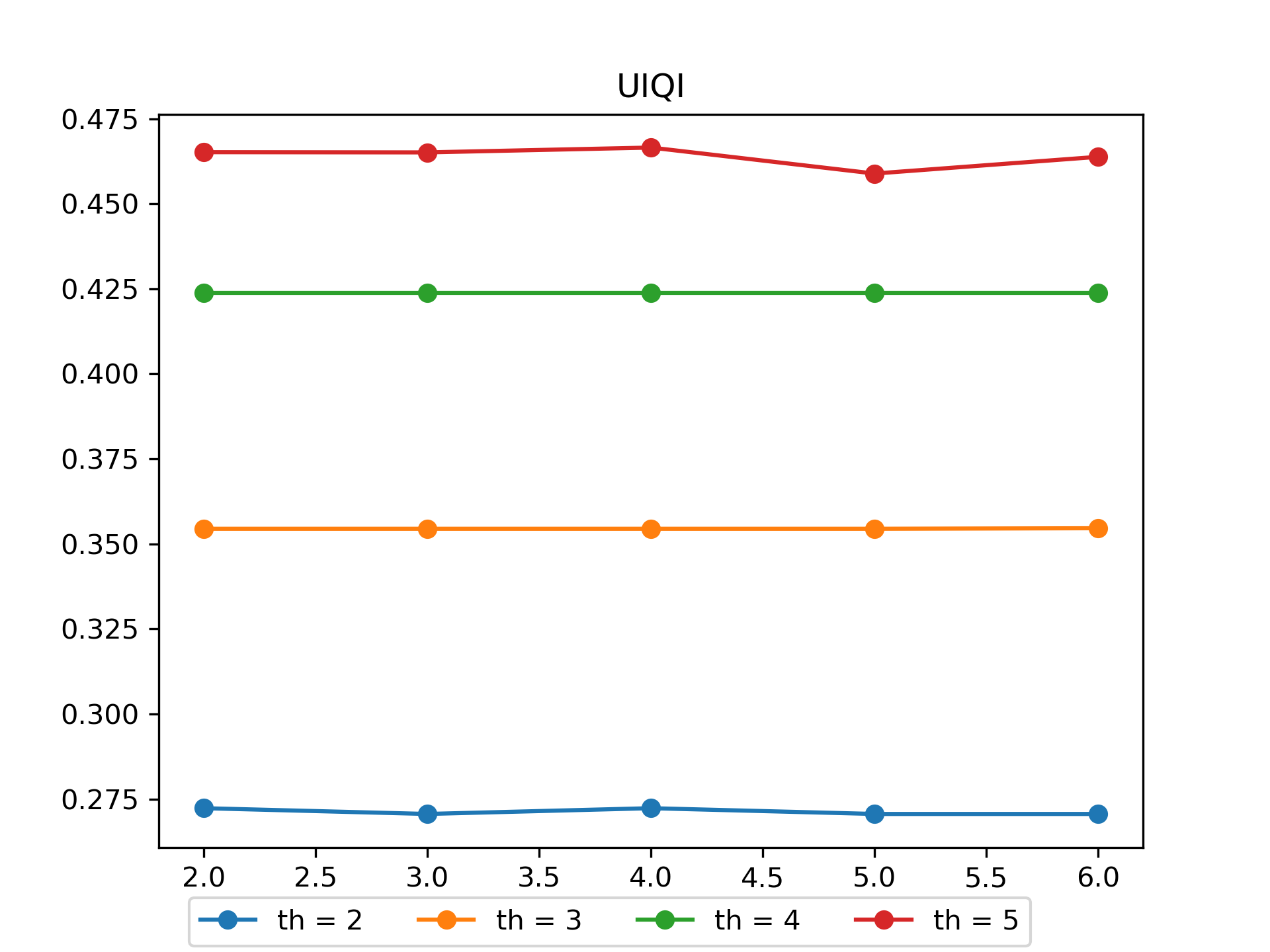}}\;\;
    \subfloat[Brainweb QILV]{\includegraphics[scale=0.3]{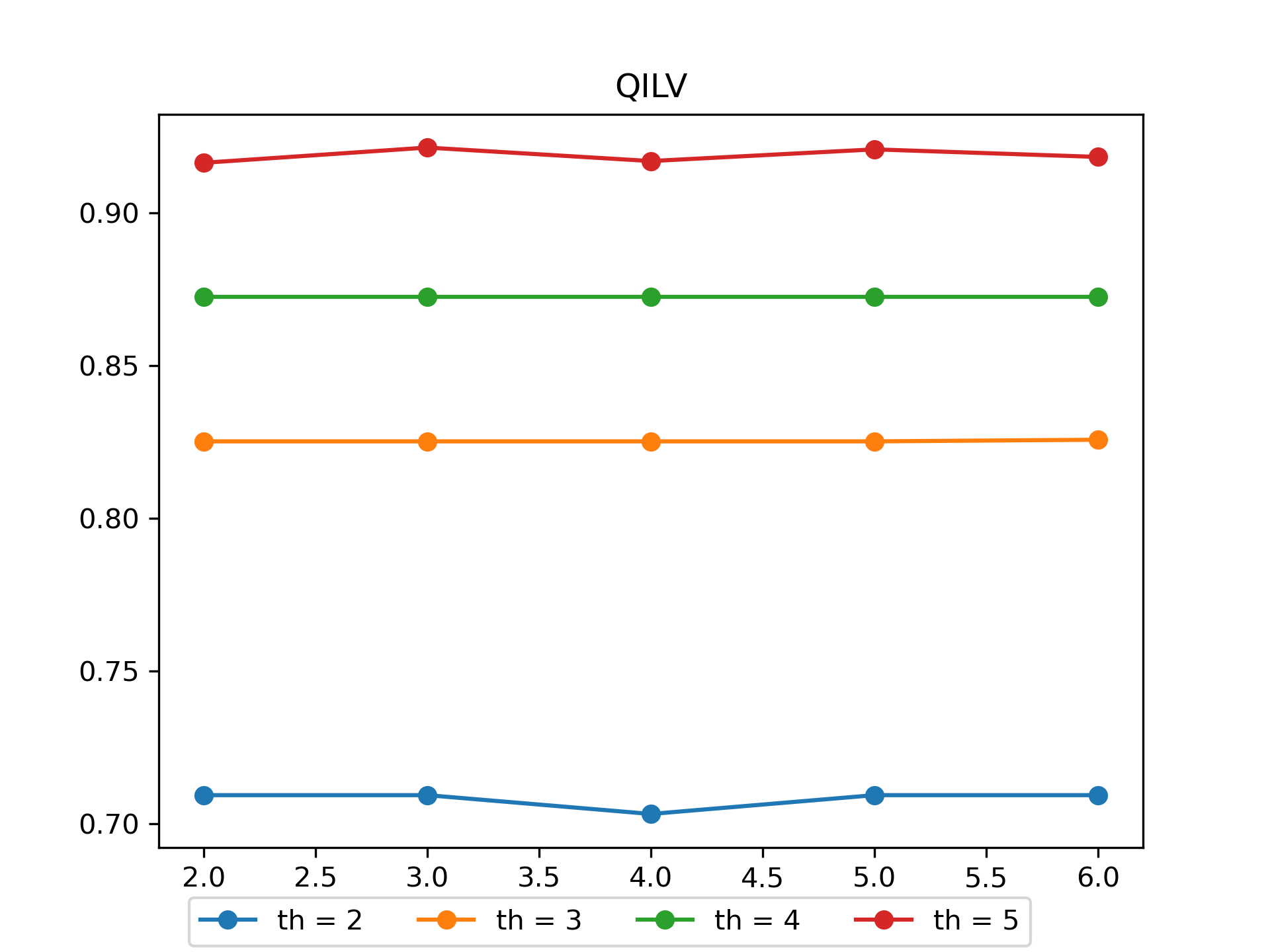}}\;\;
    \subfloat[Brainweb HPSI]{\includegraphics[scale=0.3]{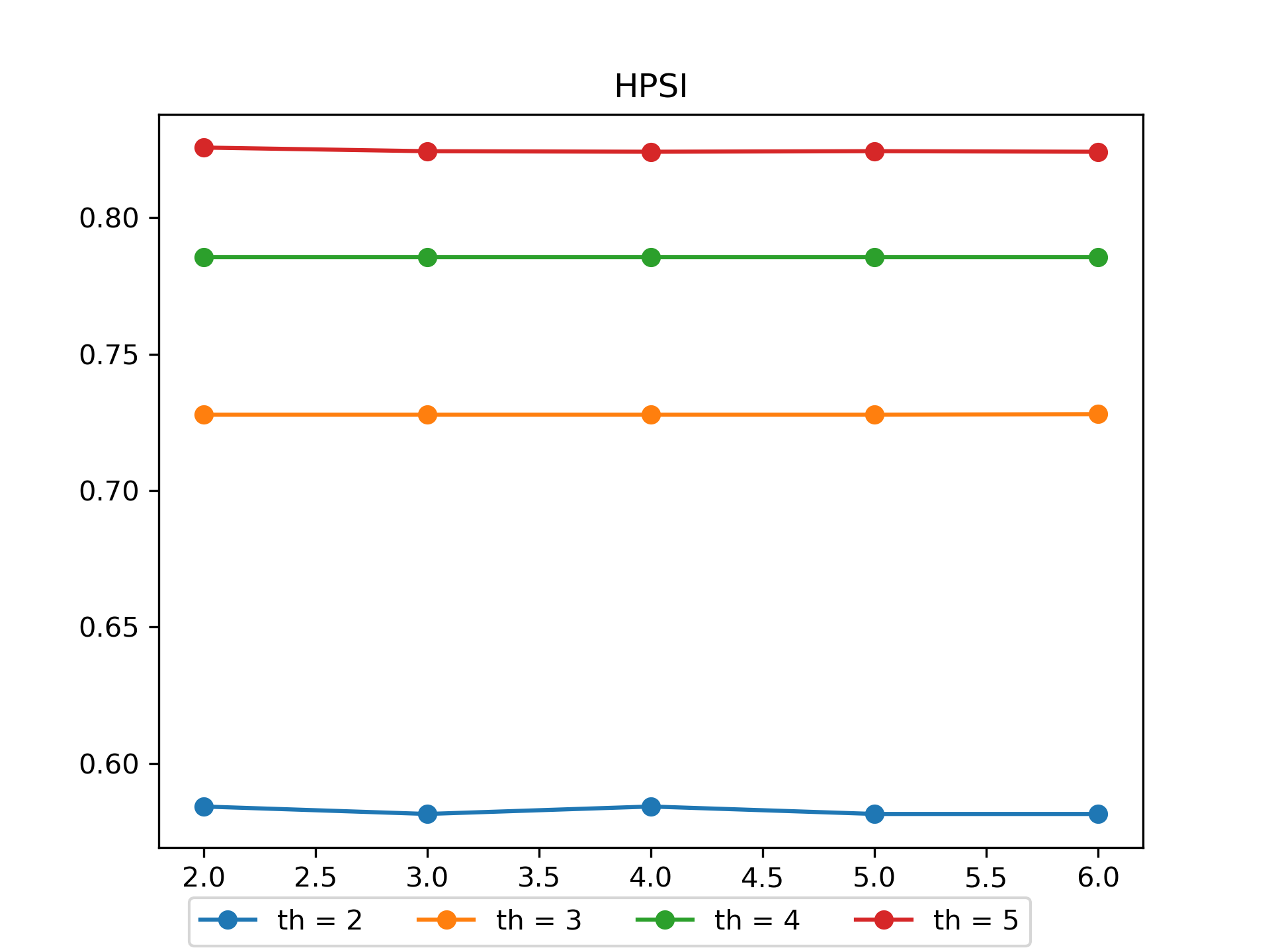}}
    \caption{Variation of performance with change in altruism in HHO (population is set to 50 in each case): (a)-(f) in the Harvard WBA dataset; (g)-(l) in the Brainweb dataset.}
    \label{alt}
\end{figure*}

\autoref{pop} shows the variation in performance of the proposed framework with respect to the population size in the HHO algorithm. A population size of 50 gives consistently better results than the others, and hence 50 is chosen as the population in the algorithm framework.

\autoref{alt} shows the variation in performance of the proposed framework with respect to the number of altruistic individuals in the population of the HHO algorithm. Altruism when set to 4 (8\% of population size) gives consistently better results than values higher or lower than 4, and hence it is chosen as the final altruism number in the algorithm for the final segmentation.

\subsection{Comparison to state-of-the-art}

\begin{table*}[]
\centering
\caption{Comparison of the proposed method with other evolutionary algorithms for multi-level thresholding on the Harvard WBA database: the mean results from all 10 images are shown. The values in bold correspond to the best results obtained for the experiment.}
\label{comp_harv}
\begin{tabular}{|c|c|c|c|c|c|c|c|}
\hline
\textbf{Algorithm} &
  \textbf{No. of Thresholds} &
  \textbf{PSNR} &
  \textbf{SSIM} &
  \textbf{FSIM} &
  \textbf{UIQI} &
  \textbf{QILV} &
  \textbf{HPSI} \\ \hline
 &
  2 &
  20.3954 &
  0.8126 &
  0.7629 &
  0.3741 &
  0.5691 &
  0.5619 \\ \cline{2-8} 
 &
  3 &
  22.5337 &
  0.8638 &
  0.8297 &
  0.4967 &
  0.7343 &
  0.7000 \\ \cline{2-8} 
 &
  4 &
  24.2621 &
  0.8964 &
  0.8700 &
  0.5839 &
  0.8190 &
  0.7745 \\ \cline{2-8} 
\multirow{-4}{*}{HHO \cite{heidari2019harris}} &
  5 &
  25.6072 &
  0.9196 &
  0.8994 &
  0.6503 &
  0.8638 &
  0.8312 \\ \hline
 &
  2 &
  17.4848 &
  0.7227 &
  0.6675 &
  0.4177 &
  0.3336 &
  0.2615 \\ \cline{2-8} 
 &
  3 &
  20.3949 &
  0.8122 &
  0.7620 &
  0.5599 &
  0.5691 &
  0.3743 \\ \cline{2-8} 
 &
  4 &
  22.5438 &
  0.8639 &
  0.8289 &
  0.6992 &
  0.7366 &
  0.4986 \\ \cline{2-8} 
\multirow{-4}{*}{LSHADE \cite{piotrowski2018shade}} &
  5 &
  24.3039 &
  0.8973 &
  0.8699 &
  0.7746 &
  0.8194 &
  0.5851 \\ \hline
 &
  2 &
  17.4334 &
  0.7217 &
  0.6648 &
  0.4170 &
  0.3292 &
  0.2610 \\ \cline{2-8} 
 &
  3 &
  20.3365 &
  0.8117 &
  0.7586 &
  0.5588 &
  0.5655 &
  0.3736 \\ \cline{2-8} 
 &
  4 &
  22.5043 &
  0.8633 &
  0.8272 &
  0.6971 &
  0.7344 &
  0.4975 \\ \cline{2-8} 
\multirow{-4}{*}{SADE \cite{qin2008differential}} &
  5 &
  24.2256 &
  0.8965 &
  0.8661 &
  0.7702 &
  0.8111 &
  0.5771 \\ \hline
 &
  2 &
  17.4246 &
  0.7222 &
  0.6671 &
  0.4132 &
  0.3283 &
  0.2608 \\ \cline{2-8} 
 &
  3 &
  20.3294 &
  0.8115 &
  0.7612 &
  0.5584 &
  0.5644 &
  0.3726 \\ \cline{2-8} 
 &
  4 &
  22.4584 &
  0.8631 &
  0.8281 &
  0.6965 &
  0.7319 &
  0.4958 \\ \cline{2-8} 
\multirow{-4}{*}{DE \cite{das2010differential}} &
  5 &
  24.2290 &
  0.8957 &
  0.8669 &
  0.7676 &
  0.8158 &
  0.5790 \\ \hline
 &
  2 &
  17.4201 &
  0.7217 &
  0.6662 &
  0.4156 &
  0.3278 &
  0.2604 \\ \cline{2-8} 
 &
  3 &
  20.3303 &
  0.8112 &
  0.7600 &
  0.5583 &
  0.5620 &
  0.3708 \\ \cline{2-8} 
 &
  4 &
  22.4765 &
  0.8617 &
  0.8266 &
  0.6952 &
  0.7126 &
  0.4893 \\ \cline{2-8} 
\multirow{-4}{*}{PSO \cite{marini2015particle}} &
  5 &
  24.2359 &
  0.8937 &
  0.8672 &
  0.7664 &
  0.7832 &
  0.5731 \\ \hline
 &
  2 &
  17.4234 &
  0.7220 &
  0.6672 &
  0.4173 &
  0.3285 &
  0.2608 \\ \cline{2-8} 
 &
  3 &
  20.3357 &
  0.8118 &
  0.7612 &
  0.5583 &
  0.5644 &
  0.3719 \\ \cline{2-8} 
 &
  4 &
  22.4405 &
  0.8622 &
  0.8272 &
  0.6928 &
  0.7313 &
  0.4938 \\ \cline{2-8} 
\multirow{-4}{*}{GWO \cite{mirjalili2014grey}} &
  5 &
  24.0956 &
  0.8939 &
  0.8644 &
  0.7634 &
  0.8092 &
  0.5719 \\ \hline
 &
  2 &
  17.2839 &
  0.7003 &
  0.6436 &
  0.3400 &
  0.2781 &
  0.2315 \\ \cline{2-8} 
 &
  3 &
  19.8034 &
  0.7694 &
  0.7108 &
  0.4587 &
  0.4467 &
  0.3327 \\ \cline{2-8} 
 &
  4 &
  21.1309 &
  0.8132 &
  0.7582 &
  0.5444 &
  0.5693 &
  0.4121 \\ \cline{2-8} 
\multirow{-4}{*}{SCA \cite{qu2018modified}} &
  5 &
  21.9658 &
  0.8387 &
  0.7892 &
  0.6008 &
  0.6478 &
  0.4662 \\ \hline
 
 &
  2 &
  20.3532 &
  0.8127 &
  0.7632 &
  0.3741 &
  0.5656 &
  0.5621 \\ \cline{2-8} 
 
 &
  3 &
  22.5069 &
  0.8637 &
  0.8297 &
  0.4982 &
  0.7339 &
  0.7014 \\ \cline{2-8} 
 
 &
  4 &
  24.1624 &
  0.8968 &
  0.8679 &
  0.5755 &
  0.8092 &
  0.7684 \\ \cline{2-8} 
 
\multirow{-4}{*}{WOA \cite{mirjalili2016whale}} &
  5 &
  25.6927 &
  0.9202 &
  0.8996 &
  0.6516 &
  0.8658 &
  0.8319 \\ \hline
 
 &
  2 &
  20.3591 &
  0.8123 &
  0.7625 &
  0.3738 &
  0.5669 &
  0.5615 \\ \cline{2-8} 
 
 &
  3 &
  22.5495 &
  0.8642 &
  0.8304 &
  0.4975 &
  0.7348 &
  0.7010 \\ \cline{2-8} 
 
 &
  4 &
  24.2011 &
  0.8966 &
  0.8678 &
  0.5773 &
  0.8105 &
  0.7714 \\ \cline{2-8} 
 
\multirow{-4}{*}{ALO \cite{mirjalili2015ant}} &
  5 &
  25.5824 &
  0.9192 &
  0.8992 &
  0.6512 &
  0.8634 &
  0.8326 \\ \hline
 
 &
  2 &
  20.3622 &
  0.8121 &
  0.7632 &
  0.3736 &
  0.5671 &
  0.5616 \\ \cline{2-8} 
 
 &
  3 &
  22.5093 &
  0.8642 &
  0.8302 &
  0.4978 &
  0.7332 &
  0.7010 \\ \cline{2-8} 
 
 &
  4 &
  24.2694 &
  0.8965 &
  0.8686 &
  0.5801 &
  0.8153 &
  0.7725 \\ \cline{2-8} 
 
\multirow{-4}{*}{MFO \cite{mirjalili2015moth}} &
  5 &
  25.5434 &
  0.9188 &
  0.8988 &
  0.6512 &
  0.8628 &
  0.8307 \\ \hline
 
 &
  2 &
  20.3491 &
  0.8124 &
  0.7637 &
  0.3732 &
  0.5655 &
  0.5621 \\ \cline{2-8} 
 
 &
  3 &
  22.4844 &
  0.8633 &
  0.8300 &
  0.4948 &
  0.7328 &
  0.6995 \\ \cline{2-8} 
 
 &
  4 &
  24.2783 &
  0.8969 &
  0.8693 &
  0.5813 &
  0.8174 &
  0.7737 \\ \cline{2-8} 
 
\multirow{-4}{*}{DA \cite{mirjalili2016dragonfly}} &
  5 &
  25.5534 &
  0.9187 &
  0.8990 &
  0.6480 &
  0.8679 &
  0.8327 \\ \hline
 &
  \textbf{2} &
  \textbf{21.6240} &
  \textbf{0.8256} &
  \textbf{0.7977} &
  \textbf{0.4117} &
  \textbf{0.7223} &
  \textbf{0.6364} \\ \cline{2-8} 
 &
  \textbf{3} &
  \textbf{23.7769} &
  \textbf{0.8722} &
  \textbf{0.8531} &
  \textbf{0.5091} &
  \textbf{0.8220} &
  \textbf{0.7399} \\ \cline{2-8} 
 &
  \textbf{4} &
  \textbf{25.3682} &
  \textbf{0.9022} &
  \textbf{0.8902} &
  \textbf{0.5894} &
  \textbf{0.8784} &
  \textbf{0.8145} \\ \cline{2-8} 
\multirow{-4}{*}{\textbf{Proposed}} &
  \textbf{5} &
  \textbf{26.6125} &
  \textbf{0.9197} &
  \textbf{0.9155} &
  \textbf{0.6515} &
  \textbf{0.9134} &
  \textbf{0.8570} \\ \hline
\end{tabular}
\end{table*}

\begin{table*}[]
\centering
\caption{Comparison of the proposed method with other evolutionary algorithms for multi-level thresholding on the Brainweb database: the mean results from all 8 images used are shown. The values in bold correspond to the best results obtained for the experiment.}
\label{comp_mri}
\begin{tabular}{|c|c|c|c|c|c|c|c|}
\hline
\textbf{Algorithm} &
  \textbf{No. of Thresholds} &
  \textbf{PSNR} &
  \textbf{SSIM} &
  \textbf{FSIM} &
  \textbf{UIQI} &
  \textbf{QILV} &
  \textbf{HPSI} \\ \hline
 &
  2 &
  18.8024 &
  0.7158 &
  0.7553 &
  0.2720 &
  0.6109 &
  0.5665 \\ \cline{2-8} 
 &
  3 &
  20.4756 &
  0.7719 &
  0.8151 &
  0.3394 &
  0.7542 &
  0.7001 \\ \cline{2-8} 
 &
  4 &
  22.3962 &
  0.8236 &
  0.8591 &
  0.4125 &
  0.8188 &
  0.7720 \\ \cline{2-8} 
\multirow{-4}{*}{HHO \cite{heidari2019harris}} &
  5 &
  23.5675 &
  0.8551 &
  0.8857 &
  0.4651 &
  0.8520 &
  0.8161 \\ \hline
 &
  2 &
  18.8024 &
  0.7158 &
  0.7553 &
  0.2720 &
  0.6109 &
  0.5665 \\ \cline{2-8} 
 &
  3 &
  20.4756 &
  0.7703 &
  0.8151 &
  0.3389 &
  0.7542 &
  0.7001 \\ \cline{2-8} 
 &
  4 &
  22.2031 &
  0.8244 &
  0.8560 &
  0.4075 &
  0.8102 &
  0.7685 \\ \cline{2-8} 
\multirow{-4}{*}{LSHADE \cite{piotrowski2018shade}} &
  5 &
  23.5619 &
  0.8549 &
  0.8856 &
  0.4649 &
  0.8519 &
  0.8161 \\ \hline
 &
  2 &
  18.7424 &
  0.7058 &
  0.7253 &
  0.2520 &
  0.5909 &
  0.5465 \\ \cline{2-8} 
 &
  3 &
  20.4156 &
  0.7503 &
  0.7851 &
  0.3189 &
  0.7242 &
  0.6801 \\ \cline{2-8} 
 &
  4 &
  22.1531 &
  0.7944 &
  0.8360 &
  0.3875 &
  0.7802 &
  0.7485 \\ \cline{2-8} 
\multirow{-4}{*}{SADE \cite{qin2008differential}} &
  5 &
  23.5219 &
  0.8249 &
  0.8656 &
  0.4549 &
  0.8319 &
  0.7861 \\ \hline
 &
  2 &
  18.7524 &
  0.6958 &
  0.7253 &
  0.2520 &
  0.5909 &
  0.5365 \\ \cline{2-8} 
 &
  3 &
  20.4356 &
  0.7503 &
  0.7951 &
  0.3189 &
  0.7342 &
  0.6801 \\ \cline{2-8} 
 &
  4 &
  22.1231 &
  0.8044 &
  0.8360 &
  0.3875 &
  0.7802 &
  0.7485 \\ \cline{2-8} 
\multirow{-4}{*}{DE \cite{das2010differential}} &
  5 &
  23.4819 &
  0.8449 &
  0.8556 &
  0.4349 &
  0.8219 &
  0.7961 \\ \hline
 &
  2 &
  18.6986 &
  0.7058 &
  0.7319 &
  0.2620 &
  0.6009 &
  0.5297 \\ \cline{2-8} 
 &
  3 &
  20.3595 &
  0.7469 &
  0.8051 &
  0.3289 &
  0.7308 &
  0.6901 \\ \cline{2-8} 
 &
  4 &
  22.1239 &
  0.7876 &
  0.8326 &
  0.3707 &
  0.7734 &
  0.7585 \\ \cline{2-8} 
\multirow{-4}{*}{PSO \cite{marini2015particle}} &
  5 &
  23.4704 &
  0.8449 &
  0.8488 &
  0.4549 &
  0.8419 &
  0.7927 \\ \hline
 &
  2 &
  18.7411 &
  0.6777 &
  0.7306 &
  0.2607 &
  0.5996 &
  0.5552 \\ \cline{2-8} 
 &
  3 &
  20.3897 &
  0.7322 &
  0.7904 &
  0.3142 &
  0.7161 &
  0.6754 \\ \cline{2-8} 
 &
  4 &
  22.1295 &
  0.8131 &
  0.8313 &
  0.3962 &
  0.7989 &
  0.7572 \\ \cline{2-8} 
\multirow{-4}{*}{GWO \cite{mirjalili2014grey}} &
  5 &
  23.5129 &
  0.8436 &
  0.8409 &
  0.4402 &
  0.8272 &
  0.7780 \\ \hline
 &
  2 &
  18.5496 &
  0.6377 &
  0.7240 &
  0.1939 &
  0.5796 &
  0.5118 \\ \cline{2-8} 
 &
  3 &
  20.3566 &
  0.7390 &
  0.7370 &
  0.2842 &
  0.6995 &
  0.6454 \\ \cline{2-8} 
 &
  4 &
  21.9280 &
  0.7931 &
  0.8247 &
  0.3294 &
  0.7789 &
  0.7138 \\ \cline{2-8} 
\multirow{-4}{*}{SCA \cite{qu2018modified}} &
  5 &
  23.3760 &
  0.8236 &
  0.8543 &
  0.4336 &
  0.7972 &
  0.7380 \\ \hline
 
 &
  2 &
  18.8024 &
  0.7158 &
  0.7553 &
  0.2720 &
  0.6109 &
  0.5665 \\ \cline{2-8} 
 
 &
  3 &
  20.4758 &
  0.7703 &
  0.8151 &
  0.3389 &
  0.7544 &
  0.7001 \\ \cline{2-8} 
 
 &
  4 &
  22.3490 &
  0.8244 &
  0.8584 &
  0.4111 &
  0.8155 &
  0.7714 \\ \cline{2-8} 
 
\multirow{-4}{*}{WOA \cite{mirjalili2016whale}} &
  5 &
  23.5825 &
  0.8549 &
  0.8866 &
  0.4659 &
  0.8527 &
  0.8161 \\ \hline
 
 &
  2 &
  18.8024 &
  0.7158 &
  0.7553 &
  0.2720 &
  0.6109 &
  0.5665 \\ \cline{2-8} 
 
 &
  3 &
  20.4756 &
  0.7703 &
  0.8151 &
  0.3389 &
  0.7542 &
  0.7001 \\ \cline{2-8} 
 
 &
  4 &
  22.2978 &
  0.8242 &
  0.8577 &
  0.4099 &
  0.8139 &
  0.7705 \\ \cline{2-8} 
 
\multirow{-4}{*}{ALO \cite{mirjalili2015ant}} &
  5 &
  23.5132 &
  0.8549 &
  0.8858 &
  0.4649 &
  0.8505 &
  0.8149 \\ \hline
 
 &
  2 &
  18.8024 &
  0.7158 &
  0.7553 &
  0.2720 &
  0.6109 &
  0.5665 \\ \cline{2-8} 
 
 &
  3 &
  20.4756 &
  0.7703 &
  0.8151 &
  0.3389 &
  0.7542 &
  0.7001 \\ \cline{2-8} 
 
 &
  4 &
  22.2978 &
  0.8242 &
  0.8577 &
  0.4099 &
  0.8139 &
  0.7705 \\ \cline{2-8} 
 
\multirow{-4}{*}{MFO \cite{mirjalili2015moth}} &
  5 &
  23.5305 &
  0.8539 &
  0.8865 &
  0.4617 &
  0.8521 &
  0.8147 \\ \hline
 
 &
  2 &
  18.8040 &
  0.7158 &
  0.7553 &
  0.2720 &
  0.6115 &
  0.5666 \\ \cline{2-8} 
 
 &
  3 &
  20.5197 &
  0.7709 &
  0.8157 &
  0.3392 &
  0.7589 &
  0.7019 \\ \cline{2-8} 
 
 &
  4 &
  22.2008 &
  0.8216 &
  0.8551 &
  0.4033 &
  0.8155 &
  0.7632 \\ \cline{2-8} 
 
\multirow{-4}{*}{DA \cite{mirjalili2016dragonfly}} &
  5 &
  23.7499 &
  0.8558 &
  0.8882 &
  0.4688 &
  0.8611 &
  0.8207 \\ \hline
 &
  \textbf{2} &
  \textbf{19.6240} &
  \textbf{0.7176} &
  \textbf{0.7652} &
  \textbf{0.2711} &
  \textbf{0.7098} &
  \textbf{0.5818} \\ \cline{2-8} 
 &
  \textbf{3} &
  \textbf{21.8767} &
  \textbf{0.7913} &
  \textbf{0.8368} &
  \textbf{0.3549} &
  \textbf{0.8258} &
  \textbf{0.7285} \\ \cline{2-8} 
 &
  \textbf{4} &
  \textbf{23.5343} &
  \textbf{0.8251} &
  \textbf{0.8741} &
  \textbf{0.4237} &
  \textbf{0.8725} &
  \textbf{0.7857} \\ \cline{2-8} 
\multirow{-4}{*}{\textbf{Proposed}} &
  \textbf{5} &
  \textbf{24.7686} &
  \textbf{0.8608} &
  \textbf{0.8959} &
  \textbf{0.4670} &
  \textbf{0.9192} &
  \textbf{0.8228} \\ \hline
\end{tabular}
\end{table*}

The following algorithms have been used to compare the results of the proposed method for brain MRI segmentation:

\begin{enumerate}
    \item HHO (basic version) by \cite{heidari2019harris}. \textcolor{black}{The algorithm is inspired by the cooperative manner in which the Harris Hawks hunt in nature. They perform the surprise pounce to engulf the prey. Myriad chasing patterns can be emulated by this algorithm as discussed in our work.}
    \item L-SHADE by Piotrowski et al. \cite{piotrowski2018shade} and utilized for brain MRI segmentation by Aranguren et al. \cite{aranguren2021improving} \textcolor{black}{The main advantage of using L-SHADE as a meta-heuristic is that the parameters internal to the algorithm are tuned according to the information that is acquired during the process of evolution. The latter has used LSHADE to perform multi-level thresholding, using the minimum cross-entropy function as the objective function. }
    \item Differential Evolution (DE) by Price et al. \cite{das2010differential}. \textcolor{black}{DE is a powerful meta-heuristic and its application to multi-objective and large scale problems have been demonstrated.}
    \item Self-Adaptive Differential Evolution (SADE) by \cite{qin2008differential}. SADE \textcolor{black}{The trial vector generation techniques, as well as the control values of parameters, are adapted by the algorithm itself using its past experiences. The performance is evaluated on constrained optimization problems, and comparison has been drawn with the other variants of adaptive DE}
    \item Particle Swarm Optimization (PSO) by Marini et al. \cite{marini2015particle} \textcolor{black}{ The solution is taken to be a swarm of particles. The prowess of the algorithm has been demonstrated in solving optimization problems, particularly in chemometrics. }
    \item Grey Wolf Optimization (GWO) by Mirjalili et al. \cite{mirjalili2014grey} \textcolor{black}{ It imitates how the grey wolves hunt in nature. It performs well on the unimodal functions. And the good results on the multimodal functions support the exploration capability of the GWO. It yields superior results on real problems as well }
    \item Sine-Cosine Algorithm (SCA) by Mirjalili et al. \cite{mirjalili2016sca}. \textcolor{black}{Initially, random solutions are there and gradually they fluctuate away from or move towards the optimal solution according to a pre-defined mathematical model that deploys sine and cosine functions. The method is tested on unimodal, multi-modal and composite functions, and, on two-dimensional test functions. }
    \item Whale Optimization Algorithm (WOA) by Mirjalili et al. \cite{mirjalili2016whale}. \textcolor{black}{It is primarily encouraged by the humpback whales, and specifically the bubble-net hunting approach. It is tested on some of the benchmarked test functions, as well as on unimodal and multimodal functions.}
    \item Ant Lion Optimizer (ALO) by Mirjalili et al. \cite{mirjalili2015ant} \textcolor{black}{It is encouraged by how the antlions hunt and is tested on unimodal, multimodal, and benchmarked test functions as well as real problems. }
    \item Moth-Flame Optimization Algorithm (MFO) by Mirjalili et al. \cite{mirjalili2015moth} \textcolor{black}{It is influenced by how moths navigate, and specifically, the transverse orientation. It is evaluated on the benchmarked problems, as well as some real engineering problems and delivers superior results. The algorithm has been considered for application in the domain of marine propeller design.}
    \item Dragonfly Algorithm by Mirjalili (DA) by Mirjalili et al. \cite{mirjalili2016dragonfly} \textcolor{black}{It is encouraged by how the dragonflies in nature swarm, namely, static and dynamic. Binary and multi-objective models of DA have been considered. Both of these have been tested on a real case study and myriad test functions.  }
\end{enumerate}
\par

 \textcolor{black}{For the algorithms, including the proposed algorithm, which require us to set the population size, the population size is set at 50. The maximum number of iterations is set at 1000 for all of the algorithms, the lower bound is set at 1 and the upper bound at 256. }\textcolor{black}{The parameters set for the state of the art algorithms are tabulated in} \autoref{param_settings}. \textcolor{black}{The names of the parameters have been taken from the commonly used literature. Apart from this, any other parameters not mentioned explicitly in} \autoref{param_settings} \textcolor{black}{are also taken up directly from the established literature.}

\begin{table}[]
\centering
\caption{\textcolor{black}{Parameter tuning settings of the other state-of-the-art algorithms}}
\label{param_settings}
\begin{tabular}{|
>{\columncolor[HTML]{FFFFFF}}c |
>{\columncolor[HTML]{FFFFFF}}c |}
\hline
\textbf{Algorithm}                               & \textbf{Parameters}                         \\ \hline
SADE                                             & $learn\_generations = 50$                   \\ \hline
\cellcolor[HTML]{FFFFFF}                         & $rate\_crossover = 0.50$                    \\ \cline{2-2} 
\multirow{-2}{*}{\cellcolor[HTML]{FFFFFF}DE}     & $value\_scale_factor = 0.20$                \\ \hline
\cellcolor[HTML]{FFFFFF}                         & $init\_population\_size = 50$               \\ \cline{2-2} 
\cellcolor[HTML]{FFFFFF}                         & $final\_population\_size = 40$              \\ \cline{2-2} 
\multirow{-3}{*}{\cellcolor[HTML]{FFFFFF}LSHADE} & $memory\_size = 5$                          \\ \hline
GWO                                              & $a$ is in range $[2, 0]$                    \\ \hline
PSO                                              & $value\_weight$ is in range $[1, 0]$        \\ \hline
WOA                                              & $a$ is in range $[2, 0]$                    \\ \hline
\cellcolor[HTML]{FFFFFF}                         & $a = 3$                                     \\ \cline{2-2} 
\multirow{-2}{*}{\cellcolor[HTML]{FFFFFF}SCA}    & $value\_r1$ is in range $[1, 0]$            \\ \hline
\cellcolor[HTML]{FFFFFF}                         & $convergence\_const$ is in range $[-1, -2]$ \\ \cline{2-2} 
\multirow{-2}{*}{\cellcolor[HTML]{FFFFFF}MFO}    & $value\_spiral\_factor = 1$                 \\ \hline
ALO                                              & No additional parameters                    \\ \hline
DA                                               & No additional parameters                    \\ \hline
\end{tabular}
\end{table}

The comparative results are shown in \autoref{comp_harv} for the images taken from Harvard WBA database images, and in \autoref{comp_mri} for the images taken from the Brainweb dataset. As evident from the comparative results, the proposed approach outperforms the other evolutionary metaheuristic algorithms by a significant margin. The values shown in the table are determined by averaging the metrics we get by running the algorithms over all the images of each dataset. It is seen that for each of the metrics, we get a significant increment using Altruism with the HHO algorithm. It is observed that for the images in the Harvard WBA database, the performance delivered by WOA is closest to the proposed approach. HHO, ALO, MFO and DA yielded values that were somewhat close to the WOA. However, there is a clear gap between the values yielded by these algorithms and the proposed method. For the Brainweb dataset, it is observed that the performance of the DA is closest to the proposed method. Some other algorithms such as HHO, L-SHADE and WOA also yield values close to the values obtained by the DA. However, the proposed algorithm outperforms these methods by a significant margin.

\subsection{Findings of the work:}

\textcolor{black}{The findings of the work can be summarized as:}
\begin{enumerate}
    \item \textcolor{black}{Evaluation of the proposed algorithm on MRI images from the WBA database developed by Harvard Medical School and the Brainweb database.}
    \item \textcolor{black}{Use of six metrics for evaluating the performance which are: PSNR, SSIM, FSIM, HPSI, QILV, UIQI. }
    \item \textcolor{black}{Performed experiments using different threshold values for multi-level thresholding. }
    \item \textcolor{black}{Output images are provided to show the qualitative results on both of the datasets. The original image is placed with the thresholded images for viewing the difference between the images. }
    \item \textcolor{black}{Compared the performances of the various chaotic maps and selected the Logistic chaotic map which delivers the best results amongst the chaotic maps.} 
    \item \textcolor{black}{Evaluated the performance of the proposed algorithm using various objective functions and demonstrated the superior performance of the proposed objective function. }
    \item \textcolor{black}{Tuned the hyperparameters of the algorithm such as the population size and value of the parameter in altruism to get the best results.} 
    \item \textcolor{black}{ Compared the performance of the proposed algorithm with other meta-heuristic algorithms such as L-SHADE, SADE, DE, PSO, GWO, SCA, WOA, ALO, MFO and DA to justify the good performance of the proposed algorithm.} 
\end{enumerate}

\section{Conclusions}\label{conclusions}
Segmentation is an important preprocessing step in medical image analysis. Brain MRI segmentation is essential for disease identification, lesion delineation and better visualization of morphological changes. In this research, we explore the unsupervised segmentation of brain MRI images using a multi-level thresholding approach. Classical methods like Otsu's thresholding, or Kapur's entropy-based methods are inefficient for segmentation using a large number of thresholds, thus we explore evolutionary optimization algorithms to approach this problem. For this, we use an improved Harris Hawks Optimization algorithm wherein we embed the concept of altruism and use a hybrid of two objective functions. As evident from the results, HHO incorporated with altruism to enhance its exploitation capability,  chaotic initialization to increase the diversity of the solutions and the hybrid objective function, yields results whose performance metrics are significantly better than the meta-heuristics that have been used earlier. The limitations of the feature selection algorithm have been mitigated to some extent by integrating altruism and a Logistic chaotic map. We evaluate the proposed algorithm on T2-weighted brain MRIs extracted from two popular publicly available databases: the Harvard WBA and the Brainweb datasets. Upon comparison, it is established that the proposed framework outperforms state-of-the-art methods on the same datasets, thus justifying the effectiveness of the model.

The main limitation of the proposed algorithm is that it consumes more time compared to other meta-heuristics due to the time complexity of the HHO algorithm and the altruism integrated into it. A balance can however be established between the time taken by the algorithm and the values of the performance metrics by decreasing the number of search agents or decreasing the number of iterations inside the feature selection algorithm. Future work will focus on overcoming these limitations and further increase the exploitation capability of the algorithm.

\section*{Acknowledgements}
The authors would like to thank the Centre for Microprocessor Applications for Training, Education and Research (CMATER) research laboratory of the Computer Science and Engineering Department, Jadavpur University, Kolkata, India for providing the infrastructural support.

\section*{Conflict of interest}
All the authors declare that there is no conflict of interest.

\bibliography{References}

\end{document}